\documentclass[twocolumn]{aastex631}

\usepackage{float}
\usepackage{multirow}
\usepackage{threeparttable}
\usepackage{tabularx}
\usepackage{amsmath}

\shorttitle{PKS 1510-089: I. Multi-Wavelength Variability}
\shortauthors{Amador-Portes et al.}
\graphicspath{{./}{figures/}}
\begin{document}

\title{Unveiling the Emission Mechanisms of Blazar PKS 1510-089: I. Multi-Wavelength Variability}

\correspondingauthor{Victor M. Patiño-Alvarez}
\email{alfre\_portess97@hotmail.com,victorm.patinoa@gmail.com}

\author[0009-0009-6341-0270]{Alfredo Amador-Portes}
\affiliation{Instituto Nacional de Astrofísica, Óptica y Electrónica, Luis Enrique Erro \#1, Tonantzintla Puebla, México, C.P. 72840}

\author[0000-0002-9896-6430]{Abigail García-Pérez}
\affiliation{Instituto Nacional de Astrofísica, Óptica y Electrónica, Luis Enrique Erro \#1, Tonantzintla Puebla, México, C.P. 72840}
\affiliation{Dipartimento di Fisica, Università degli Studi di Torino, via Pietro Giuria 1, I-10125 Torino, Italy}

\author[0000-0002-2558-0967]{Vahram Chavushyan}
\affiliation{Instituto Nacional de Astrofísica, Óptica y Electrónica, Luis Enrique Erro \#1, Tonantzintla Puebla, México, C.P. 72840}

\author[0000-0002-5442-818X]{Victor M. Patiño-Alvarez}
\affiliation{Instituto Nacional de Astrofísica, Óptica y Electrónica, Luis Enrique Erro \#1, Tonantzintla Puebla, México, C.P. 72840}
\affiliation{Max-Planck-Institut für Radioastronomie, Auf dem Hügel 69, D-53121 Bonn, Germany}

\begin{abstract}
The flat spectrum radio quasar PKS 1510-089 is one of the most active blazars in $\gamma$-rays, exhibiting phases of very high activity. This study investigates its variability over a decade across a wide range of wavelengths, from radio to $\gamma$-rays. Utilizing the non-thermal dominance parameter, we analyze the H$\beta$, H$\gamma$, and $\lambda5100\text{ \AA}$ continuum light curves to discern the primary source of continuum emission, either from the accretion disk or the jet, during different activity phases. Our findings underscore the dominance of jet emission in the continuum during flare-like events. We observed an approximately 80-day delay between the H$\beta$ and continuum emissions, which we attribute to the spatial separation between the optical emission zone and the broad-line region. Near-zero delays between optical and near-infrared emissions suggest that the emitting regions within the jet are co-spatial. Synchrotron self-Compton was identified as the primary mechanism for $\gamma$-ray emission during flares, supported by the minimal delay observed between optical/near-infrared emissions and $\gamma$-rays. Additionally, we found a delay of about 60 days between the leading optical/near-infrared emissions and X-rays, indicating that inverse Compton scattering within the jet predominantly drives X-ray emission. However, distinguishing between synchrotron self-Compton and external inverse Compton mechanism was not feasible. Shifts in the spectral index across the 15-230 GHz range corresponded with ejections from the radio core, suggesting changes in the physical conditions of the jet.
\end{abstract}


\section{Introduction}\label{sec:intro}
Blazars are a class of radio-loud active galactic nuclei (AGNs) characterized by relativistic jets oriented close to our line of sight \citep{UrryAndPadovani1995}, which leads to Doppler boosting and relativistic amplification of the flux as observed from the earth \citep{Sher1968}. They exhibit substantial variability across all wavelengths on a wide range of timescales \citep{Fan2018, Gupta2018}. Flat-spectrum radio quasars (FSRQs), a subgroup of blazars, demonstrate variability classified into intra-day variability (IDV), short-term (days to weeks), and long-term (months to years). FSRQs typically feature prominent emission lines in their optical spectra. Their spectral energy distribution (SED) shows two peaks at distinct frequencies: a low-energy peak attributed to synchrotron emission from the jet and thermal emission from the accretion disk, spanning from radio to X-rays. The high-energy peak results from inverse Compton (IC) scattering of low-energy photons, ranging from X-rays to $\gamma$-rays (e.g. \citealp{Bottcher2007, Bottcher2013, Romero2017}). The origin of seed photons for IC scattering varies: if they are of synchrotron origin, the process is known as synchrotron-self Compton (SSC; \citealp{Maraschi1992}). Alternatively, if seed photons originate from sources external to the jet, such as the accretion disk, the broad-line region (BLR), or the dusty torus, the process is termed external inverse Compton (EC; \citealp{Sikora1994}).

The object of interest, PKS 1510-089, (redshift $z=0.361$; \citealp{BurbidgeAndKinman1966}), is classified as an FSRQ due to its prominent broad emission lines in the optical spectrum \citep{Tadhunter1993} and its high variability across the entire electromagnetic spectrum. This variability has made it the target of numerous multi-wavelength observing campaigns (e.g. \citealp{Marscher2010, Rani2010, Aleksic2014, Fuhrmann2016, Prince2019, Yuan2023}). PKS 1510-089 was initially detected in high energies (HE, $E>100$ MeV) by \citet{Hartman1999} and in very high energies (VHE, $E>100$ GeV) by the \citet{HESS2013}, with persistent VHE emission reported by the \citet{MAGIC2018} and \citet{Dzhatdoev2022} spanning the period from 2008 to 2020. The angle between the relativistic jet of PKS 1510-089 and our line of sight is approximately 3 degrees \citep{Homan2002}, enabling the observation of knots with apparent velocities of up to 20 times the speed of light ($20c$) along the jet \citep{Jorstad2005}. For instance, \citet{Orienti2011}, using Very Long Baseline Array (VLBA) observations, concluded that superluminal knots were ejected from the core of PKS 1510-089 approximately every year, at least from 1995 to 2010.

Correlated variability between multi-waveband flux, linear polarization, and parsec-scale structure has facilitated the modeling of jet kinematics. \citet{Marscher2010} proposed a spine-sheath structure for the jet, supported by a 720-degree rotation of the electric vector polarization angle observed over 50 days, coinciding with six $\gamma$-ray flares. These flares culminated in a bright optical and $\gamma$-ray event as a knot passed through the 43 GHz core. \citet{Ahnen2017} similarly described such a scenario during a flare-like activity period in 2015. In contrast, \cite{Park2019}  suggested an alternative hypothesis where the jet comprises multiple layers exhibiting different speeds and time-dependent behaviors. This interpretation arose from observations of two components emerging from the core with similar apparent velocities, which contradicts the spine-sheath model.

The precise location of the $\gamma$-ray emission zone for this source remains a topic of ongoing investigation. Some studies suggest it could be situated close to the supermassive black hole (SMBH), within the BLR (e.g. \citealp{Poutanen2010, Tavecchio2010, Brown2013}), while others propose it is downstream the jet, beyond the BLR, and therefore the central parsec (e.g. \citealp{Tavecchio2010, Orienti2013, Dotson2015, HESS2021}). This discrepancy raises questions about the dominant mechanism responsible for the $\gamma$-ray emission. For this source, there is currently no consensus, with suggestions that the emission could involve a combination of EC and SSC mechanisms (e.g. \citealp{Kataoka2008, DAmmando2009, Castignani2017}). However, models exclusively proposing SSC or EC \citep{Aleksic2014}, exclusively SSC \citep{HESS2021}, and lepto-hadronic models \citep{Dzhatdoev2022} are also considered and not entirely ruled out.

The mass of the black hole (M$_{\text{BH}}$) in PKS 1510-089 has been estimated to be in the order of $10^{8}\text{ M}_{\odot}$ using various methodologies such as the temperature profile of the accretion disk \citep{Abdo2010, Castignani2017}, single-epoch spectra \citep{Oshlack2002, Xie2005}, and reverberation mapping (RM; \citealp{Rakshit2020}).

In this study, we examine the variability of $\gamma$-rays, X-rays, optical polarization degree, V-band, J-band, 1 mm, and 15 GHz light curves, as well as the flux of the broad emission lines H$\beta$ and H$\gamma$, and the optical continuum at $\lambda5100\text{ \AA}$. These latter fluxes were derived from spectra obtained at the Observatorio Astrofísico Guillermo Haro (OAGH), and the Steward Observatory (SO; \citealp{Smith2009C}). Our objective is to investigate the time delays between multi-wavelength light curves over approximately 10 years using cross-correlation analysis. This allows us to stratify the different emission bands and their dominant emission mechanisms. Furthermore, we analyze the spectroscopic light curves by separating the continuum based on its dominant source, whether the accretion disk or jet, utilizing the non-thermal dominance (NTD) parameter \citep{Shaw2012, PatinoAlvarez2016}.

\section{Observations and Data}\label{sec:obs}
\subsection{Optical Spectra}\label{sec:spectra}

The optical spectra come from two observatories. A total of 34 optical spectra were taken at the Observatorio Astrofísico Guillermo Haro (OAGH), while 353 spectra were taken at the Steward Observatory (SO). The OAGH\footnote{\url{ https://astro.inaoep.mx/observatorios/oagh/}} spectra were obtained under the spectroscopic monitoring program of bright $\gamma$-ray sources \citep{PatinoAlvarez2013Monitoring}. The wavelength range for these spectra is $3800–7100\text{ \AA}$. He–Ar lamp spectra were taken after each object exposure to enable wavelength calibration. \autoref{tab:obslog} shows the observation log for the spectra taken at OAGH. All observations were conducted using the 2.1m telescope equipped with a Boller \& Chivens long-slit spectrograph, installed at the Cassegrain Focus, with a grating of 150 l/mm, resulting in a resolution of approximately $15\text{ \AA}$. The spectroscopic data reduction was performed utilizing the \texttt{IRAF} package\footnote{\url{https://iraf-community.github.io}} \citep{Tody1986, Tody1993}, following standard procedures for bias and flat-field correction, cosmic-rays removal, 2D wavelength calibration, sky spectrum subtraction, and spectrophotometric calibration using standard stars that were observed each night.

\begin{table}[t]
\caption{ObsLog of spectra observed at OAGH.}
\label{tab:obslog}
\begin{tabularx}{\columnwidth}{lcccccr}
\toprule
\multirow{2}{*}{UT Date} & & \multirow{2}{*}{JD$_{245}$} & & Seeing & & Exposure \\
 & & & & (arcsec) & & Time (s)\\
\hline
2011 Jan 30 & & 5591.52 & \hspace{0.1in} & 2.5 & \hspace{0.2in} & $2\times1800$ \\
2011 Feb 12 & & 5604.44 & \hspace{0.1in} & 1.5 & \hspace{0.2in} & $3\times1800$ \\
2011 Feb 13 & & 5605.50 & \hspace{0.1in} & 1.9 & \hspace{0.2in} & $3\times1200$ \\
2011 Mar 27 & & 5647.40 & \hspace{0.1in} & 1.2 & \hspace{0.2in} & $3\times1800$ \\
2011 Mar 29 & & 5649.44 & \hspace{0.1in} & 2.4 & \hspace{0.2in} & $3\times1800$ \\
2011 Mar 30 & & 5650.47 & \hspace{0.1in} & 2.0 & \hspace{0.2in} & $3\times1800$ \\
2011 Mar 31 & & 5651.41 & \hspace{0.1in} & 2.1 & \hspace{0.2in} & $3\times1800$ \\
2011 Apr 01 & & 5652.39 & \hspace{0.1in} & 1.9 & \hspace{0.2in} & $3\times1800$ \\
2011 Apr 06 & & 5657.40 & \hspace{0.1in} & 1.8 & \hspace{0.2in} & $3\times1800$ \\
2011 Apr 07 & & 5658.45 & \hspace{0.1in} & 2.4 & \hspace{0.2in} & $3\times1800$ \\
2011 Apr 08 & & 5659.30 & \hspace{0.1in} & 2.0 & \hspace{0.2in} & $3\times1800$ \\
2011 May 30 & & 5711.22 & \hspace{0.1in} & 2.0 & \hspace{0.2in} & $3\times1800$ \\
2011 May 31 & & 5712.30 & \hspace{0.1in} & 1.5 & \hspace{0.2in} & $3\times1800$ \\
2011 Jun 26 & & 5738.16 & \hspace{0.1in} & 2.0 & \hspace{0.2in} & $3\times1800$ \\
2013 Apr 08 & & 6390.40 & \hspace{0.1in} & 1.4 & \hspace{0.2in} & $3\times1800$ \\
2013 Apr 12 & & 6424.31 & \hspace{0.1in} & 1.8 & \hspace{0.2in} & $3\times1800$ \\
2013 Jun 06 & & 6449.24 & \hspace{0.1in} & 1.6 & \hspace{0.2in} & $3\times1800$ \\
2013 Jun 07 & & 6450.29 & \hspace{0.1in} & 1.4 & \hspace{0.2in} & $2\times1800$ \\
2013 Jun 08 & & 6451.27 & \hspace{0.1in} & 1.7 & \hspace{0.2in} & $3\times1200$ \\
2014 Mar 26 & & 6742.47 & \hspace{0.1in} & 2.3 & \hspace{0.2in} & $3\times1200$ \\
2014 Mar 28 & & 6744.40 & \hspace{0.1in} & 2.3 & \hspace{0.2in} & $3\times1200$ \\
2014 Apr 28 & & 6775.35 & \hspace{0.1in} & 2.0 & \hspace{0.2in} & $3\times1800$ \\
2014 May 30 & & 6807.27 & \hspace{0.1in} & 1.8 & \hspace{0.2in} & $3\times1800$ \\
2014 Jun 01 & & 6809.25 & \hspace{0.1in} & 2.0 & \hspace{0.2in} & $3\times1800$ \\
2014 Jun 24 & & 6832.20 & \hspace{0.1in} & 2.0 & \hspace{0.2in} & $3\times1800$ \\
2015 Apr 19 & & 7131.40 & \hspace{0.1in} & 2.1 & \hspace{0.2in} & $3\times1800$ \\
2015 Apr 20 & & 7162.50 & \hspace{0.1in} & 1.8 & \hspace{0.2in} & $3\times1800$ \\
2015 Jun 19 & & 7192.25 & \hspace{0.1in} & 1.7 & \hspace{0.2in} & $3\times1800$ \\
2015 Jun 20 & & 7193.22 & \hspace{0.1in} & 2.2 & \hspace{0.2in} & $3\times1800$ \\
2015 Jun 21 & & 7194.25 & \hspace{0.1in} & 2.8 & \hspace{0.2in} & $3\times1800$ \\
2016 Mar 02 & & 7541.45 & \hspace{0.1in} & 1.3 & \hspace{0.2in} & $3\times1800$ \\
2016 Apr 09 & & 7487.35 & \hspace{0.1in} & 1.3 & \hspace{0.2in} & $3\times1800$ \\
2016 May 04 & & 7512.28 & \hspace{0.1in} & 1.5 & \hspace{0.2in} & $3\times1800$ \\
2016 Jun 07 & & 7546.23 & \hspace{0.1in} & 1.8 & \hspace{0.2in} & $3\times1800$ \\
\hline
\end{tabularx}
\begin{tablenotes}
\small
\item \textbf{Notes}. Hereinafter JD$_{245}$ represents JD-2450000
\end{tablenotes}
\end{table}

The observations at the SO were conducted as part of the Ground-based Observational Support of the Fermi Gamma-ray Space Telescope at the University of Arizona monitoring program\footnote{\url{http://james.as.arizona.edu/~psmith/Fermi/}}. These observations used the SPOL spectropolarimeter with slit widths of $3\farcs0$, $4\farcs1$, and $5\farcs1$. The spectra were re-calibrated against the V-band magnitude and cover a period of 10 years, from 2008 to 2018. The wavelength range for these spectra is $4000-7500\text{ \AA}$. Further details about observational setup and data reduction can be found in \citet{Smith2009C}.

\begin{figure}[b]
\centering
\includegraphics[width=\columnwidth]{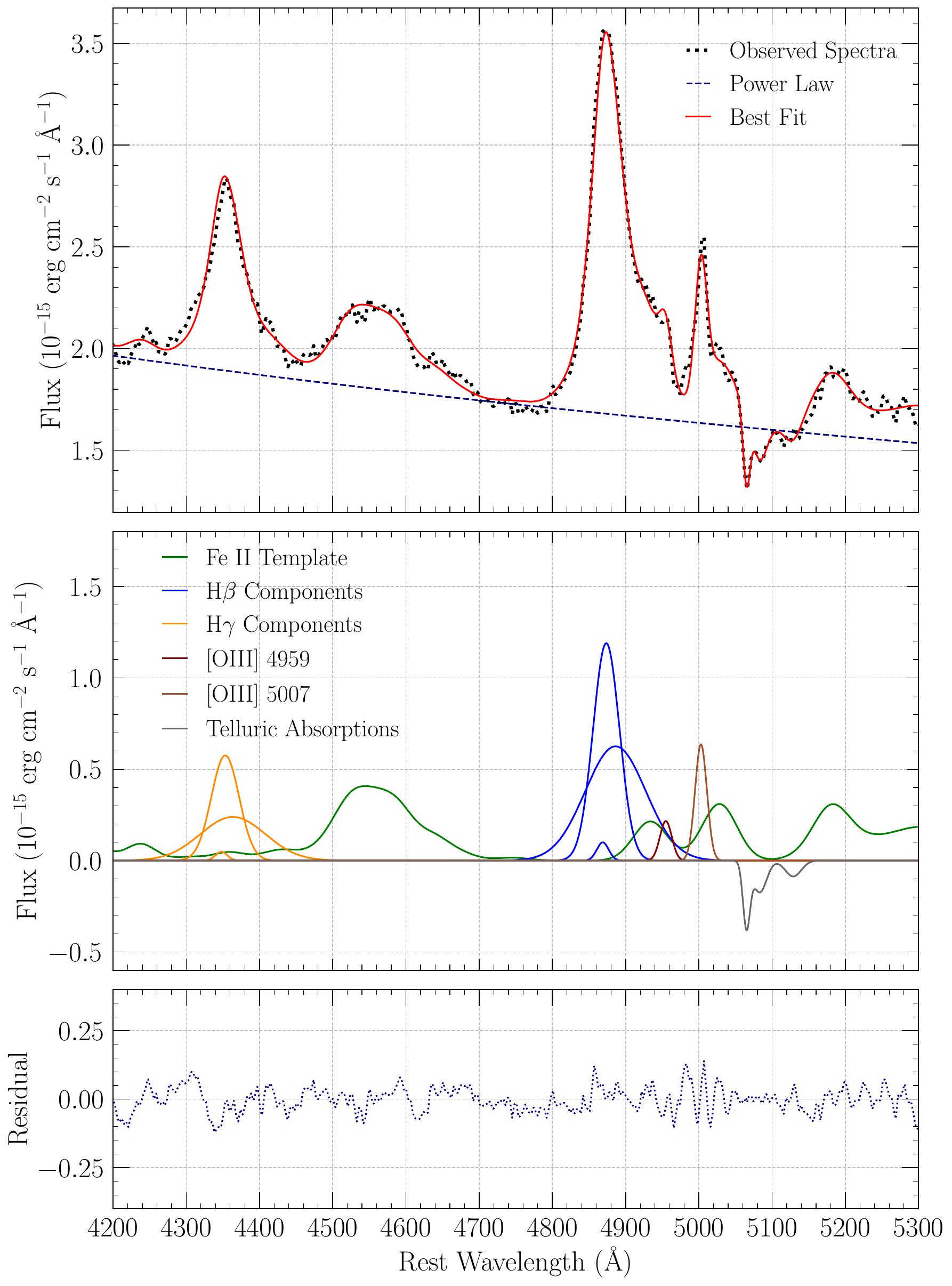}
\caption{Example decomposition of the H$\beta$ and H$\gamma$ emission lines from a spectrum observed at SO on March 19, 2010. Top panel: The rest-frame spectrum with the best-fit model overlaid. The continuum is represented by a power-law function. Middle panel: The broad and narrow components used to fit H$\beta$ and H$\gamma$, along with the Fe II template, [O III] doublet, and telluric absorptions. Bottom panel: Residuals from the subtraction of the best-fit model from the observed spectrum.}
\label{fig:spectrum}
\end{figure}

We adjusted the spectra to the rest-frame of the object and applied a cosmological correction to the flux, taking the form $(1+z)^{3}$, following \citet{Peterson1997}. The galactic reddening correction was applied using the dust maps from \citet{SchlaflyAndFinkbeiner2011}, adopting a color excess of $E(B-V)=0.09$. A galactic reddening law with $R_{v}=3.1$ was employed, following the prescription of \citet{Cardelli1989}. To retrieve the H$\beta$, H$\gamma$, and $\lambda5100\text{ \AA}$ continuum flux, the spectra were fitted with several components. First, the local continuum was represented using a power-law function and subtracted from the spectrum. Second, the multiplet Fe II emission was modeled and subtracted utilizing the template from \citet{Kovacevic2010}, covering the wavelength range of $4000-5500\text{ \AA}$. Finally, all emission and absorption lines were fitted with the aid of the astropy.modeling\footnote{\url{https://docs.astropy.org/en/stable/modeling/index.html}} framework. For the telluric absorptions around $5075\text{ \AA}$ in the rest-frame, we used three Gaussians to model their shape. The H$\beta$ and H$\gamma$ emission lines were fitted using three different Gaussians: narrow, broad, and very broad components. For the forbidden lines [O III]$\lambda\lambda4959,5007\text{ \AA}$ one Gaussian was used for each one. An example of spectrum decomposition can be seen in \autoref{fig:spectrum}. 

\begin{table}[b]
\centering
\caption{Median values from each of the sources of uncertainty; The dispersion of the spectrum and S/N ($\sigma_{\text{S/N}}$). The subtraction of Fe II ($\sigma_{\text{Fe II}}$). The flux calibration ($\sigma_{\text{Cal}}$). And the resulting reported uncertainty ($\sigma$).}
\label{tab:errors}
\begin{tabularx}{\columnwidth}{ccccccccc}
\toprule
\multirow{2}{*}{Emission Line} & \hspace{0.02in} & $\sigma_{\text{S/N}}$ & \hspace{0.1in} & $\sigma_{\text{Fe II}}$ & \hspace{0.1in} & $\sigma_{\text{Cal}}$ & \hspace{0.1in} & $\sigma$ \\
 & \multicolumn{8}{c}{($\times10^{-15}$ erg s$^{-1}$ cm$^{-2}$)} \\
\hline
H$\beta$ & \hspace{0.02in} & 2.15 & \hspace{0.1in} & 2.67 & \hspace{0.1in} & 13.1 & \hspace{0.1in} & 13.7 \\
H$\gamma$ & \hspace{0.02in} & 1.52 & \hspace{0.1in} & 2.33 & \hspace{0.1in} & 4.82 & \hspace{0.1in} & 5.67 \\
\hline
\end{tabularx}
\end{table}

\begin{table*}[t]
\caption{Sample of flux measurements for the $\lambda5100\text{ \AA}$ continuum, H$\gamma$, and H$\beta$ emission lines.}
\label{tab:fluxes}
\begin{tabularx}{2\columnwidth}{lcccccccr}
\toprule
\multirow{2}{*}{JD$_{245}$} & & Flux $\lambda5100\text{ \AA}$ & & Flux H$\gamma$ & & Flux H$\beta$ & & \multirow{2}{*}{Observatory} \\
 & & ($\times10^{-15}$ erg s$^{-1}$ cm$^{-2}$ \AA\ $^{-1}$) & & ($\times10^{-15}$ erg s$^{-1}$ cm$^{-2}$) & & ($\times10^{-14}$ erg s$^{-1}$ cm$^{-2}$) & & \\
\hline
4830.03	& & $1.41\pm0.15$ &	& $43.37\pm5.49$ & & $10.48\pm1.10$ & \hspace{0.1in} & SO \\
4831.03	& & $1.17\pm0.12$ &	& $40.99\pm4.93$ & & $9.26\pm0.97$ & \hspace{0.1in} & SO \\
4832.02	& & $1.22\pm0.13$ &	& $38.53\pm4.56$ & & $9.88\pm1.03$ & \hspace{0.1in} & SO \\
4833.03	& & $1.27\pm0.13$ &	& $35.82\pm4.22$ & & $8.94\pm0.94$ & \hspace{0.1in} & SO \\
4860.02	& & $1.54\pm0.16$ &	& $38.93\pm4.78$ & & $9.23\pm0.97$ & \hspace{0.1in} & SO \\
4861.02	& & $1.42\pm0.15$ &	& $38.20\pm4.18$ & & $8.86\pm0.92$ & \hspace{0.1in} & SO \\
4862.03	& & $1.56\pm0.16$ &	& $35.90\pm4.24$ & & $9.52\pm0.99$ & \hspace{0.1in} & SO \\
4863.03	& & $1.54\pm0.16$ &	& $35.34\pm4.33$ & & $9.99\pm1.03$ & \hspace{0.1in} & SO \\
4864.02	& & $1.35\pm0.14$ &	& $36.79\pm4.06$ & & $9.68\pm0.99$ & \hspace{0.1in} & SO \\
4881.97	& & $1.23\pm0.13$ &	& $36.28\pm3.97$ & & $9.69\pm1.01$ & \hspace{0.1in} & SO \\
4882.97 & & $1.29\pm0.13$ &	& $35.35\pm3.83$ & & $8.52\pm0.88$ & \hspace{0.1in} & SO \\
\hline
\end{tabularx}
\begin{tablenotes}
\small
\item (This table is available in its entirety in machine-readable form.)
\end{tablenotes}
\end{table*}

The [O III] emission lines originate in the Narrow-Line Region (NLR), which remains stable over significant periods of time \citep{Bennert2002}. Therefore, any differences in their flux will be due to calibration effects, and different observational conditions. To homogenize all the spectral measurements between observations made at the SO and the OAGH, a flux re-calibration was performed. Given that the SO spectra set are re-calibrated by V-band photometry, we use their spectra to calculate the mean flux of the [O III]$\lambda5007\text{ \AA}$, obtaining a value of $1.273\times10^{-14}$ erg s$^{-1}$ cm$^{-2}$. Hence, the calibration factor is calculated by dividing the measured flux by this mean flux, and this factor is applied to all OAGH spectra to ensure consistent calibration across observations.

\begin{figure*}[t]
\centering
\includegraphics[width=2\columnwidth]{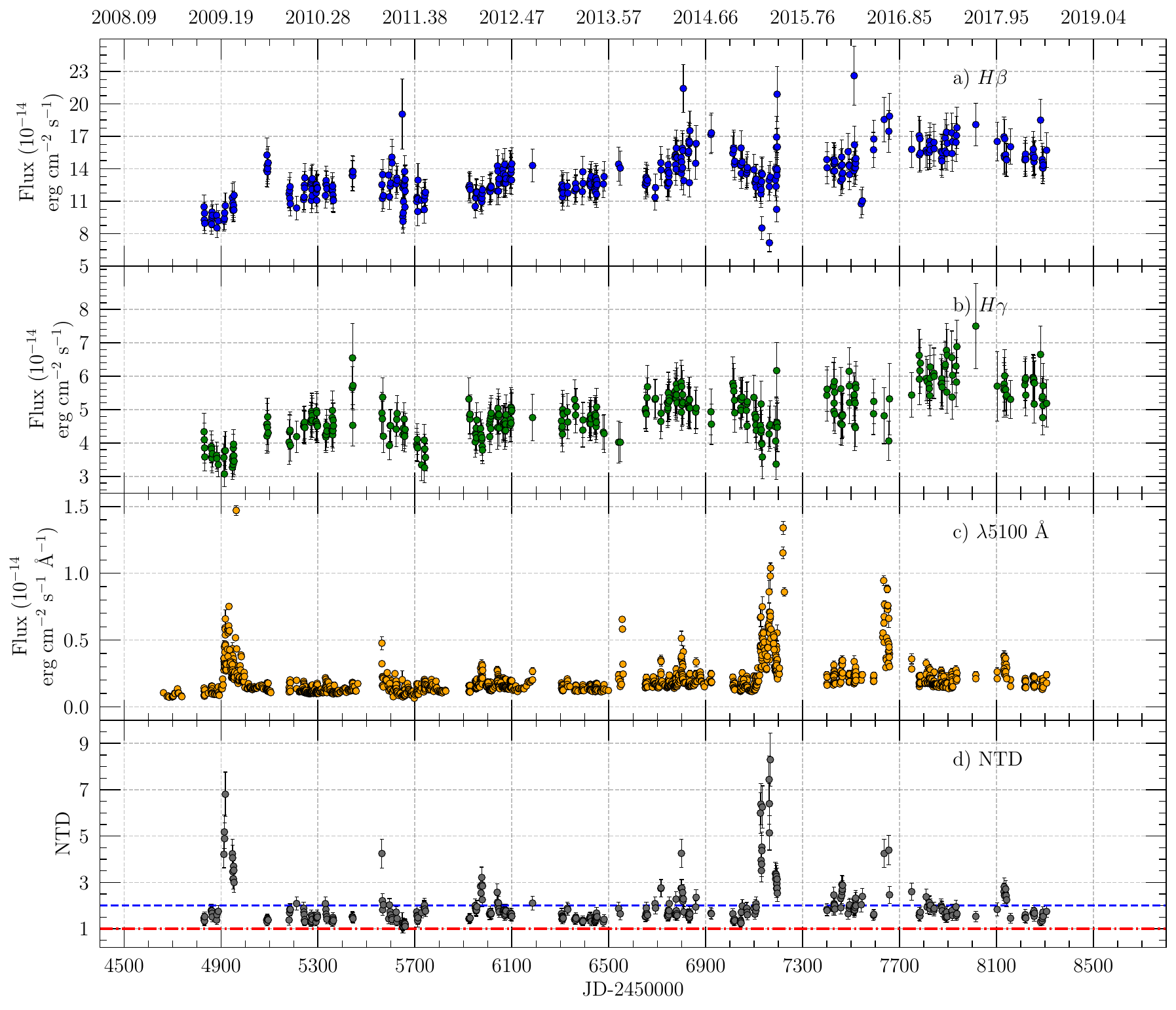}
\caption{Spectroscopic Light curves. (a) H$\beta$ emission line flux, (b) H$\gamma$ emission line flux, (c) $\lambda5100\text{ \AA}$ continuum flux, and (d) NTD parameter. The red dash-dotted and blue dashed lines in panel (d) represent NTD$=1$ and NTD$=2$, respectively.}
\label{fig:speccurves}
\end{figure*}

The H$\beta$ and H$\gamma$ emission line fluxes were calculated by subtracting all components of the original spectrum, leaving only the respective line emissions, and then integrating the remaining component. The integration range for H$\beta$ was $4781-5044\text{ \AA}$, and for H$\gamma$, it was $4272-4435\text{ \AA}$. To determine the flux of the $\lambda5100\text{ \AA}$ continuum, the telluric absorptions and Fe II emission were subtracted from the original spectrum. The mean value and its standard deviation were then measured in the range $5050-5150\text{ \AA}$. The uncertainty in the continuum flux calculation is derived from the quadratic sum of the standard deviation and the flux calibration error, with the latter accounted for as $10\%$ of the flux (Paul Smith, private communication). There are three unique contributions to the uncertainty in the H$\beta$ and H$\gamma$ flux measurements. The first arises from the random error caused by the dispersion of the spectra and the signal-to-noise ratio (S/N), calculated as in \citet{Tresse1999}. The second contribution is introduced by the subtraction of Fe II emission, estimated as in \citet{LeonTavares2013}. The final contribution arises from flux calibration, considered as $10\%$ of the total flux. The reported uncertainty in the emission line flux calculations for both lines results from the quadratic sum of the three sources of uncertainty. The median values for each source, along with the total reported uncertainty, are shown in \autoref{tab:errors}. A sample of the $\lambda5100\text{ \AA}$ continuum, H$\gamma$, and H$\beta$ fluxes are shown in \autoref{tab:fluxes}.

We investigated the influence of the narrow component of the H$\beta$ line by constraining its width and central wavelength based on the [O III]$\lambda5007\text{ \AA}$ line during profile fitting, allowing only its flux to vary. From this, we estimated the flux of the narrow component of H$\beta$, obtaining a value of $3.58\pm1.11\times10^{-15}$ erg s$^{-1}$. Our analysis showed that the narrow component contributes $5.2\%$  of the total H$\beta$ flux at its minimum and $1.5\%$ at its maximum. Notably, this contribution is smaller than the uncertainty in the total H$\beta$ flux ($10.1\%$ and $17.1\%$ for minimum and maximum flux respectively). Given its imperceptible impact, we used the total H$\beta$ flux for all subsequent calculations. Consequently, since a similar behavior is expected for the H$\gamma$ emission line, we also use the their total line profile flux for the cross-correlations analysis.

We calculated the NTD parameter \citep{Shaw2012} using the H$\beta$ and $\lambda5100\text{ \AA}$ continuum luminosities\footnote{Assuming a luminosity distance of 1906.9 Mpc, with cosmological parameters of H$_{0}=71$ km s$^{-1}$ Mpc$^{-1}$, $\Omega_{\Lambda}=0.73$, $\Omega_{M}=0.27$} as follows:
\begin{equation}
    NTD = \frac{L_{obs}}{L_{pred}}
\label{eq:NTD1}
\end{equation}

Where $L_{obs}$ is the observed continuum luminosity, and $L_{pred}$ is the predicted luminosity estimated from the H$\beta$ luminosity using the non-blazar relation of \citet{GreeneAndHo2005}. We calculated $L_{pred}$ and its error by solving Equation 2 from \citep{GreeneAndHo2005}.

The NTD parameter measures the relative contribution of the non-thermal emission (synchrotron emission from the jet) to the total continuum emission from an AGN source. Following the procedure outlined in \citet{PatinoAlvarez2016}, when NTD $=1$, the continuum emission is purely thermal, originating from the accretion disk. If $1<$ NTD $<2$, the accretion disk remains the primary source, but the jet also contributes to the emission. At NTD $=2$, both the disk and jet contribute equally. When NTD $>2$, the continuum emission is dominated by nonthermal processes in the jet. Two regimes can be characterized: Jet-Dominance emission when NTD $>2$, and Disk-Dominance emission when NTD $<2$.

The light curves of H$\beta$, H$\gamma$, and $\lambda 5100\text{ \AA}$ continuum fluxes along with the NTD, are displayed in \autoref{fig:speccurves}.

\subsection{Photometric Observations}\label{sec:photo}

We obtained $\gamma$-ray data from the public database of the LAT instrument aboard the Fermi Gamma-ray Space Telescope\footnote{\url{https://fermi.gsfc.nasa.gov/cgi-bin/ssc/LAT/LATDataQuery.cgi}} \citep{Abdo2010}, covering the range of $0.1-300$ GeV. Weekly light curves were constructed utilizing \texttt{Fermitools version 2.0.8}. All sources within $15^{o}$ radius around PKS 1510-089 were included in the model, based on the Data Release 3 of the Fourth Fermi-LAT Source Catalog (4FGL-DR3, \citealp{Abdollahi2022}). The spectral parameters of all the sources within $5^{o}$ are kept free to vary, while for the rest of the sources, only the prefactor/normalization is kept free. The diffuse isotropic background, along with the Galactic interstellar emission and extragalactic background, are modeled using the latest available templates: \texttt{iso\_P8R3\_SOURCE\_V2} and \texttt{gll\_iem\_v07}, respectively. The X-ray data in the range of $0.3-10$ keV were acquired from the Swift-XRT Monitoring of Fermi-LAT Sources of Interest database\footnote{\url{https://www.swift.psu.edu/monitoring/}} (XRT; \citealp{StrohAndFalcone2013}). These data were processed using standard SWIFT tools. Swift Software version 3.9, \texttt{FTOOLS version 6.12} \citep{Blackburn1995}  and \texttt{XSPEC version 12.7.1} \citep{Arnaud1996}, and the X-ray light curve was generated with \texttt{xrtgrblc version 1.6}.

The near-infrared (NIR) J-band data were obtained from the Small and Moderate Aperture Research Telescope System (SMARTS; \citealp{Bonning2012}). Optical V-band data were retrieved from two sources, SMARTS and SO. Photometric observations from SO were utilized to augment the data points of the $\lambda5100\text{ \AA}$ continuum light curve by establishing a linear relationship between spectra and quasi-simultaneous (within 24 hours) V-band observations. The fitting procedure employed orthogonal distance regression utilizing the \texttt{SciPy ODR package}. The fit demonstrated statistical significance with a p-value ($p_{v}$) of 0 (to machine precision). A p-value indicates the probability that our null hypothesis (i.e., no relationship between the lineal fit and the data) is true, with values below 0.05 typically considered statistically significant.

\begin{figure}[t]
\centering
\includegraphics[width=\columnwidth]{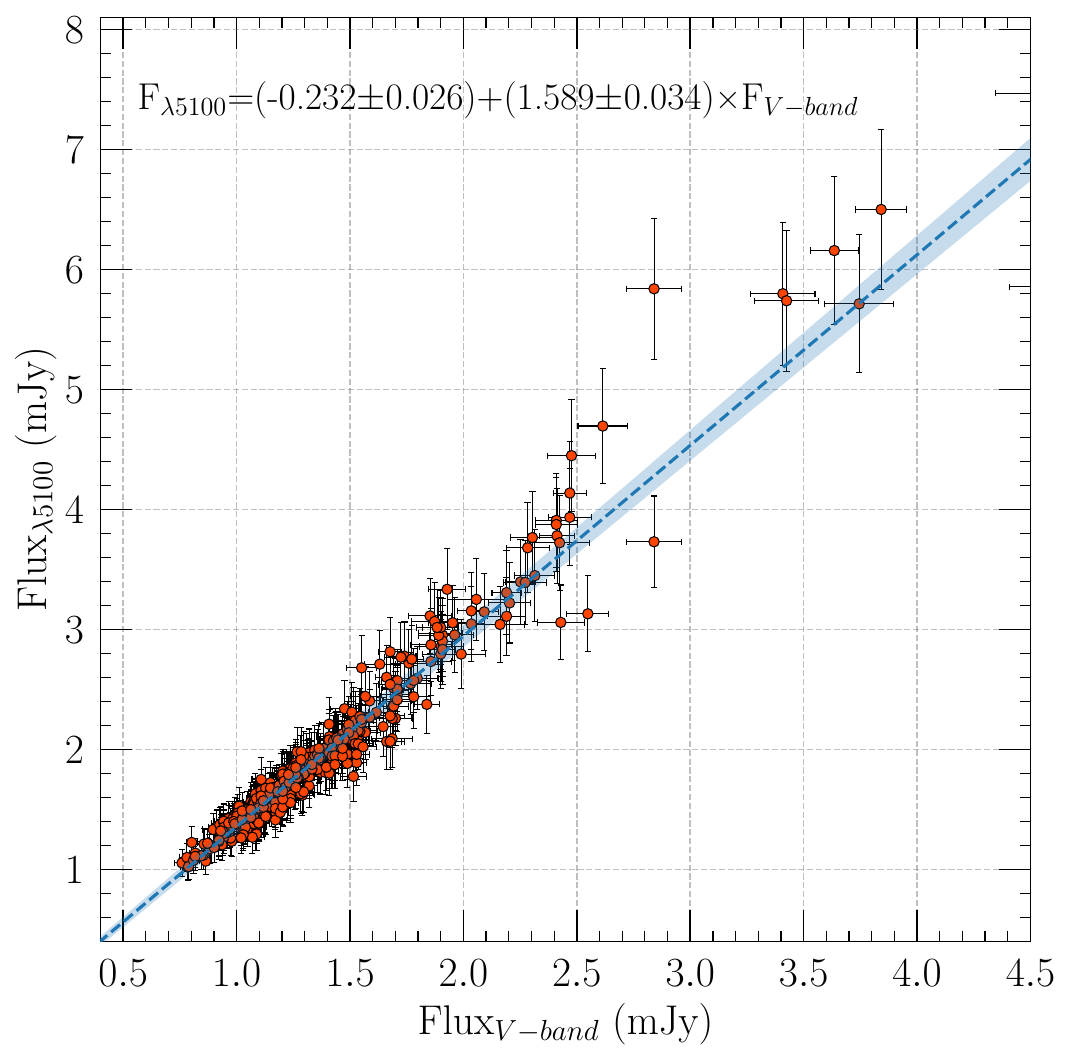}
\caption{Relation between $\lambda 5100\text{ \AA}$ and V-Band flux for quasi-simultaneous observations at the SO. The blue dashed line represents the best linear fit, and the shaded area has its uncertainty at $1\sigma$}.
\label{fig:relationVC}
\end{figure}

The resulting correlation is illustrated in \autoref{fig:relationVC}. Additionally, optical linear polarization degree and optical polarization angle data were obtained from the SO monitoring program. To address $180^{o}$ wraps in the polarization angle ($\theta$), we selected data points with $\theta/\sigma_{\theta}>5$ and allowed changes of $\leq90^{o}$ between consecutive observations, assuming minimal variability.

The 1mm data were sourced from the Submillimeter Array (SMA) public database\footnote{\url{http://sma1.sma.hawaii.edu/callist/callist.html}} \citep{Gurwell2007}. Data at 15 GHz were obtained from the Owens Valley Radio Observatory (OVRO). Further information regarding the instruments and data processing can be found in \citet{Richards2011}.

\begin{figure*}[t]
\centering
\includegraphics[width=2\columnwidth]{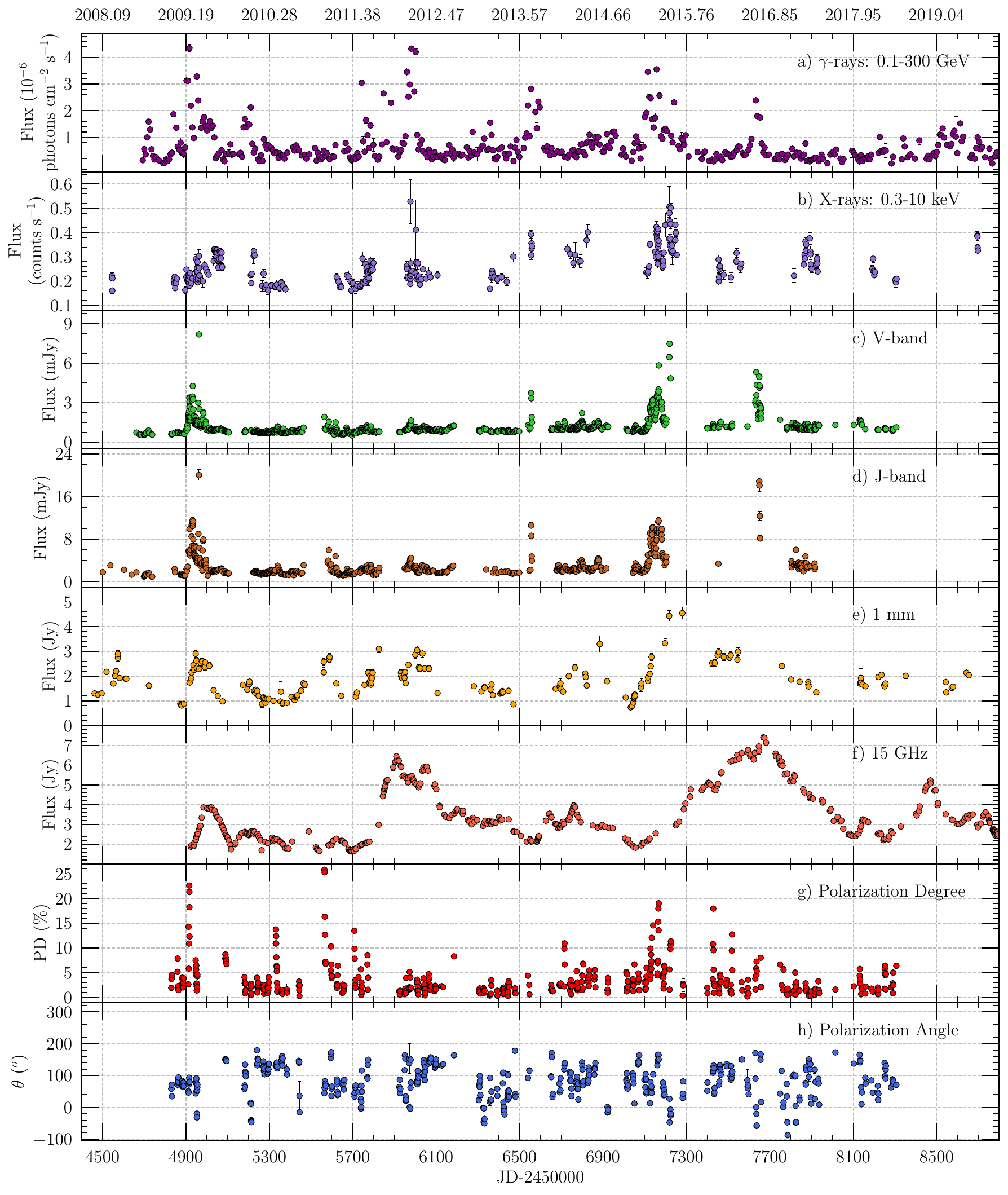}
\caption{Photometric Light curves. (a) $\gamma$-ray data from Fermi/LAT ($0.1-300$ GeV), (b) X-ray data from Swift/XRT ($0.3-10$ keV), (c) V-band from SO and SMARTS, (d) J-band from SMARTS, (e) 1-mm data from SMA, (f) Radio data at 15 GHz from OVRO, and (g) Optical polarization degree from SO and (h) Optical polarization angle from SO.}
\label{fig:photcurves}
\end{figure*}

To visualize the variability of this source, \autoref{fig:photcurves} presents the light curves of all photometric data described above.

\section{Variability}\label{sec:variability}

We conducted cross-correlations among all multi-wavelength light curves previously described. Three different methods were employed to ensure robustness: the interpolated cross-correlation function (ICCF; \citealp{GaskellAndSparke1986}), the discrete cross-correlation function (DCCF; \citealp{EdelsonAndKrolik1988}), and the Z-transformed discrete cross-correlation function (ZDCF; \citealp{Alexander1997}). These methods followed the established protocols outlined in \citet{PatinoAlvarez2013Variability}, and \citet{AmayaAlmazan2022}. Since each of the cross-correlation methods used has different sensitivities to sampling, noise, and data irregularities, a certain degree of flexibility is required in evaluating how these factors affect the results. Therefore, only time delays with correlation coefficients significant at levels above $99\%$, and consistent across at least two of the three methods, were considered. Further details on significance levels can be found in \citet{Emmanoulopoulos2013} and \citet{AmayaAlmazan2022}. The reported delays represent the average results obtained from either two or all three methods, depending on the case.

We examined the possibility of spurious correlations by analyzing the Fourier transform and power spectrum of each individual light curve. According to \citet{Bath1974}, the cross-correlation function (CCF) and power spectrum form a Fourier pair, implying that significant spectral power in the power spectrum could lead to high correlation coefficients in the CCF, influenced by the behavior of a single light curve. After identifying all delays with a correlation coefficient $\geq99\%$ significance, we scrutinized the Fourier transform and power spectra to detect notable power at these identified delays. Additionally, to investigate potential spurious correlations arising from observation cadence, we constructed unitary light curves. These are interpolated light curves where observation times are assigned a value of 1, with all other points set to 0. This approach ensures that instances of high spectral power are attributed solely to the sampling of the light curve, rather than its intrinsic variability. In other words, any significant delay coinciding with a peak in the unitary light curve is discarded (see \citealp{AmayaAlmazan2022} for more details).

\begin{figure*}[t]
\centering
\includegraphics[width=2.1\columnwidth]{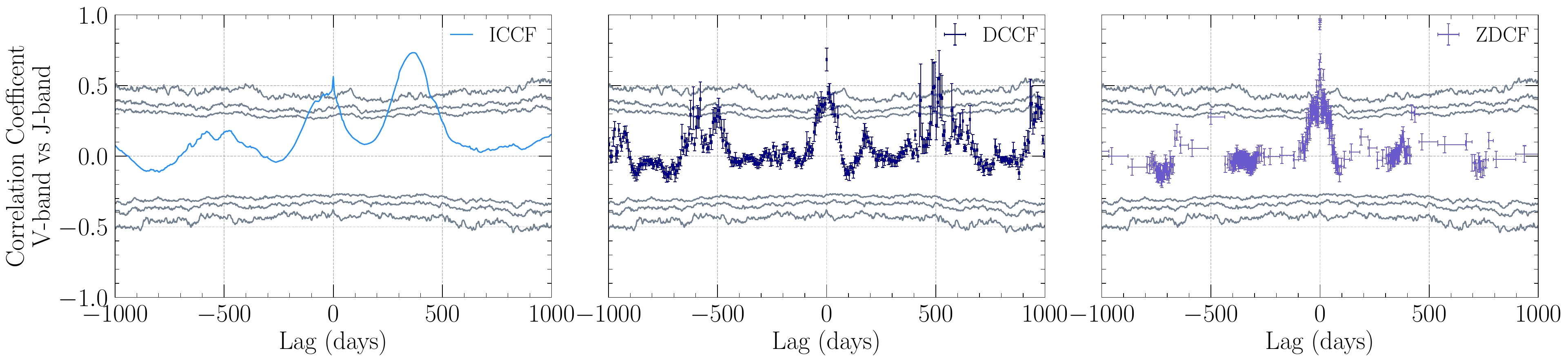}
\caption{Cross-correlation functions for V-band and J-band light curves. Left panel: Interpolated cross-correlation function (ICCF). Middle panel: Discrete cross-correlation function (DCCF). Right panel: Z-transformed cross-correlation function (ZDCF). Gray lines depict significance levels at $90\%$, $95\%$, and $99\%$. The complete set of cross-correlation function figures are available as online material}.
\label{fig:CCexample}
\end{figure*}

The uncertainty in the reported delay corresponds to the maximum uncertainty value among the three methods employed. The sequence of cross-correlations performed corresponds to the order of results presented. A positive delay indicates that the first band leads the second, whereas a negative delay signifies the first band lags behind the second. Detailed results of the cross-correlation analysis for the complete set of light curves are summarized in \autoref{tab:delays}. An illustrative example of cross-correlation functions is depicted in \autoref{fig:CCexample}, while the complete set is available as online material.

\begin{table}[t]
\centering
\caption{Time delays obtained from the cross-correlation analysis for the full light curves.}
\label{tab:delays}
\begin{tabularx}{\columnwidth}{lcr}
\toprule
Band & \hspace{1.2in} & Delay (days) \\
\hline
15 GHz vs 1mm & \hspace{1.2in} & $-40.3\pm18.7$ \\
\multirow{2}{*}{$5100 \text{ \AA}$ vs 1 mm} & \hspace{1.2in} & $18.1\pm18.7$ \\
 & \hspace{1.2in} & $62.7\pm18.7$ \\
$5100 \text{ \AA}$ vs $\gamma$-rays & \hspace{1.2in} & $-6.8\pm6.4$ \\
$5100 \text{ \AA}$ vs H$\beta$ & \hspace{1.2in} & $82.4\pm6.1$ \\
$5100 \text{ \AA}$ vs H$\gamma$ & \hspace{1.2in} & $87.7\pm6.6$ \\ 
$5100 \text{ \AA}$ vs J-band & \hspace{1.2in} & $0.0\pm3.7$ \\
$5100 \text{ \AA}$ vs NTD & \hspace{1.2in} & $0.1\pm6.7$ \\
$5100 \text{ \AA}$ vs PD & \hspace{1.2in} & $5.4\pm5.0$ \\
$5100 \text{ \AA}$ vs V-band & \hspace{1.2in} & $0.2\pm2.5$ \\
$5100 \text{ \AA}$ vs X-rays & \hspace{1.2in} & $65.4\pm13.4$ \\
$\gamma$-rays vs NTD & \hspace{1.2in} & $6.0\pm6.7$ \\
\multirow{2}{*}{J-band vs 1 mm} & \hspace{1.2in} & $12.6\pm18.7$ \\
& \hspace{1.2in} & $47.3\pm18.7$ \\
J-band vs $\gamma$-rays & \hspace{1.2in} & $-8.8\pm6.4$ \\
J-band vs H$\beta$ & \hspace{1.2in} & $81.8\pm6.1$ \\
J-band vs NTD & \hspace{1.2in} & $-0.5\pm6.7$ \\
J-band vs PD & \hspace{1.2in} & $-0.95\pm5.0$ \\
J-band vs X-rays & \hspace{1.2in} & $62.8\pm24.0$ \\ 
PD vs NTD & \hspace{1.2in} & $6.2\pm7.3$ \\
\multirow{2}{*}{V-band vs 1 mm} & \hspace{1.2in} & $12.6\pm18.7$ \\
 & \hspace{1.2in} & $67.2\pm18.7$ \\
V-band vs $\gamma$-rays & \hspace{1.2in} & $-9.0\pm11.3$ \\
V-band vs H$\beta$ & \hspace{1.2in} & $78.9^{+14.9}_{-4.2}$ \\ 
V-band vs J-band & \hspace{1.2in} & $0.0\pm3.7$  \\
V-band vs NTD & \hspace{1.2in} & $0.6\pm6.7$ \\
V-band vs PD & \hspace{1.2in} & $3.2\pm5.0$ \\
V-band vs X-rays & \hspace{1.2in} & $65.8\pm53.6$ \\ 
\hline
\end{tabularx}
\begin{tablenotes}
\small
\item \textbf{Notes}. All delays show correlations at $\geq99\%$ significance level. Results are only presented for cross-correlation analyses where non-spurious correlations were found. All cross-correlations were performed in the order stated in this table. Comparison between the optical position angle (PA) and any other band yielded no correlation.
\end{tablenotes}
\end{table}

Delays identified between optical/NIR emissions ($\lambda5100\text{ \AA}$ continuum, J-band, V-band, and NTD), are consistent with zero, indicating quasi-simultaneous emissions and implying that the emitting regions are co-spatial. The variability in the continuum emission is primarily driven by periods of flare-like activity, which are in turn caused by high jet emission, as indicated by the NTD light curve with values exceeding 2. Thus, the variability in the $\lambda5100\text{ \AA}$ continuum emission predominantly originates from synchrotron emission within the jet. Given the co-spatiality of the optical and near-infrared emission regions, the V-band and J-band emissions during flare-like events are also primarily emitted within the jet. This determination supports synchrotron emission as the primary mechanism driving variability in the $\lambda5100\text{ \AA}$ continuum, J-band, and V-band. Additionally, a correlation was found between the $\lambda5100\text{ \AA}$ continuum and 1 mm emissions, which reinforces the scenario where optical continuum emission is jet-dominated, given the known association of 1 mm emission with the jet. 

All the time delays between each optical/NIR band and X-rays are consistent within 1 sigma, which is expected due to their co-spatial origin. The fact that the optical/NIR emission leads to the X-ray emission by approximately 65 days excludes the possibility of thermal emission from the inner accretion disk or synchrotron emission from the jet. In those cases, the X-rays would be expected to lead the optical/NIR emission. In the scenario where IC scattering occurs in the hot corona, X-rays are generated by up-scattering UV/soft X-ray photons from the accretion disk by high-energy electrons in the corona. According to this model, the X-ray emission should either lead or coincide with the optical variations, not lag behind them. However, for IC scattering in the jet, X-ray emission can naturally lag behind optical/NIR emission. Therefore, IC scattering from the jet emerges as the most plausible dominant mechanism for the observed X-ray emission.

The delay of $7.0\pm7.7$ days between the optical/NIR bands leading the $\gamma$-rays emission suggests a scenario where the optical emission region and the source of seed photons are quasi-cospatial. This near-zero delay implies that the emission regions are closely aligned spatially. Since the variability in the optical light curves is mainly driven by the jet, delays between the optical light curves and any other light curves will also trace the jet emission during periods of flare-like activity. Therefore, the nearly zero time lag between the optical/NIR and $\gamma$-ray emissions indicates co-spatiality of these emission regions within the jet during flaring periods.

\begin{figure*}[t]
\centering
\includegraphics[width=2\columnwidth]{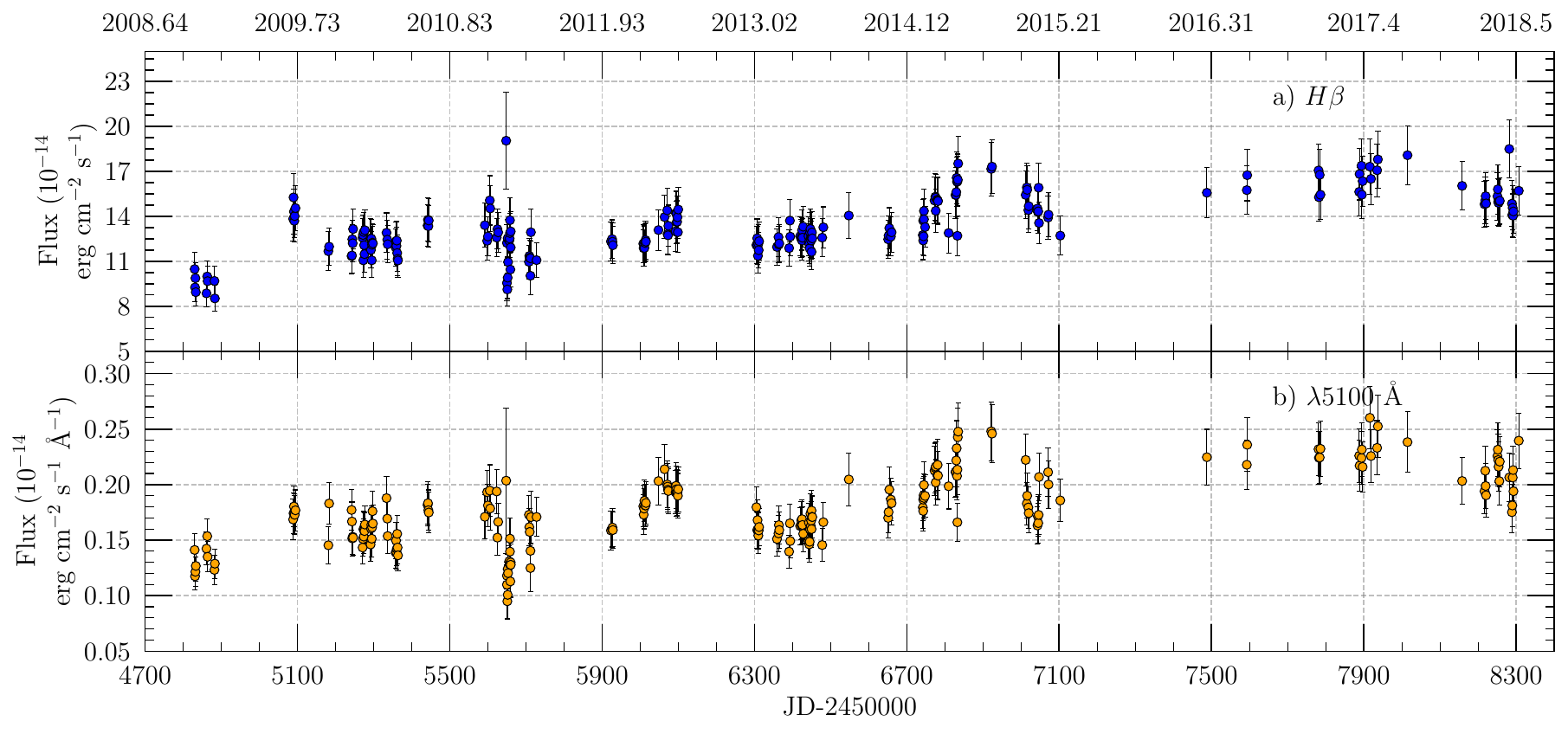}
\caption{Spectroscopic light curves for the disk dominance regime. (a) H$\beta$ emission line flux, and (b) the $\lambda5100\text{ \AA}$ continuum flux. The $\lambda5100\text{ \AA}$ continuum data points added from the V-band observations (\autoref{sec:photo}) are excluded from this light curve, focusing solely on spectroscopic measurements in the disk dominance regime.}
\label{fig:regcurves}
\end{figure*}

In the IC process, the energy of the up-scattered photons can be approximated by $E_{\gamma}\approx\gamma^{2}E_{\text{seed}}$, where $E_{\gamma}$ is the energy of the $\gamma$-ray photon, $E_{\text{seed}}$ is the energy of the seed photon, and $\gamma$ is the Lorentz factor of the electrons. The LAT detection range for $\gamma$-rays spans from $0.1$ to $300$ GeV. For PKS 1510-089, the Lorentz factor has been estimated to be as high as $\gamma=10^{4}$ \citep{Paliya2018} and $\gamma=10^{6}$ \citep{Rani2011} based on SED modeling. Consequently, the energy of the seed photons $E_{\text{seed}}$ for $\gamma=10^{4}$ ranges from approximately $1.0$ to $3000.0$ eV, which falls within the NIR to near-ultraviolet (NUV) spectrum. Therefore, for the $\gamma$-ray energies detected by Fermi/LAT, the seed photons must be within the NIR-NUV range, corresponding to observable flux in the V-band, J-band, and $\lambda5100\text{ \AA}$ continuum. Given the quasi-cospatial relationship observed between the optical/NIR emission region and the seed photon region, both within the jet, we presume SSC as the dominant emission mechanism for the flare events. In SSC, the $\gamma$-rays are produced through the scattering of synchrotron photons by the same population of relativistic electrons. However, this does not exclude the possibility of EC processes since both SSC and EC should be occurring at the same time. Further analysis is needed to ascertain which is the dominant contribution, either SSC or EC.

The approximately 80-day delay observed between the emission lines and the continuum emission likely reflects the light travel time between the continuum source and the  BLR. This delay is evident in the H$\beta$ and H$\gamma$ light curves, where significant flare events appear in the optical bands and emission lines around 2015-2016 (JD$_{245}\approx7200$). However, some optical flares do not correspond to similar flares in the H$\beta$ and H$\gamma$ light curves. Instead, these optical flares exhibit increasing trends in flux, as observed around JD$_{245}\approx4900$ and JD$_{245}\approx7700$. Cross-correlation functions were computed for both jet dominance and disk dominance regime light curves, considering uncertainties (i.e. NTD$+\sigma<2$) between the $\lambda5100\text{ \AA}$ continuum and the H$\beta$ emission line. The results for these subsets did not reveal any significant correlation, indicating no physical delay between the light curves. In the jet dominance regime, the observed delays appeared to be spurious correlations likely stemming from the sparse observation cadence and a limited number of data points in the light curve. Conversely, in the disk dominance regime, substantial segments of the cross-correlation functions exceeded the $99\%$ significance level (although with a clear peak at zero delay in the ICCF and the ZDCF), leading to inconclusive outcomes (extended discussion on \autoref{sec:spectralfeat}). The light curve for the disk dominance regime is depicted in \autoref{fig:regcurves}.

The polarization degree (PD) reflects the proportion of the total synchrotron emission that exhibits polarization. At the SO, PD is determined by integrating the polarized spectrum within the observed range of $5000-7000\text{ \AA}$ (details can be found in \citealp{Smith2009C}). The variability of the optical/NIR emission arises from jet activity and synchrotron emission, which is anisotropic and highly polarized. Consequently, any changes in the synchrotron emission, which manifest as variations in the optical/NIR flux, should be simultaneously reflected in the optical polarization degree (PD). This results in near-zero delays between the optical/NIR bands and the PD, consistent with our expectations. This observation reflects that both the optical/NIR flux and the polarization degree are driven by the same physical processes within the jet.

\begin{figure*}[t]
\centering
\includegraphics[width=2\columnwidth]{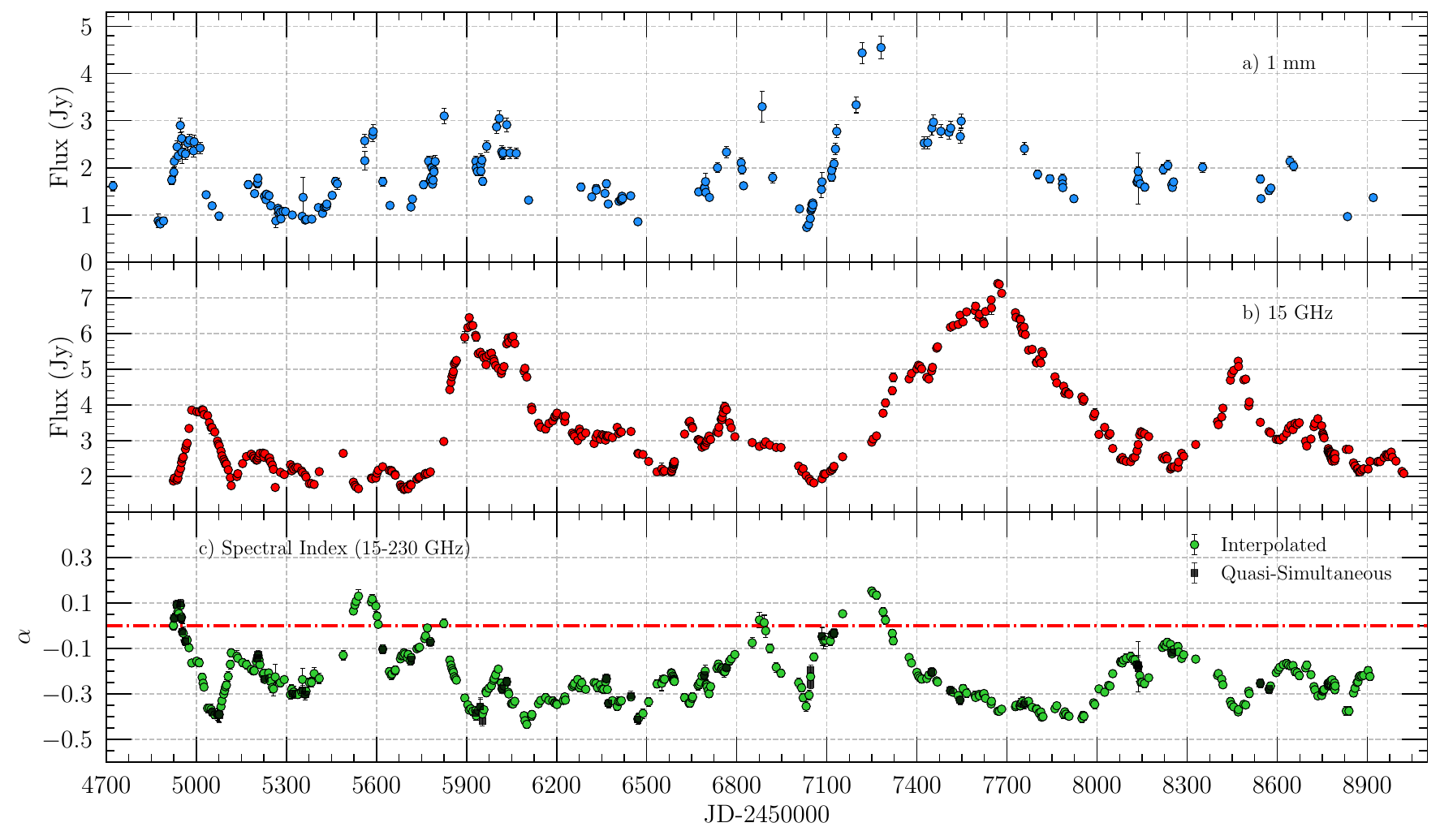}
\caption{Radio light curves. (a) 1 mm, (b) 15 GHz, and (c) Spectral index between 15 GHz and 1 mm (230 GHz). In panel (c), green points indicate estimations through interpolation, while black squares denote available quasi-simultaneous epochs. The horizontal dot-dashed line represents $\alpha=0$.}
\label{fig:spectralindex}
\end{figure*}

We observed a delay of $-40.3\pm18.7$ days between the 15 GHz and 1 mm emissions, indicating that the 1 mm emission precedes the 15 GHz emission. This delay aligns with the expected gradual energy loss (electron cooling) through synchrotron radiation. High-energy electrons emit synchrotron radiation at higher frequencies as they travel downstream in the jet, and subsequently cool down to emit at lower frequencies. This process creates a time lag between emissions at different frequencies. The lag between radio flares at various frequencies roughly corresponds to the radiative cooling time of the relativistic electrons responsible for the lower-frequency emission \citep{Urry1997, Urry1999, BaiAndLee2003}. Using the expression from \citet{Deng2008}, we can estimate the time delay as follows:
\begin{equation}
\tau\approx2\times10^{4}[(1+z)/\delta]^{-1/2}B^{-3/2}(\nu_{2}^{-1/2}-\nu_{1}^{-1/2})
\end{equation}
Here, $z$ represents the redshift, $\delta$  is the Doppler factor, $B$ is the magnetic field in Gauss, and $\nu$ is the frequency in units of $10^{15}$ Hz, with $\nu_{1}$ being the higher frequency and $\nu_{2}$ the lower frequency. Given the redshift of PKS 1510-089 ($z=0.361$) and the radio frequencies involved ($\nu_{1}=230\text{ GHz}$, and $\nu_{2}=15\text{ GHz}$), we can estimate the time delay between these bands. This estimate should closely align with the results obtained from cross-correlation analysis. There is a wide range of magnetic field estimates for PKS 1510-089. \citet{Malmrose2011} reported values between $0.1$ and $1$ G, while \citet{Kataoka2008} found values of $0.86$ and $1.3$ G, also \citet{Castignani2017} estimated the magnetic field to be between $2.2$ and $3.8$ G, and \citet{Ghisellini1998} reported values ranging from $0.061$ G during a low-activity state to $5.89$ G during a high-activity state. Assuming a Doppler factor of $\delta=10$ \citep{Deng2008} and a mean magnetic field strength of $2$ G, we estimate a time delay of $\tau\approx48.63$ days between the emissions at 230 GHz (1 mm) and 15 GHz, which is consistent with the delay determined through cross-correlation analysis.

To gain further insights into this phenomenon, we calculated the spectral index using flux densities measured at 15 GHz and 1 mm (230 GHz) during quasi-simultaneous epochs. Additionally, we interpolated the 15 GHz light curve to align with epochs of 1 mm observations, leveraging the higher sampling frequency of the 15 GHz data. For both scenarios, the spectral index was derived as follows.

\begin{equation}
    \alpha=\frac{\log(F_{15\text{ GHz}}/F_{230\text{ GHz}})}{\log(15\text{ GHz}/230\text{ GHz})}
\label{eq:SpectralIndex}
\end{equation}

\begin{figure*}[t]
\centering
\includegraphics[width=2\columnwidth]{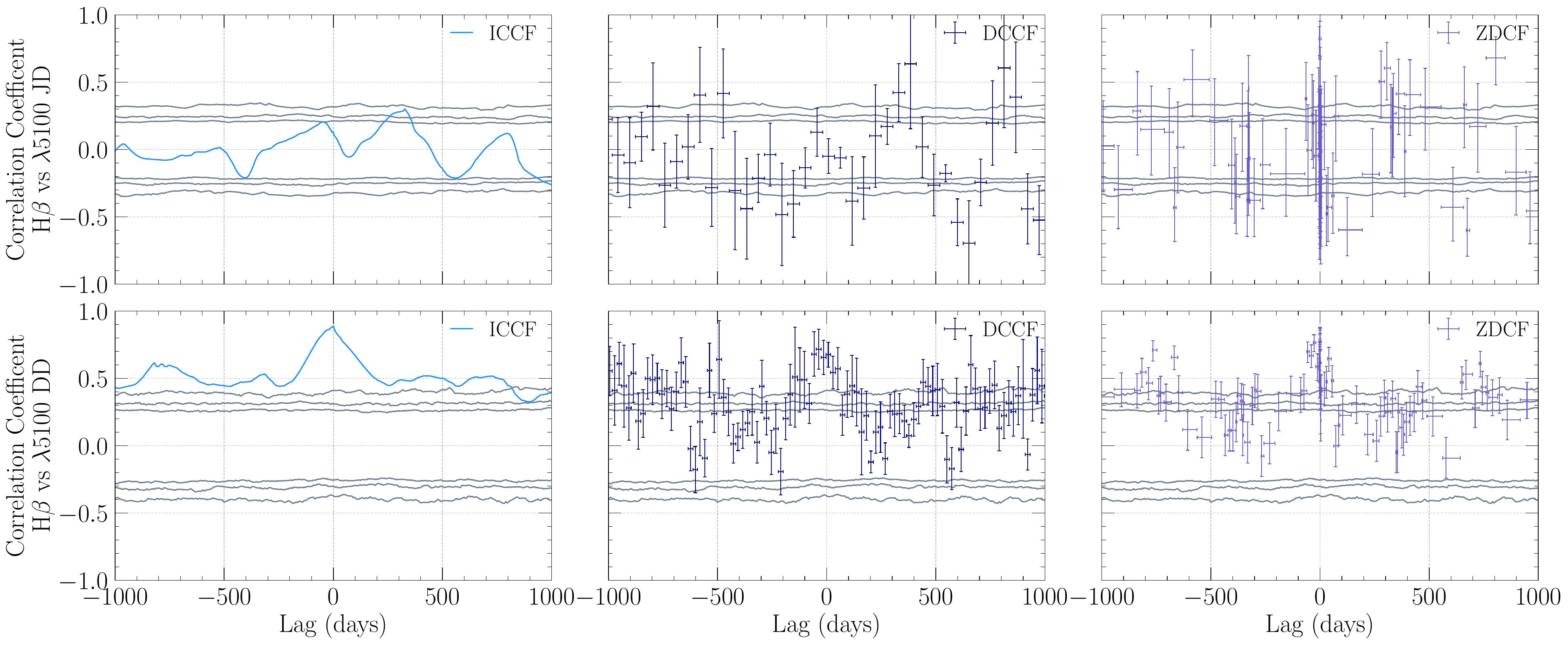}
\caption{Cross-correlation plots for the H$\beta$ emission line and $\lambda5100$ Å continuum light curves, categorized by disk dominance (DD) and jet dominance (JD) regimes as detailed in \autoref{sec:spectra}. Top row: Jet dominance light curves. Bottom row: Disk dominance (DD) light curves. Columns: Left column: Interpolation Method (ICCF). Middle column: Discrete Cross-Correlation Function (DCCF). Right column: Z-Transformed Cross-Correlation Function (ZDCF). Gray lines depict significance levels at $90\%$, $95\%$, and $99\%$.}
\label{fig:CCforJDandDD}
\end{figure*}

Where $F_{15\text{ GHz}}$ is the 15 GHz flux and $F_{230\text{ GHz}}$ is the 230 GHz (1 mm) flux. This parameter characterizes the flux distribution across frequencies, specifically between the 15 GHz and 230 GHz emission regions. A spectral index value of $\alpha=0$ typically delineates the boundary between an optically thin ($\alpha<0$) and optically thick ($\alpha>0$) synchrotron emission regions \citep{Fromm2011, ParkAndTrippe2014}. Throughout nearly the entire observation period, the spectral index remains in the optically thin regime ($\alpha<0$, see \autoref{fig:spectralindex}c).

Instances, where a positive spectral index occurs, coincide with flare-like events, notably observed in the NTD (jet dominance regime), X-rays, and $\gamma$-rays light curves during periods in mid-2009, early-2011, early-2012, late-2014, and mid-2015. The fluctuations in the spectral index indicate variations in the physical properties of the emission region, primarily reflecting changes in optical depth \citep{Trippe2011}. According to \citet{Park2019}, components approaching the core exhibit optically thick radio spectra, whereas those further away typically show optically thin spectra. This scenario correlates well with the observed flare-like events. Mobile components passing through the radio core have been observed during epochs when the radio spectrum becomes optically thick, such as in mid-2009 \citep{Marscher2010}, early-2011 \citep{Orienti2013}, early-2012 where $\alpha$ values hover near zero \citep{Aleksic2014}, and mid-2015 \citep{Ahnen2017}.

\section{Discussion}\label{sec:discussion}
\subsection{$\gamma$-ray and X-ray Dominant Emission Mechanism}\label{sec:highenergies}

In \autoref{sec:variability}, we found that the different bands exhibited varying degrees of variability, with $\gamma$-rays being the most variable. We identified seven flare-like events in the $\gamma$-ray light curve, occurring around JD$_{245}\approx4900, 5200, 5700, 6000, 6600, 7150, 7650$. Notably, at least four detections of components being ejected from the radio core coincide with these $\gamma$-ray flare intervals: mid-2009 \citep{Marscher2010}, early-2011 \citep{Orienti2013}, early-2012 \citep{Aleksic2014}, and mid-2015 \citep{Ahnen2017}. This suggests a link between component ejections and brief increases in $\gamma$-ray flux (flare events). The dominant emission mechanism during these flare-like events appears to be SSC, due to the quasi-co-spatiality (near-zero delay of $7.0\pm7.7$ days) between the optical/NIR emission region and the source region of the seed photons responsible for the $\gamma$-ray emission detected by Fermi/LAT.

However, both SSC and EC should be happening at the same time, since none of the photon fields are zero at any time. The key question is which emission mechanism dominates during flare events. Several scenarios could explain the observed near-zero delay between the optical/NIR and $\gamma$-ray emissions. If the primary source of optical emission during high-variability states (i.e., synchrotron emission from the jet) is located in the same region as the seed photons for $\gamma$-ray production. The co-spatiality would suggest that SSC scattering is the dominant mechanism during these flare events. Alternatively, the near-zero delay might indicate the ejection of relativistic particles into the jet (such as a component seen in VLBI maps). In this case, the electrons within the jet could be interacting with external photon fields near the base of the jet, pointing to EC scattering as the dominant mechanism. To distinguish between these possibilities, a detailed review of VLBI studies of this source (particularly those focusing on ejection events and interactions between VLBI components) would be valuable. Additionally, studying quasi-simultaneous spectral energy distributions (SEDs) during both flaring and quiescent states would provide further insights into the physical processes at play. Similar delays have been observed for PKS 1510-089 between the leading optical/NIR bands and $\gamma$-ray emission: $13\pm1$ days employing the R-band \citep{Abdo2010}, $1-2$ days \citep{Nalewajko2012}, zero delay \citep{Castignani2017} using the V-band, and \citet{Prince2019} found three delays: $-20$ days, zero delay and $52$ days utilizing the B-band.

Both $\gamma$-ray and radio emissions are anticipated to originate from the jet, suggesting that a correlation between them would indicate ongoing activity within the jet. For instance, \citet{LeonTavares2011} observed a trend in a sample of blazars observed by Fermi/LAT, where $\gamma$-ray flares coincided with the initiation of millimeter radio flares as a component traversed the radio core. However, \citet{Prince2019} reported no such correlation for PKS 1510-089 in their analysis of $\gamma$-ray versus 15 GHz and 1 mm data, aligning with our results. One challenge in establishing a correlation between the $\gamma$-rays and the radio is the difference in the duration of the flares, which is primarily attributed to the size of the emission regions. This is evident in \autoref{fig:photcurves}; although there are at least 4 flares (JD$_{245}\sim4950,5900,8500$, and throughout 2016) in the radio light curve that could be associated with $\gamma$-ray activity, their durations range from months to years, most likely diluting possible correlations.

Our findings regarding the lack of correlation between $\gamma$-rays and X-rays are consistent with previous studies by \citet{Abdo2010}, and \citet{Prince2019}. The observed delay between the leading optical emission and the X-rays rules out several emission mechanisms such as synchrotron radiation, black body emission from the inner part of the accretion disk, and IC scattering from the hot corona as the main emission mechanisms. This is because, in the case of synchrotron emission, X-rays would be expected to occur before optical emission; for the other two mechanisms, the correlation is not clearly justifiable, since we know the optical variability is dominated by synchrotron from the jet. With our analysis, we could not determine the origin of the seed photons, therefore, we can only deduce that the dominant emission mechanism is IC scattering within the jet. \citet{Kataoka2008} and \citet{Marscher2010} suggest that the electrons could scatter infrared (IR) seed photons from the sheath of the jet or an external source, such as a dusty torus, to produce the X-ray emission. This scenario aligns with our findings, reinforcing the role of the jet in the complex emission dynamics observed.

\subsection{Differences in Spectral Features for Different Regimes}\label{sec:spectralfeat}

The cross-correlations of the full data set between the $\lambda5100\text{ \AA}$ continuum and the H$\beta$, and H$\gamma$ emission line fluxes revealed a delay of approximately $80\pm6$ days. Initially, this delay could not be straightforwardly interpreted as indicative of the size of a virialized BLR. If indeed related to the canonical BLR, such a delay would typically reflect the outer radius of the BLR, from the accretion disk. Subsequently, we stratified the dataset into disk dominance and jet dominance regimes for further analysis. When performing separate cross-correlations for each regime, we were unable to reproduce the delay observed in the full dataset. In the disk dominance regime, the findings were inconclusive, largely due to extensive sections of the cross-correlation function exceeding $99\%$ significance levels (refer to \autoref{fig:CCforJDandDD}). This behavior stemmed from the similarity in the light curves, characterized by prolonged periods of quiescence with minimal variability, possibly suggesting a nearly constant accretion rate over the observation period. In contrast, in the jet dominance regime, any delays identified were deemed spurious correlations, likely attributable to either the limited number of data points, the sparse frequency of observations, or a combination thereof. 

We suspect that the delay is instead tracing the distance between the edge of the BLR, and the source of the optical synchrotron emission during flares. Another possibility is the existence of BLR material outside of the canonical BLR (e.g. \citealp{LeonTavares2013, Isler2013, Chavushyan2020}), that is ionized whenever there is a flare in the jet. A deeper study of the behavior of the emission line is needed in order to determine what scenario is more likely, which is underway right now for the second paper of this series.

\section{Summary}\label{sec:summary}

PKS 1510-089, a highly active blazar known for frequent flare-like events across various wavelengths, was the focus of our study involving multi-wavelength light curves spanning approximately 10 years. Our analysis primarily utilized optical spectroscopic data from the OAGH and the Steward Observatory. These optical spectra enabled measurements of prominent emission lines such as H$\beta$, and H$\gamma$, as well as the $\lambda5100\text{ \AA}$ continuum. In conjunction with these spectroscopic data, we integrated and analyzed public photometric observations spanning multiple wavelength bands, ranging from radio to $\gamma$-rays. Our comprehensive analyses yielded the following key findings:

\begin{enumerate}
    \item Delays between optical/IR emissions ($\lambda5100\text{ \AA}$ continuum, J-band, V-band, and NTD) are consistent with zero, indicating nearly simultaneous emissions from co-spatial regions. The variability in the continuum emission is primarily driven by periods of flare-like activity, indicated by NTD values greater than 2. This association suggests that $\lambda5100\text{ \AA}$ continuum, J-band, and V-band emissions predominantly stem from synchrotron processes within the jet.
    \item The observed delay of $7.0\pm7.7$ days between optical/IR bands and $\gamma$-ray emissions suggests a close spatial relationship between the optical emission region and the source of seed photons for $\gamma$-rays. The inverse Compton (IC) emission detected by Fermi/LAT aligns with seed photons originating from optical/NIR energies, which during flares are primarily emitted within the jet. Further analysis is needed to determine whether the dominant emission mechanism is SSC or EC.
    \item The observed delay where optical emissions precede X-ray emissions excludes several emission mechanisms such as synchrotron radiation and black body emission from the inner accretion disk or inverse Compton from a hot corona. This points towards inverse Compton from the jet as the dominant emission mechanism. The lack of correlation between $\gamma$-rays and X-rays across our study and literature suggests that both the jet sheath and external regions may serve as sources for seed photons.
    \item Spectral index shifts between optically thin ($\alpha<0$) and thick ($\alpha>0$) epochs are noted, particularly during flare-like events in mid-2009, early-2011, early-2012, late-2014, and mid-2015. These periods coincide with ejections from the radio core and mobile components traversing it, evident in optically thick radio spectra during flare-like events.
    \item Cross-correlations across the full dataset between the $\lambda5100\text{ \AA}$ continuum and H$\beta$, and H$\gamma$ emission line fluxes revealed a delay of approximately $80\pm6$ days. This delay signifies the separation between the continuum source and the BLR, in which we cannot assure the influence of the jet on the emission. Thus, this delay cannot be interpreted as the size of a virialized BLR. In the disk dominance regime, inconclusive results were obtained due to significant portions of cross-correlation functions exceeding $99\%$ significance, reflecting similar light curves and quiescent periods with minimal variability.
\end{enumerate}

In forthcoming papers (II and III), we will present our studies of PKS 1510-089, investigating the connection between the jet and the broad-line region (BLR) and pinpointing the $\gamma$-ray emission zone using high-resolution Very Long Baseline Interferometry (VLBI) observations. These studies aim to enhance our understanding of PKS 1510-089 multi-wavelength variability and contribute to a broader knowledge of jet-BLR interactions and the origins of $\gamma$-ray emissions (and dominant emissions mechanism) in blazars.

\section*{Acknowledgments}
We thank the anonymous referee for the constructive comments that helped to improve the manuscript. A. A.-P. and A. G.-P. gratefully acknowledge the support received from the CONAHCYT (Consejo Nacional de Humanidades, Ciencia y Tecnología) program for their Ph.D. studies. This work was supported by CONAHCYT research grants 280789 and 320987. Furthermore, this research was made possible thanks to the generous assistance provided by the Max Planck Institute for Radio Astronomy (MPIfR) - Mexico Max Planck Partner Group led by V.M.P.-A. Data from the Steward Observatory spectropolarimetric monitoring project were used. This program is supported by Fermi Guest Investigator grants NNX08AW56G, NNX09AU10G, NNX12AO93G, and NNX15AU81G. This publication is based on data collected at the Observatorio Astrofísico Guillermo Haro (OAGH), Cananea, Sonora, Mexico, operated by the Instituto Nacional de Astrofísica, Óptica y Electrónica (INAOE). Funding for the OAGH has been provided by CONAHCYT. This paper has made use of up-to-date SMARTS optical/near-infrared light curves that are available at \url{www.astro.yale.edu/smarts/glast/home.php}. Special thanks to Mark A. Gurwell for providing the 1 mm flux density light-curve data from the Submillimeter Array, a collaborative effort between the Smithsonian Astrophysical Observatory and the Academia Sinica Institute of Astronomy and Astrophysics, funded by the Smithsonian Institution and the Academia Sinica. This research has made use of data from the OVRO 40 m monitoring program, supported by private funding from the California Insitute of Technology and the Max Planck Institute for Radio Astronomy, and by NASA grants NNX08AW31G, NNX11A043G, and NNX14AQ89G and NSF grants AST-0808050 and AST-1109911. Sebastian Kiehlmann graciously provided these specific OVRO data, not publicly accessible on the OVRO website through private communication.

\software{Astropy \citep{Astropy2013, Astropy2018, Astropy2022},
          IRAF \citep{Tody1986, Tody1993},
          FTOOLS \citep{Blackburn1995},
          XSPEC \citep{Arnaud1996},
          Fermitools (v 2.0.8),
          SciPy \citep{Virtanen2020}.
          }

\bibliography{ManuscriptPKS_I}{}

\begin{thebibliography}{}
\expandafter\ifx\csname natexlab\endcsname\relax\def\natexlab#1{#1}\fi
\providecommand{\url}[1]{\href{#1}{#1}}
\providecommand{\dodoi}[1]{doi:~\href{http://doi.org/#1}{\nolinkurl{#1}}}
\providecommand{\doeprint}[1]{\href{http://ascl.net/#1}{\nolinkurl{http://ascl.net/#1}}}
\providecommand{\doarXiv}[1]{\href{https://arxiv.org/abs/#1}{\nolinkurl{https://arxiv.org/abs/#1}}}

\bibitem[{Abdo {et~al.}(2010)Abdo, Ackermann, Agudo, Ajello, Allafort, Aller,
  Aller, Antolini, Arkharov, Axelsson, Bach, Baldini, Ballet, Barbiellini,
  Bastieri, Bechtol, Bellazzini, Berdyugin, Berenji, Blandford, Blinov, Bloom,
  Boettcher, Bonamente, Borgland, Bouvier, Bregeon, Brez, Brigida, Bruel,
  Buehler, Buemi, Burnett, Buson, Caliandro, Cameron, Caraveo, Carosati,
  Carrigan, Casandjian, Cavazzuti, Cecchi, Çelik, Chekhtman, Chen, Cheung,
  Chiang, Ciprini, Claus, Cohen-Tanugi, Conrad, Corbel, Costamante, Dermer,
  de~Angelis, de~Palma, Donato, do~Couto~e Silva, Drell, Dubois, Dumora,
  Farnier, Favuzzi, Fegan, Ferrara, Focke, Forné, Fortin, Fukazawa, Funk,
  Fusco, Gargano, Gasparrini, Gehrels, Germani, Giebels, Giglietto, Giordano,
  Giroletti, Glanzman, Godfrey, Grenier, Grove, Guiriec, Gurwell, Gusbar,
  Gómez, Hadasch, Hagen-Thorn, Hayashida, Hays, Horan, Hughes, Jóhannesson,
  Johnson, Johnson, Kamae, Katagiri, Kataoka, Kawai, Kimeridze, Knödlseder,
  Konstantinova, Kopatskaya, Koptelova, Kovalev, Kurtanidze, Kuss, Lahteenmaki,
  Lande, Larionov, Larionova, Larionova, Larsson, Latronico, Lee, Leto, Lister,
  Longo, Loparco, Lott, Lovellette, Lubrano, Madejski, Makeev, Massaro,
  Mazziotta, McConville, McEnery, McHardy, Michelson, Mitthumsiri, Mizuno,
  Moiseev, Monte, Monzani, Morozova, Morselli, Moskalenko, Murgia,
  Naumann-Godo, Nikolashvili, Nolan, Norris, Nuss, Ohno, Ohsugi, Okumura,
  Omodei, Orlando, Ormes, Ozaki, Paneque, Panetta, Parent, Pasanen, Pelassa,
  Pepe, Pesce-Rollins, Piron, Porter, Pushkarev, Rainò, Raiteri, Rando,
  Razzano, Reimer, Reimer, Reinthal, Ripken, Ritz, Roca-Sogorb, Rodriguez,
  Roth, Roustazadeh, Ryde, Sadrozinski, Sander, Scargle, Sgrò, Sigua, Smith,
  Sokolovsky, Spandre, Spinelli, Starck, Strickman, Suson, Takahashi,
  Takahashi, Takalo, Tanaka, Taylor, Thayer, Thayer, Thompson, Tibaldo,
  Tornikoski, Torres, Tosti, Tramacere, Trigilio, Troitsky, Umana, Usher,
  Vandenbroucke, Vasileiou, Vilchez, Villata, Vitale, Waite, Wang, Winer, Wood,
  Yang, Ylinen, \& Ziegler}]{Abdo2010}
Abdo, A.~A., Ackermann, M., Agudo, I., {et~al.} 2010, The Astrophysical
  Journal, 721, 1425, \dodoi{10.1088/0004-637X/721/2/1425}

\bibitem[{{Abdollahi} {et~al.}(2022){Abdollahi}, {Acero}, {Baldini}, {Ballet},
  {Bastieri}, {Bellazzini}, {Berenji}, {Berretta}, {Bissaldi}, {Blandford},
  {Bloom}, {Bonino}, {Brill}, {Britto}, {Bruel}, {Burnett}, {Buson}, {Cameron},
  {Caputo}, {Caraveo}, {Castro}, {Chaty}, {Cheung}, {Chiaro}, {Cibrario},
  {Ciprini}, {Coronado-Bl{\'a}zquez}, {Crnogorcevic}, {Cutini}, {D'Ammando},
  {De Gaetano}, {Digel}, {Di Lalla}, {Dirirsa}, {Di Venere}, {Dom{\'\i}nguez},
  {Fallah Ramazani}, {Fegan}, {Ferrara}, {Fiori}, {Fleischhack}, {Franckowiak},
  {Fukazawa}, {Funk}, {Fusco}, {Galanti}, {Gammaldi}, {Gargano}, {Garrappa},
  {Gasparrini}, {Giacchino}, {Giglietto}, {Giordano}, {Giroletti}, {Glanzman},
  {Green}, {Grenier}, {Grondin}, {Guillemot}, {Guiriec}, {Gustafsson},
  {Harding}, {Hays}, {Hewitt}, {Horan}, {Hou}, {J{\'o}hannesson}, {Karwin},
  {Kayanoki}, {Kerr}, {Kuss}, {Landriu}, {Larsson}, {Latronico},
  {Lemoine-Goumard}, {Li}, {Liodakis}, {Longo}, {Loparco}, {Lott}, {Lubrano},
  {Maldera}, {Malyshev}, {Manfreda}, {Mart{\'\i}-Devesa}, {Mazziotta}, {Mereu},
  {Meyer}, {Michelson}, {Mirabal}, {Mitthumsiri}, {Mizuno}, {Moiseev},
  {Monzani}, {Morselli}, {Moskalenko}, {Negro}, {Nuss}, {Omodei}, {Orienti},
  {Orlando}, {Paneque}, {Pei}, {Perkins}, {Persic}, {Pesce-Rollins},
  {Petrosian}, {Pillera}, {Poon}, {Porter}, {Principe}, {Rain{\`o}}, {Rando},
  {Rani}, {Razzano}, {Razzaque}, {Reimer}, {Reimer}, {Reposeur},
  {S{\'a}nchez-Conde}, {Saz Parkinson}, {Scotton}, {Serini}, {Sgr{\`o}},
  {Siskind}, {Smith}, {Spandre}, {Spinelli}, {Sueoka}, {Suson}, {Tajima},
  {Tak}, {Thayer}, {Thompson}, {Torres}, {Troja}, {Valverde}, {Wood}, \&
  {Zaharijas}}]{Abdollahi2022}
{Abdollahi}, S., {Acero}, F., {Baldini}, L., {et~al.} 2022, \apjs, 260, 53,
  \dodoi{10.3847/1538-4365/ac6751}

\bibitem[{{Ahnen} {et~al.}(2017){Ahnen}, {Ansoldi, S.}, {Antonelli, L. A.},
  {Arcaro, C.}, {Babi\'{}c, A.}, {Banerjee, B.}, {Bangale, P.}, {Barres de
  Almeida, U.}, {Barrio, J. A.}, {Bednarek, W.}, {Bernardini, E.}, {Berti, A.},
  {Biasuzzi, B.}, {Biland, A.}, {Blanch, O.}, {Bonnefoy, S.}, {Bonnoli, G.},
  {Borracci, F.}, {Bretz, T.}, {Carosi, R.}, {Carosi, A.}, {Chatterjee, A.},
  {Colin, P.}, {Colombo, E.}, {Contreras, J. L.}, {Cortina, J.}, {Covino, S.},
  {Cumani, P.}, {Da Vela, P.}, {Dazzi, F.}, {De Angelis, A.}, {De Lotto, B.},
  {de O\~na Wilhelmi, E.}, {Di Pierro, F.}, {Doert, M.}, {Dom\'{\i}nguez, A.},
  {Dominis Prester, D.}, {Dorner, D.}, {Doro, M.}, {Einecke, S.}, {Eisenacher
  Glawion, D.}, {Elsaesser, D.}, {Engelkemeier, M.}, {Fallah Ramazani, V.},
  {Fern\'andez-Barral, A.}, {Fidalgo, D.}, {Fonseca, M. V.}, {Font, L.},
  {Fruck, C.}, {Galindo, D.}, {Garc\'{\i}a L\'opez, R. J.}, {Garczarczyk, M.},
  {Gaug, M.}, {Giammaria, P.}, {Godinovi\'{}c, N.}, {Gora, D.}, {Guberman, D.},
  {Hadasch, D.}, {Hahn, A.}, {Hassan, T.}, {Hayashida, M.}, {Herrera, J.},
  {Hose, J.}, {Hrupec, D.}, {Hughes, G.}, {Ishio, K.}, {Konno, Y.}, {Kubo, H.},
  {Kushida, J.}, {Kuvezdi\'{}c, D.}, {Lelas, D.}, {Lindfors, E.}, {Lombardi,
  S.}, {Longo, F.}, {L\'opez, M.}, {Majumdar, P.}, {Makariev, M.}, {Maneva,
  G.}, {Manganaro, M.}, {Mannheim, K.}, {Maraschi, L.}, {Mariotti, M.},
  {Mart\'{\i}nez, M.}, {Mazin, D.}, {Menzel, U.}, {Mirzoyan, R.}, {Moralejo,
  A.}, {Moretti, E.}, {Nakajima, D.}, {Neustroev, V.}, {Niedzwiecki, A.},
  {Nievas Rosillo, M.}, {Nilsson, K.}, {Nishijima, K.}, {Noda, K.}, {Nogu\'es,
  L.}, {Paiano, S.}, {Palacio, J.}, {Palatiello, M.}, {Paneque, D.}, {Paoletti,
  R.}, {Paredes, J. M.}, {Paredes-Fortuny, X.}, {Pedaletti, G.}, {Peresano,
  M.}, {Perri, L.}, {Persic, M.}, {Poutanen, J.}, {Prada Moroni, P. G.},
  {Prandini, E.}, {Puljak, I.}, {Garcia, J. R.}, {Reichardt, I.}, {Rhode, W.},
  {Rib\'o, M.}, {Rico, J.}, {Saito, T.}, {Satalecka, K.}, {Schroeder, S.},
  {Schweizer, T.}, {Shore, S. N.}, {Sillanp\"a\"a, A.}, {Sitarek, J.},
  {Snidari\'{}c, I.}, {Sobczynska, D.}, {Stamerra, A.}, {Strzys, M.},
  {Suri\'{}c, T.}, {Takalo, L.}, {Tavecchio, F.}, {Temnikov, P.}, {Terzi\'{}c,
  T.}, {Tescaro, D.}, {Teshima, M.}, {Torres, D. F.}, {Torres-Alb\`a, N.},
  {Toyama, T.}, {Treves, A.}, {Vanzo, G.}, {Vazquez Acosta, M.}, {Vovk, I.},
  {Ward, J. E.}, {Will, M.}, {Wu, M. H.}, {Zari\'{}c, D.}, {Desiante, R.},
  {Becerra Gonz\'alez, J.}, {D\'{}Ammando, F.}, {Larsson, S.}, {Raiteri, C.
  M.}, {Reinthal, R.}, {L\"ahteenm\"aki, A.}, {J\"arvel\"a, E.}, {Tornikoski,
  M.}, {Ramakrishnan, V.}, {Jorstad, S. G.}, {Marscher, A. P.}, {Bala, V.},
  {MacDonald, N. R.}, {Kaur, N.}, {Sameer}, {Baliyan, K.}, {Acosta-Pulido, J.
  A.}, {Lazaro, C.}, {Mart\'{\i}-nez-Lombilla, C.}, {Grinon-Marin, A. B.},
  {Pastor Yabar, A.}, {Protasio, C.}, {Carnerero, M. I.}, {Jermak, H.},
  {Steele, I. A.}, {Larionov, V. M.}, {Borman, G. A.}, \& {Grishina, T.
  S.}}]{Ahnen2017}
{Ahnen}, M.~L., {Ansoldi, S.}, {Antonelli, L. A.}, {et~al.} 2017, A\&A, 603,
  A29, \dodoi{10.1051/0004-6361/201629960}

\bibitem[{{Aleksić} {et~al.}(2014){Aleksić}, {Ansoldi, S.}, {Antonelli, L.
  A.}, {Antoranz, P.}, {Babic, A.}, {Bangale, P.}, {Barres de Almeida, U.},
  {Barrio, J. A.}, {Becerra Gonz\'alez, J.}, {Bednarek, W.}, {Bernardini, E.},
  {Biland, A.}, {Blanch, O.}, {Bonnefoy, S.}, {Bonnoli, G.}, {Borracci, F.},
  {Bretz, T.}, {Carmona, E.}, {Carosi, A.}, {Carreto Fidalgo, D.}, {Colin, P.},
  {Colombo, E.}, {Contreras, J. L.}, {Cortina, J.}, {Covino, S.}, {Da Vela,
  P.}, {Dazzi, F.}, {De Angelis, A.}, {De Caneva, G.}, {De Lotto, B.}, {Delgado
  Mendez, C.}, {Doert, M.}, {Dom\'{\i}nguez, A.}, {Dominis Prester, D.},
  {Dorner, D.}, {Doro, M.}, {Einecke, S.}, {Eisenacher, D.}, {Elsaesser, D.},
  {Farina, E.}, {Ferenc, D.}, {Fonseca, M. V.}, {Font, L.}, {Frantzen, K.},
  {Fruck, C.}, {Garc\'{\i}a L\'opez, R. J.}, {Garczarczyk, M.}, {Garrido
  Terrats, D.}, {Gaug, M.}, {Godinovi\'{}c, N.}, {Gonz\'alez Mu\~noz, A.},
  {Gozzini, S. R.}, {Hadasch, D.}, {Hayashida, M.}, {Herrera, J.}, {Herrero,
  A.}, {Hildebrand, D.}, {Hose, J.}, {Hrupec, D.}, {Idec, W.}, {Kadenius, V.},
  {Kellermann, H.}, {Kodani, K.}, {Konno, Y.}, {Krause, J.}, {Kubo, H.},
  {Kushida, J.}, {La Barbera, A.}, {Lelas, D.}, {Lewandowska, N.}, {Lindfors,
  E.}, {Lombardi, S.}, {L\'opez, M.}, {L\'opez-Coto, R.}, {L\'opez-Oramas, A.},
  {Lorenz, E.}, {Lozano, I.}, {Makariev, M.}, {Mallot, K.}, {Maneva, G.},
  {Mankuzhiyil, N.}, {Mannheim, K.}, {Maraschi, L.}, {Marcote, B.}, {Mariotti,
  M.}, {Mart\'{\i}nez, M.}, {Mazin, D.}, {Menzel, U.}, {Meucci, M.}, {Miranda,
  J. M.}, {Mirzoyan, R.}, {Moralejo, A.}, {Munar-Adrover, P.}, {Nakajima, D.},
  {Niedzwiecki, A.}, {Nilsson, K.}, {Nishijima, K.}, {Noda, K.}, {Nowak, N.},
  {Orito, R.}, {Overkemping, A.}, {Paiano, S.}, {Palatiello, M.}, {Paneque,
  D.}, {Paoletti, R.}, {Paredes, J. M.}, {Paredes-Fortuny, X.}, {Partini, S.},
  {Persic, M.}, {Prada, F.}, {Prada Moroni, P. G.}, {Prandini, E.}, {Preziuso,
  S.}, {Puljak, I.}, {Reinthal, R.}, {Rhode, W.}, {Rib\'o, M.}, {Rico, J.},
  {Rodriguez Garcia, J.}, {R\"ugamer, S.}, {Saggion, A.}, {Saito, T.}, {Saito,
  K.}, {Satalecka, K.}, {Scalzotto, V.}, {Scapin, V.}, {Schultz, C.},
  {Schweizer, T.}, {Shore, S. N.}, {Sillanp\"a\"a, A.}, {Sitarek, J.},
  {Snidaric, I.}, {Sobczynska, D.}, {Spanier, F.}, {Stamatescu, V.}, {Stamerra,
  A.}, {Steinbring, T.}, {Storz, J.}, {Strzys, M.}, {Sun, S.}, {Suri\'{}c, T.},
  {Takalo, L.}, {Takami, H.}, {Tavecchio, F.}, {Temnikov, P.}, {Terzi\'{}c,
  T.}, {Tescaro, D.}, {Teshima, M.}, {Thaele, J.}, {Tibolla, O.}, {Torres, D.
  F.}, {Toyama, T.}, {Treves, A.}, {Uellenbeck, M.}, {Vogler, P.}, {Wagner, R.
  M.}, {Zandanel, F.}, {Zanin, R.}, {(the MAGIC Collaboration)}, {Lucarelli,
  F.}, {Pittori, C.}, {Vercellone, S.}, {Verrecchia, F.}, {(for the AGILE
  Collaboration)}, {Buson, S.}, {D\'{}Ammando, F.}, {Stawarz, L.}, {Giroletti,
  M.}, {Orienti, M.}, {(for the Fermi-LAT Collaboration)}, {Mundell, C.},
  {Steele, I.}, {Zarpudin, B.}, {Raiteri, C. M.}, {Villata, M.}, {Sandrinelli,
  A.}, {L\"ahteenm\"aki, A.}, {Tammi, J.}, {Tornikoski, M.}, {Hovatta, T.},
  {Readhead, A. C. S.}, {Max-Moerbeck, W.}, {Richards, J. L.}, {Jorstad, S.},
  {Marscher, A.}, {Gurwell, M. A.}, {Larionov, V. M.}, {Blinov, D. A.},
  {Konstantinova, T. S.}, {Kopatskaya, E. N.}, {Larionova, L. V.}, {Larionova,
  E. G.}, {Morozova, D. A.}, {Troitsky, I. S.}, {Mokrushina, A. A.}, {Pavlova,
  Yu. V.}, {Chen, W. P.}, {Lin, H. C.}, {Panwar, N.}, {Agudo, I.}, {Casadio,
  C.}, {G\'omez, J. L.}, {Molina, S. N.}, {Kurtanidze, O. M.}, {Nikolashvili,
  M. G.}, {Kurtanidze, S. O.}, {Chigladze, R. A.}, {Acosta-Pulido, J. A.},
  {Carnerero, M. I.}, {Manilla-Robles, A.}, {Ovcharov, E.}, {Bozhilov, V.},
  {Metodieva, I.}, {Aller, M. F.}, {Aller, H. D.}, {Fuhrman, L.}, {Angelakis,
  E.}, {Nestoras, I.}, {Krichbaum, T. P.}, {Zensus, J. A.}, {Ungerechts, H.},
  \& {Sievers, A.}}]{Aleksic2014}
{Aleksić}, J., {Ansoldi, S.}, {Antonelli, L. A.}, {et~al.} 2014, A\&A, 569,
  A46, \dodoi{10.1051/0004-6361/201423484}

\bibitem[{Alexander(1997)}]{Alexander1997}
Alexander, T. 1997, in Astronomical Time Series, ed. D.~Maoz, A.~Sternberg, \&
  E.~M. Leibowitz (Dordrecht: Springer Netherlands), 163--166

\bibitem[{Amaya-Almaz\'an {et~al.}(2022)Amaya-Almaz\'an, Chavushyan, \&
  {Pati{\~n}o-{\'A}lvarez}}]{AmayaAlmazan2022}
Amaya-Almaz\'an, R.~A., Chavushyan, V., \& {Pati{\~n}o-{\'A}lvarez}, V.~M.
  2022, The Astrophysical Journal, 929, 14, \dodoi{10.3847/1538-4357/ac5741}

\bibitem[{{Arnaud}(1996)}]{Arnaud1996}
{Arnaud}, K.~A. 1996, in Astronomical Society of the Pacific Conference Series,
  Vol. 101, Astronomical Data Analysis Software and Systems V, ed. G.~H.
  {Jacoby} \& J.~{Barnes}, 17

\bibitem[{{Astropy Collaboration} {et~al.}(2013){Astropy Collaboration},
  {Robitaille}, {Tollerud}, {Greenfield}, {Droettboom}, {Bray}, {Aldcroft},
  {Davis}, {Ginsburg}, {Price-Whelan}, {Kerzendorf}, {Conley}, {Crighton},
  {Barbary}, {Muna}, {Ferguson}, {Grollier}, {Parikh}, {Nair}, {Unther},
  {Deil}, {Woillez}, {Conseil}, {Kramer}, {Turner}, {Singer}, {Fox}, {Weaver},
  {Zabalza}, {Edwards}, {Azalee Bostroem}, {Burke}, {Casey}, {Crawford},
  {Dencheva}, {Ely}, {Jenness}, {Labrie}, {Lim}, {Pierfederici}, {Pontzen},
  {Ptak}, {Refsdal}, {Servillat}, \& {Streicher}}]{Astropy2013}
{Astropy Collaboration}, {Robitaille}, T.~P., {Tollerud}, E.~J., {et~al.} 2013,
  \aap, 558, A33, \dodoi{10.1051/0004-6361/201322068}

\bibitem[{{Astropy Collaboration} {et~al.}(2018){Astropy Collaboration},
  {Price-Whelan}, {Sip{\H{o}}cz}, {G{\"u}nther}, {Lim}, {Crawford}, {Conseil},
  {Shupe}, {Craig}, {Dencheva}, {Ginsburg}, {VanderPlas}, {Bradley},
  {P{\'e}rez-Su{\'a}rez}, {de Val-Borro}, {Aldcroft}, {Cruz}, {Robitaille},
  {Tollerud}, {Ardelean}, {Babej}, {Bach}, {Bachetti}, {Bakanov}, {Bamford},
  {Barentsen}, {Barmby}, {Baumbach}, {Berry}, {Biscani}, {Boquien}, {Bostroem},
  {Bouma}, {Brammer}, {Bray}, {Breytenbach}, {Buddelmeijer}, {Burke},
  {Calderone}, {Cano Rodr{\'\i}guez}, {Cara}, {Cardoso}, {Cheedella}, {Copin},
  {Corrales}, {Crichton}, {D'Avella}, {Deil}, {Depagne}, {Dietrich}, {Donath},
  {Droettboom}, {Earl}, {Erben}, {Fabbro}, {Ferreira}, {Finethy}, {Fox},
  {Garrison}, {Gibbons}, {Goldstein}, {Gommers}, {Greco}, {Greenfield},
  {Groener}, {Grollier}, {Hagen}, {Hirst}, {Homeier}, {Horton}, {Hosseinzadeh},
  {Hu}, {Hunkeler}, {Ivezi{\'c}}, {Jain}, {Jenness}, {Kanarek}, {Kendrew},
  {Kern}, {Kerzendorf}, {Khvalko}, {King}, {Kirkby}, {Kulkarni}, {Kumar},
  {Lee}, {Lenz}, {Littlefair}, {Ma}, {Macleod}, {Mastropietro}, {McCully},
  {Montagnac}, {Morris}, {Mueller}, {Mumford}, {Muna}, {Murphy}, {Nelson},
  {Nguyen}, {Ninan}, {N{\"o}the}, {Ogaz}, {Oh}, {Parejko}, {Parley}, {Pascual},
  {Patil}, {Patil}, {Plunkett}, {Prochaska}, {Rastogi}, {Reddy Janga},
  {Sabater}, {Sakurikar}, {Seifert}, {Sherbert}, {Sherwood-Taylor}, {Shih},
  {Sick}, {Silbiger}, {Singanamalla}, {Singer}, {Sladen}, {Sooley},
  {Sornarajah}, {Streicher}, {Teuben}, {Thomas}, {Tremblay}, {Turner},
  {Terr{\'o}n}, {van Kerkwijk}, {de la Vega}, {Watkins}, {Weaver}, {Whitmore},
  {Woillez}, {Zabalza}, \& {Astropy Contributors}}]{Astropy2018}
{Astropy Collaboration}, {Price-Whelan}, A.~M., {Sip{\H{o}}cz}, B.~M., {et~al.}
  2018, \aj, 156, 123, \dodoi{10.3847/1538-3881/aabc4f}

\bibitem[{{Astropy Collaboration} {et~al.}(2022){Astropy Collaboration},
  {Price-Whelan}, {Lim}, {Earl}, {Starkman}, {Bradley}, {Shupe}, {Patil},
  {Corrales}, {Brasseur}, {N{\"o}the}, {Donath}, {Tollerud}, {Morris},
  {Ginsburg}, {Vaher}, {Weaver}, {Tocknell}, {Jamieson}, {van Kerkwijk},
  {Robitaille}, {Merry}, {Bachetti}, {G{\"u}nther}, {Aldcroft},
  {Alvarado-Montes}, {Archibald}, {B{\'o}di}, {Bapat}, {Barentsen},
  {Baz{\'a}n}, {Biswas}, {Boquien}, {Burke}, {Cara}, {Cara}, {Conroy},
  {Conseil}, {Craig}, {Cross}, {Cruz}, {D'Eugenio}, {Dencheva}, {Devillepoix},
  {Dietrich}, {Eigenbrot}, {Erben}, {Ferreira}, {Foreman-Mackey}, {Fox},
  {Freij}, {Garg}, {Geda}, {Glattly}, {Gondhalekar}, {Gordon}, {Grant},
  {Greenfield}, {Groener}, {Guest}, {Gurovich}, {Handberg}, {Hart},
  {Hatfield-Dodds}, {Homeier}, {Hosseinzadeh}, {Jenness}, {Jones}, {Joseph},
  {Kalmbach}, {Karamehmetoglu}, {Ka{\l}uszy{\'n}ski}, {Kelley}, {Kern},
  {Kerzendorf}, {Koch}, {Kulumani}, {Lee}, {Ly}, {Ma}, {MacBride}, {Maljaars},
  {Muna}, {Murphy}, {Norman}, {O'Steen}, {Oman}, {Pacifici}, {Pascual},
  {Pascual-Granado}, {Patil}, {Perren}, {Pickering}, {Rastogi}, {Roulston},
  {Ryan}, {Rykoff}, {Sabater}, {Sakurikar}, {Salgado}, {Sanghi}, {Saunders},
  {Savchenko}, {Schwardt}, {Seifert-Eckert}, {Shih}, {Jain}, {Shukla}, {Sick},
  {Simpson}, {Singanamalla}, {Singer}, {Singhal}, {Sinha}, {Sip{\H{o}}cz},
  {Spitler}, {Stansby}, {Streicher}, {{\v{S}}umak}, {Swinbank}, {Taranu},
  {Tewary}, {Tremblay}, {de Val-Borro}, {Van Kooten}, {Vasovi{\'c}}, {Verma},
  {de Miranda Cardoso}, {Williams}, {Wilson}, {Winkel}, {Wood-Vasey}, {Xue},
  {Yoachim}, {Zhang}, {Zonca}, \& {Astropy Project Contributors}}]{Astropy2022}
{Astropy Collaboration}, {Price-Whelan}, A.~M., {Lim}, P.~L., {et~al.} 2022,
  \apj, 935, 167, \dodoi{10.3847/1538-4357/ac7c74}

\bibitem[{{Bai} \& {Lee}(2003)}]{BaiAndLee2003}
{Bai}, J.~M., \& {Lee}, M.~G. 2003, \apjl, 585, L113, \dodoi{10.1086/374304}

\bibitem[{B{\aa}th(1974)}]{Bath1974}
B{\aa}th, M. 1974, Spectral analysis in geophysics. Amsterdam,  Oxford-New York

\bibitem[{Bennert {et~al.}(2002)Bennert, Falcke, Schulz, Wilson, \&
  Wills}]{Bennert2002}
Bennert, N., Falcke, H., Schulz, H., Wilson, A.~S., \& Wills, B.~J. 2002, The
  Astrophysical Journal, 574, L105, \dodoi{10.1086/342420}

\bibitem[{{Blackburn}(1995)}]{Blackburn1995}
{Blackburn}, J.~K. 1995, in Astronomical Society of the Pacific Conference
  Series, Vol.~77, Astronomical Data Analysis Software and Systems IV, ed.
  R.~A. {Shaw}, H.~E. {Payne}, \& J.~J.~E. {Hayes}, 367

\bibitem[{Bonning {et~al.}(2012)Bonning, Urry, Bailyn, Buxton, Chatterjee,
  Coppi, Fossati, Isler, \& Maraschi}]{Bonning2012}
Bonning, E., Urry, C.~M., Bailyn, C., {et~al.} 2012, The Astrophysical Journal,
  756, 13, \dodoi{10.1088/0004-637X/756/1/13}

\bibitem[{{B{\"o}ttcher}(2007)}]{Bottcher2007}
{B{\"o}ttcher}, M. 2007, \apss, 309, 95, \dodoi{10.1007/s10509-007-9404-0}

\bibitem[{{B{\"o}ttcher} {et~al.}(2013){B{\"o}ttcher}, {Reimer}, {Sweeney}, \&
  {Prakash}}]{Bottcher2013}
{B{\"o}ttcher}, M., {Reimer}, A., {Sweeney}, K., \& {Prakash}, A. 2013, \apj,
  768, 54, \dodoi{10.1088/0004-637X/768/1/54}

\bibitem[{{Brown}(2013)}]{Brown2013}
{Brown}, A.~M. 2013, \mnras, 431, 824, \dodoi{10.1093/mnras/stt218}

\bibitem[{{Burbidge} \& {Kinman}(1966)}]{BurbidgeAndKinman1966}
{Burbidge}, E.~M., \& {Kinman}, T.~D. 1966, \apj, 145, 654,
  \dodoi{10.1086/148808}

\bibitem[{{Cardelli} {et~al.}(1989){Cardelli}, {Clayton}, \&
  {Mathis}}]{Cardelli1989}
{Cardelli}, J.~A., {Clayton}, G.~C., \& {Mathis}, J.~S. 1989, \apj, 345, 245,
  \dodoi{10.1086/167900}

\bibitem[{{Castignani} {et~al.}(2017){Castignani}, {Pian, E.}, {Belloni, T.
  M.}, {D\'{}Ammando, F.}, {Foschini, L.}, {Ghisellini, G.}, {Pursimo, T.},
  {Bazzano, A.}, {Beckmann, V.}, {Bianchin, V.}, {Fiocchi, M. T.}, {Impiombato,
  D.}, {Raiteri, C. M.}, {Soldi, S.}, {Tagliaferri, G.}, {Treves, A.}, \&
  {T\"urler, M.}}]{Castignani2017}
{Castignani}, G., {Pian, E.}, {Belloni, T. M.}, {et~al.} 2017, A\&A, 601, A30,
  \dodoi{10.1051/0004-6361/201629775}

\bibitem[{{Chavushyan} {et~al.}(2020){Chavushyan}, {Patiño-\'Alvarez},
  {Amaya-Almaz\'an}, \& {Carrasco}}]{Chavushyan2020}
{Chavushyan}, V., {Patiño-\'Alvarez}, V.~M., {Amaya-Almaz\'an}, R.~A., \&
  {Carrasco}, L. 2020, The Astrophysical Journal, 891, 68,
  \dodoi{10.3847/1538-4357/ab6ef6}

\bibitem[{{D'Ammando} {et~al.}(2009){D'Ammando}, {Pucella, G.}, {Raiteri, C.
  M.}, {Villata, M.}, {Vittorini, V.}, {Vercellone, S.}, {Donnarumma, I.},
  {Longo, F.}, {Tavani, M.}, {Argan, A.}, {Barbiellini, G.}, {Boffelli, F.},
  {Bulgarelli, A.}, {Caraveo, P.}, {Cattaneo, P. W.}, {Chen, A. W.}, {Cocco,
  V.}, {Costa, E.}, {Del Monte, E.}, {De Paris, G.}, {Di Cocco, G.},
  {Evangelista, Y.}, {Feroci, M.}, {Ferrari, A.}, {Fiorini, M.}, {Froysland,
  T.}, {Fuschino, F.}, {Galli, M.}, {Gianotti, F.}, {Giuliani, A.}, {Labanti,
  C.}, {Lapshov, I.}, {Lazzarotto, F.}, {Lipari, P.}, {Marisaldi, M.},
  {Mereghetti, S.}, {Morselli, A.}, {Pacciani, L.}, {Pellizzoni, A.}, {Perotti,
  F.}, {Piano, G.}, {Picozza, P.}, {Pilia, M.}, {Prest, M.}, {Rapisarda, M.},
  {Rappoldi, A.}, {Sabatini, S.}, {Soffitta, P.}, {Trifoglio, M.}, {Trois, A.},
  {Vallazza, E.}, {Zambra, A.}, {Zanello, D.}, {Agudo, I.}, {Aller, M. F.},
  {Aller, H. D.}, {Arkharov, A. A.}, {Bach, U.}, {Benitez, E.}, {Berdyugin,
  A.}, {Blinov, D. A.}, {Buemi, C. S.}, {Chen, W. P.}, {Di Paola, A.}, {Di
  Rico, G.}, {Dultzin, D.}, {Fuhrmann, L.}, {G\'omez, J. L.}, {Gurwell, M. A.},
  {Jorstad, S. G.}, {Heidt, J.}, {Hiriart, D.}, {Hsiao, H. Y.}, {Kimeridze,
  G.}, {Konstantinova, T. S.}, {Kopatskaya, E. N.}, {Koptelova, E.},
  {Kurtanidze, O.}, {Larionov, V. M.}, {Leto, P.}, {Lindfors, E.}, {Lopez, J.
  M.}, {Marscher, A. P.}, {McHardy, I. M.}, {Melnichuk, D. A.}, {Mommert, M.},
  {Mujica, R.}, {Nilsson, K.}, {Pasanen, M.}, {Roca-Sogorb, M.}, {Sorcia, M.},
  {Takalo, L. O.}, {Taylor, B.}, {Trigilio, C.}, {Troitsky, I. S.}, {Umana,
  G.}, {Antonelli, L. A.}, {Colafrancesco, S.}, {Cutini, S.}, {Gasparrini, D.},
  {Pittori, C.}, {Preger, B.}, {Santolamazza, P.}, {Verrecchia, F.}, {Giommi,
  P.}, \& {Salotti, L.}}]{DAmmando2009}
{D'Ammando}, F., {Pucella, G.}, {Raiteri, C. M.}, {et~al.} 2009, A\&A, 508,
  181, \dodoi{10.1051/0004-6361/200912560}

\bibitem[{{Deng} {et~al.}(2008){Deng}, {Bai}, {Zhang}, \& {Yang}}]{Deng2008}
{Deng}, W.-G., {Bai}, J.-M., {Zhang}, L., \& {Yang}, X. 2008, \cjaa, 8, 195,
  \dodoi{10.1088/1009-9271/8/2/06}

\bibitem[{{Dotson} {et~al.}(2015){Dotson}, {Georganopoulos}, {Meyer}, \&
  {McCann}}]{Dotson2015}
{Dotson}, A., {Georganopoulos}, M., {Meyer}, E.~T., \& {McCann}, K. 2015, \apj,
  809, 164, \dodoi{10.1088/0004-637X/809/2/164}

\bibitem[{Dzhatdoev {et~al.}(2022)Dzhatdoev, Khalikov, Latypova, Podlesnyi, \&
  Vaiman}]{Dzhatdoev2022}
Dzhatdoev, T.~A., Khalikov, E.~V., Latypova, V.~S., Podlesnyi, E.~I., \&
  Vaiman, I.~A. 2022, Monthly Notices of the Royal Astronomical Society, 515,
  5242, \dodoi{10.1093/mnras/stac2094}

\bibitem[{{Edelson} \& {Krolik}(1988)}]{EdelsonAndKrolik1988}
{Edelson}, R.~A., \& {Krolik}, J.~H. 1988, \apj, 333, 646,
  \dodoi{10.1086/166773}

\bibitem[{{Emmanoulopoulos} {et~al.}(2013){Emmanoulopoulos}, {McHardy},
  {Papadakis}, \& {Emmanoulopoulos}}]{Emmanoulopoulos2013}
{Emmanoulopoulos}, D., {McHardy}, I.~M., {Papadakis}, I.~E., \&
  {Emmanoulopoulos}, D. 2013, \mnras, 433, 907, \dodoi{10.1093/mnras/stt764}

\bibitem[{Fan {et~al.}(2018)Fan, Tao, Liu, Yuan, Sawangwit, Yang, Huang, Zhang,
  Zhang, Zhang, \& Zhu}]{Fan2018}
Fan, J.~H., Tao, J., Liu, Y., {et~al.} 2018, The Astronomical Journal, 155, 90,
  \dodoi{10.3847/1538-3881/aaa547}

\bibitem[{{Fromm} {et~al.}(2011){Fromm}, {Perucho}, {Ros}, {Savolainen},
  {Lobanov}, {Zensus}, {Aller}, {Aller}, {Gurwell}, \&
  {L{\"a}hteenm{\"a}ki}}]{Fromm2011}
{Fromm}, C.~M., {Perucho}, M., {Ros}, E., {et~al.} 2011, \aap, 531, A95,
  \dodoi{10.1051/0004-6361/201116857}

\bibitem[{{Fuhrmann} {et~al.}(2016){Fuhrmann}, {Angelakis}, {Zensus},
  {Nestoras}, {Marchili}, {Pavlidou}, {Karamanavis}, {Ungerechts}, {Krichbaum},
  {Larsson}, {Lee}, {Max-Moerbeck}, {Myserlis}, {Pearson}, {Readhead},
  {Richards}, {Sievers}, \& {Sohn}}]{Fuhrmann2016}
{Fuhrmann}, L., {Angelakis}, E., {Zensus}, J.~A., {et~al.} 2016, \aap, 596,
  A45, \dodoi{10.1051/0004-6361/201528034}

\bibitem[{{Gaskell} \& {Sparke}(1986)}]{GaskellAndSparke1986}
{Gaskell}, C.~M., \& {Sparke}, L.~S. 1986, \apj, 305, 175,
  \dodoi{10.1086/164238}

\bibitem[{{Ghisellini} {et~al.}(1998){Ghisellini}, {Celotti}, {Fossati},
  {Maraschi}, \& {Comastri}}]{Ghisellini1998}
{Ghisellini}, G., {Celotti}, A., {Fossati}, G., {Maraschi}, L., \& {Comastri},
  A. 1998, \mnras, 301, 451, \dodoi{10.1046/j.1365-8711.1998.02032.x}

\bibitem[{Greene \& Ho(2005)}]{GreeneAndHo2005}
Greene, J.~E., \& Ho, L.~C. 2005, The Astrophysical Journal, 630, 122,
  \dodoi{10.1086/431897}

\bibitem[{Gupta(2018)}]{Gupta2018}
Gupta, A.~C. 2018, Galaxies, 6, \dodoi{10.3390/galaxies6010001}

\bibitem[{{Gurwell} {et~al.}(2007){Gurwell}, {Peck}, {Hostler}, {Darrah}, \&
  {Katz}}]{Gurwell2007}
{Gurwell}, M.~A., {Peck}, A.~B., {Hostler}, S.~R., {Darrah}, M.~R., \& {Katz},
  C.~A. 2007, in Astronomical Society of the Pacific Conference Series, Vol.
  375, From Z-Machines to ALMA: (Sub)Millimeter Spectroscopy of Galaxies, ed.
  A.~J. {Baker}, J.~{Glenn}, A.~I. {Harris}, J.~G. {Mangum}, \& M.~S. {Yun},
  234

\bibitem[{Hartman {et~al.}(1999)Hartman, Bertsch, Bloom, Chen, Deines-Jones,
  Esposito, Fichtel, Friedlander, Hunter, McDonald, Sreekumar, Thompson, Jones,
  Lin, Michelson, Nolan, Tompkins, Kanbach, Mayer-Hasselwander, Mücke, Pohl,
  Reimer, Kniffen, Schneid, von Montigny, Mukherjee, \& Dingus}]{Hartman1999}
Hartman, R.~C., Bertsch, D.~L., Bloom, S.~D., {et~al.} 1999, The Astrophysical
  Journal Supplement Series, 123, 79, \dodoi{10.1086/313231}

\bibitem[{{H.E.S.S. Collaboration} {et~al.}(2013){H.E.S.S. Collaboration},
  {Abramowski, A.}, {Acero, F.}, {Aharonian, F.}, {Akhperjanian, A. G.},
  {Anton, G.}, {Balenderan, S.}, {Balzer, A.}, {Barnacka, A.}, {Becherini, Y.},
  {Becker Tjus, J.}, {Behera, B.}, {Bernl\"ohr, K.}, {Birsin, E.}, {Biteau,
  J.}, {Bochow, A.}, {Boisson, C.}, {Bolmont, J.}, {Bordas, P.}, {Brucker, J.},
  {Brun, F.}, {Brun, P.}, {Bulik, T.}, {Carrigan, S.}, {Casanova, S.},
  {Cerruti, M.}, {Chadwick, P. M.}, {Chaves, R. C. G.}, {Cheesebrough, A.},
  {Colafrancesco, S.}, {Cologna, G.}, {Conrad, J.}, {Couturier, C.}, {Dalton,
  M.}, {Daniel, M. K.}, {Davids, I. D.}, {Degrange, B.}, {Deil, C.}, {deWilt,
  P.}, {Dickinson, H. J.}, {Djannati-Ata\"{\i}, A.}, {Domainko, W.}, {Drury, L.
  O\'{}C.}, {Dubus, G.}, {Dutson, K.}, {Dyks, J.}, {Dyrda, M.}, {Egberts, K.},
  {Eger, P.}, {Espigat, P.}, {Fallon, L.}, {Farnier, C.}, {Fegan, S.},
  {Feinstein, F.}, {Fernandes, M. V.}, {Fernandez, D.}, {Fiasson, A.},
  {Fontaine, G.}, {F\"orster, A.}, {F\"u\ss{}ling, M.}, {Gajdus, M.}, {Gallant,
  Y. A.}, {Garrigoux, T.}, {Gast, H.}, {Giebels, B.}, {Glicenstein, J. F.},
  {Gl\"uck, B.}, {G\"oring, D.}, {Grondin, M.-H.}, {Grudzi\'{}nska, M.},
  {H\"affner, S.}, {Hague, J. D.}, {Hahn, J.}, {Hampf, D.}, {Harris, J.},
  {Hauser, M.}, {Heinz, S.}, {Heinzelmann, G.}, {Henri, G.}, {Hermann, G.},
  {Hillert, A.}, {Hinton, J. A.}, {Hofmann, W.}, {Hofverberg, P.}, {Holler,
  M.}, {Horns, D.}, {Jacholkowska, A.}, {Jahn, C.}, {Jamrozy, M.}, {Jung, I.},
  {Kastendieck, M. A.}, {Katarzy\'{}nski, K.}, {Katz, U.}, {Kaufmann, S.},
  {Kh\'elifi, B.}, {Klepser, S.}, {Klochkov, D.}, {Klu\'{}zniak, W.}, {Kneiske,
  T.}, {Kolitzus, D.}, {Komin, Nu.}, {Kosack, K.}, {Kossakowski, R.}, {Krayzel,
  F.}, {Kr\"uger, P. P.}, {Laffon, H.}, {Lamanna, G.}, {Lefaucheur, J.},
  {Lemoine-Goumard, M.}, {Lenain, J.-P.}, {Lennarz, D.}, {Lohse, T.}, {Lopatin,
  A.}, {Lu, C.-C.}, {Marandon, V.}, {Marcowith, A.}, {Masbou, J.}, {Maurin,
  G.}, {Maxted, N.}, {Mayer, M.}, {McComb, T. J. L.}, {Medina, M. C.},
  {M\'ehault, J.}, {Menzler, U.}, {Moderski, R.}, {Mohamed, M.}, {Moulin, E.},
  {Naumann, C. L.}, {Naumann-Godo, M.}, {de Naurois, M.}, {Nedbal, D.},
  {Nguyen, N.}, {Niemiec, J.}, {Nolan, S. J.}, {Ohm, S.}, {de O\~na Wilhelmi,
  E.}, {Opitz, B.}, {Ostrowski, M.}, {Oya, I.}, {Panter, M.}, {Parsons, R. D.},
  {Paz Arribas, M.}, {Pekeur, N. W.}, {Pelletier, G.}, {Perez, J.}, {Petrucci,
  P.-O.}, {Peyaud, B.}, {Pita, S.}, {P\"uhlhofer, G.}, {Punch, M.},
  {Quirrenbach, A.}, {Raab, S.}, {Raue, M.}, {Reimer, A.}, {Reimer, O.},
  {Renaud, M.}, {de los Reyes, R.}, {Rieger, F.}, {Ripken, J.}, {Rob, L.},
  {Rosier-Lees, S.}, {Rowell, G.}, {Rudak, B.}, {Rulten, C. B.}, {Sahakian,
  V.}, {Sanchez, D. A.}, {Santangelo, A.}, {Schlickeiser, R.}, {Schulz, A.},
  {Schwanke, U.}, {Schwarzburg, S.}, {Schwemmer, S.}, {Sheidaei, F.}, {Skilton,
  J. L.}, {Sol, H.}, {Spengler, G.}, {Stawarz, L.}, {Steenkamp, R.}, {Stegmann,
  C.}, {Stinzing, F.}, {Stycz, K.}, {Sushch, I.}, {Szostek, A.}, {Tavernet,
  J.-P.}, {Terrier, R.}, {Tluczykont, M.}, {Trichard, C.}, {Valerius, K.}, {van
  Eldik, C.}, {Vasileiadis, G.}, {Venter, C.}, {Viana, A.}, {Vincent, P.},
  {V\"olk, H. J.}, {Volpe, F.}, {Vorobiov, S.}, {Vorster, M.}, {Wagner, S. J.},
  {Ward, M.}, {White, R.}, {Wierzcholska, A.}, {Wouters, D.}, {Zacharias, M.},
  {Zajczyk, A.}, {Zdziarski, A. A.}, {Zech, A.}, \& {Zechlin, H.
  S.}}]{HESS2013}
{H.E.S.S. Collaboration}, {Abramowski, A.}, {Acero, F.}, {et~al.} 2013, A\&A,
  554, A107, \dodoi{10.1051/0004-6361/201321135}

\bibitem[{{H.E.S.S. Collaboration} {et~al.}(2021){H.E.S.S. Collaboration},
  {Abdalla, H.}, {Adam, R.}, {Aharonian, F.}, {Ait Benkhali, F.}, {Ang\"uner,
  E. O.}, {Arcaro, C.}, {Armand, C.}, {Armstrong, T.}, {Ashkar, H.}, {Backes,
  M.}, {Baghmanyan, V.}, {Barbosa Martins, V.}, {Barnacka, A.}, {Barnard, M.},
  {Becherini, Y.}, {Berge, D.}, {Bernl\"ohr, K.}, {Bi, B.}, {B\"ottcher, M.},
  {Boisson, C.}, {Bolmont, J.}, {Bonnefoy, S.}, {de Bony de Lavergne, M.},
  {Bregeon, J.}, {Breuhaus, M.}, {Brun, F.}, {Brun, P.}, {Bryan, M.},
  {B\"uchele, M.}, {Bulik, T.}, {Bylund, T.}, {Caroff, S.}, {Carosi, A.},
  {Casanova, S.}, {Chand, T.}, {Chandra, S.}, {Chen, A.}, {Cotter, G.},
  {Curylo, M.}, {Damascene Mbarubucyeye, J.}, {Davids, I. D.}, {Davies, J.},
  {Deil, C.}, {Devin, J.}, {deWilt, P.}, {Dirson, L.}, {Djannati-Ata\"{\i},
  A.}, {Dmytriiev, A.}, {Donath, A.}, {Doroshenko, V.}, {Dyks, J.}, {Egberts,
  K.}, {Eichhorn, F.}, {Einecke, S.}, {Emery, G.}, {Ernenwein, J.-P.}, {Feijen,
  K.}, {Fegan, S.}, {Fiasson, A.}, {Fichet de Clairfontaine, G.}, {Filipovic,
  M.}, {Fontaine, G.}, {Funk, S.}, {F\"u\ss{}ling, M.}, {Gabici, S.}, {Gallant,
  Y. A.}, {Giavitto, G.}, {Giunti, L.}, {Glawion, D.}, {Glicenstein, J. F.},
  {Gottschall, D.}, {Grondin, M.-H.}, {Hahn, J.}, {Haupt, M.}, {Hermann, G.},
  {Hinton, J. A.}, {Hofmann, W.}, {Hoischen, C.}, {Holch, T. L.}, {Holler, M.},
  {H\"orbe, M.}, {Horns, D.}, {Huber, D.}, {Jamrozy, M.}, {Jankowsky, D.},
  {Jankowsky, F.}, {Jardin-Blicq, A.}, {Joshi, V.}, {Jung-Richardt, I.},
  {Kastendieck, M. A.}, {Katarzy\'{}nski, K.}, {Katz, U.}, {Khangulyan, D.},
  {Kh\'elifi, B.}, {Klepser, S.}, {Klu\'{}zniak, W.}, {Komin, Nu.}, {Konno,
  R.}, {Kosack, K.}, {Kostunin, D.}, {Kreter, M.}, {Lamanna, G.}, {Lemi\`ere,
  A.}, {Lemoine-Goumard, M.}, {Lenain, J.-P.}, {Levy, C.}, {Lohse, T.},
  {Lypova, I.}, {Mackey, J.}, {Majumdar, J.}, {Malyshev, D.}, {Malyshev, D.},
  {Marandon, V.}, {Marchegiani, P.}, {Marcowith, A.}, {Mares, A.},
  {Mart\'{\i}-Devesa, G.}, {Marx, R.}, {Maurin, G.}, {Meintjes, P. J.}, {Meyer,
  M.}, {Mitchell, A. M. W.}, {Moderski, R.}, {Mohamed, M.}, {Mohrmann, L.},
  {Montanari, A.}, {Moore, C.}, {Morris, P.}, {Moulin, E.}, {Muller, J.},
  {Murach, T.}, {Nakashima, K.}, {Nayerhoda, A.}, {de Naurois, M.}, {Ndiyavala,
  H.}, {Niederwanger, F.}, {Niemiec, J.}, {Oakes, L.}, {O\'{}Brien, P.},
  {Odaka, H.}, {Ohm, S.}, {Olivera-Nieto, L.}, {de Ona Wilhelmi, E.},
  {Ostrowski, M.}, {Panter, M.}, {Panny, S.}, {Parsons, R. D.}, {Peron, G.},
  {Peyaud, B.}, {Piel, Q.}, {Pita, S.}, {Poireau, V.}, {Priyana Noel, A.},
  {Prokhorov, D. A.}, {Prokoph, H.}, {P\"uhlhofer, G.}, {Punch, M.},
  {Quirrenbach, A.}, {Raab, S.}, {Rauth, R.}, {Reichherzer, P.}, {Reimer, A.},
  {Reimer, O.}, {Remy, Q.}, {Renaud, M.}, {Rieger, F.}, {Rinchiuso, L.},
  {Romoli, C.}, {Rowell, G.}, {Rudak, B.}, {Ruiz-Velasco, E.}, {Sahakian, V.},
  {Sailer, S.}, {Sanchez, D. A.}, {Santangelo, A.}, {Sasaki, M.}, {Scalici,
  M.}, {Sch\"ussler, F.}, {Schutte, H. M.}, {Schwanke, U.}, {Schwemmer, S.},
  {Seglar-Arroyo, M.}, {Senniappan, M.}, {Seyffert, A. S.}, {Shafi, N.},
  {Shiningayamwe, K.}, {Simoni, R.}, {Sinha, A.}, {Sol, H.}, {Specovius, A.},
  {Spencer, S.}, {Spir-Jacob, M.}, {Stawarz, L.}, {Sun, L.}, {Steenkamp, R.},
  {Stegmann, C.}, {Steinmassl, S.}, {Steppa, C.}, {Takahashi, T.}, {Tavernier,
  T.}, {Taylor, A. M.}, {Terrier, R.}, {Tiziani, D.}, {Tluczykont, M.},
  {Tomankova, L.}, {Trichard, C.}, {Tsirou, M.}, {Tuffs, R.}, {Uchiyama, Y.},
  {van der Walt, D. J.}, {van Eldik, C.}, {van Rensburg, C.}, {van Soelen, B.},
  {Vasileiadis, G.}, {Veh, J.}, {Venter, C.}, {Vincent, P.}, {Vink, J.},
  {V\"olk, H. J.}, {Vuillaume, T.}, {Wadiasingh, Z.}, {Wagner, S. J.}, {Watson,
  J.}, {Werner, F.}, {White, R.}, {Wierzcholska, A.}, {Wong, Yu. W.},
  {Yusafzai, A.}, {Zacharias, M.}, {Zanin, R.}, {Zargaryan, D.}, {Zdziarski, A.
  A.}, {Zech, A.}, {Zhu, S. J.}, {Zorn, J.}, {Zouari, S.}, {Zywucka, N.},
  {MAGIC Collaboration}, {Acciari, V. A.}, {Ansoldi, S.}, {Antonelli, L. A.},
  {Arbet Engels, A.}, {Asano, K.}, {Baack, D.}, {Babi\'{}c, A.}, {Baquero, A.},
  {Barres de Almeida, U.}, {Barrio, J. A.}, {Becerra Gonz\'alez, J.},
  {Bednarek, W.}, {Bellizzi, L.}, {Bernardini, E.}, {Berti, A.}, {Besenrieder,
  J.}, {Bhattacharyya, W.}, {Bigongiari, C.}, {Biland, A.}, {Blanch, O.},
  {Bonnoli, G.}, {Bosnjak, Z.}, {Busetto, G.}, {Carosi, R.}, {Ceribella, G.},
  {Cerruti, M.}, {Chai, Y.}, {Chilingarian, A.}, {Cikota, S.}, {Colak, S. M.},
  {Colin, U.}, {Colombo, E.}, {Contreras, J. L.}, {Cortina, J.}, {Covino, S.},
  {D\'{}Amico, G.}, {D\'{}Elia, V.}, {Da Vela, P.}, {Dazzi, F.}, {De Angelis,
  A.}, {De Lotto, B.}, {Delfino, M.}, {Delgado, J.}, {Depaoli, D.}, {Di Pierro,
  F.}, {Di Venere, L.}, {Do Souto Espi\~neira, E.}, {Dominis Prester, D.},
  {Donini, A.}, {Dorner, D.}, {Doro, M.}, {Elsaesser, D.}, {Fallah Ramazani,
  V.}, {Fattorini, A.}, {Ferrara, G.}, {Foffano, L.}, {Fonseca, M. V.}, {Font,
  L.}, {Fruck, C.}, {Fukami, S.}, {Garc\'{\i}a L\'opez, R. J.}, {Garczarczyk,
  M.}, {Gasparyan, S.}, {Gaug, M.}, {Giglietto, N.}, {Giordano, F.}, {Gliwny,
  P.}, {Godinovi\'{}c, N.}, {Green, D.}, {Hadasch, D.}, {Hahn, A.}, {Heckmann,
  L.}, {Herrera, J.}, {Hoang, J.}, {Hrupec, D.}, {H\"utten, M.}, {Inada, T.},
  {Inoue, S.}, {Ishio, K.}, {Iwamura, Y.}, {Jouvin, L.}, {Kajiwara, Y.},
  {Karjalainen, M.}, {Kerszberg, D.}, {Kobayashi, Y.}, {Kubo, H.}, {Kushida,
  J.}, {Lamastra, A.}, {Lelas, D.}, {Leone, F.}, {Lindfors, E.}, {Lombardi,
  S.}, {Longo, F.}, {L\'opez, M.}, {L\'opez-Coto, R.}, {L\'opez-Oramas, A.},
  {Loporchio, S.}, {Machado de Oliveira Fraga, B.}, {Maggio, C.}, {Majumdar,
  P.}, {Makariev, M.}, {Mallamaci, M.}, {Maneva, G.}, {Manganaro, M.},
  {Mannheim, K.}, {Maraschi, L.}, {Mariotti, M.}, {Mart\'{\i}nez, M.}, {Mazin,
  D.}, {Mender, S.}, {Mi\'{}canovi\'{}c, S.}, {Miceli, D.}, {Miener, T.},
  {Minev, M.}, {Miranda, J. M.}, {Mirzoyan, R.}, {Molina, E.}, {Moralejo, A.},
  {Morcuende, D.}, {Moreno, V.}, {Moretti, E.}, {Munar-Adrover, P.},
  {Neustroev, V.}, {Nigro, C.}, {Nilsson, K.}, {Ninci, D.}, {Nishijima, K.},
  {Noda, K.}, {Nozaki, S.}, {Ohtani, Y.}, {Oka, T.}, {Otero-Santos, J.},
  {Palatiello, M.}, {Paneque, D.}, {Paoletti, R.}, {Paredes, J. M.},
  {Pavleti\'{}c, L.}, {Pe\~nil, P.}, {Perennes, C.}, {Persic, M.}, {Prada
  Moroni, P. G.}, {Prandini, E.}, {Priyadarshi, C.}, {Puljak, I.}, {Rhode, W.},
  {Rib\'o, M.}, {Rico, J.}, {Righi, C.}, {Rugliancich, A.}, {Saha, L.},
  {Sahakyan, N.}, {Saito, T.}, {Sakurai, S.}, {Satalecka, K.}, {Schleicher,
  B.}, {Schmidt, K.}, {Schweizer, T.}, {Sitarek, J.}, {Snidari\'{}c, I.},
  {Sobczynska, D.}, {Spolon, A.}, {Stamerra, A.}, {Strom, D.}, {Strzys, M.},
  {Suda, Y.}, {Suri\'{}c, T.}, {Takahashi, M.}, {Tavecchio, F.}, {Temnikov,
  P.}, {Terzi\'{}c, T.}, {Teshima, M.}, {Torres-Alb\`a, N.}, {Tosti, L.},
  {Truzzi, S.}, {van Scherpenberg, J.}, {Vanzo, G.}, {Vazquez Acosta, M.},
  {Ventura, S.}, {Verguilov, V.}, {Vigorito, C. F.}, {Vitale, V.}, {Vovk, I.},
  {Will, M.}, {Zari\'{}c, D.}, {Jorstad, S. G.}, {Marscher, A. P.}, {Boccardi,
  B.}, {Casadio, C.}, {Hodgson, J.}, {Kim, J.-Y.}, {Krichbaum, T. P.},
  {L\"ahteenm\"aki, A.}, {Tornikoski, M.}, {Traianou, E.}, \& {Weaver, Z.
  R.}}]{HESS2021}
{H.E.S.S. Collaboration}, {Abdalla, H.}, {Adam, R.}, {et~al.} 2021, A\&A, 648,
  A23, \dodoi{10.1051/0004-6361/202038949}

\bibitem[{Homan {et~al.}(2002)Homan, Wardle, Cheung, Roberts, \&
  Attridge}]{Homan2002}
Homan, D.~C., Wardle, J. F.~C., Cheung, C.~C., Roberts, D.~H., \& Attridge,
  J.~M. 2002, The Astrophysical Journal, 580, 742, \dodoi{10.1086/343894}

\bibitem[{{Isler} {et~al.}(2013){Isler}, {Urry}, {Coppi}, {Bailyn},
  {Chatterjee}, {Fossati}, {Bonning}, {Maraschi}, \& {Buxton}}]{Isler2013}
{Isler}, J.~C., {Urry}, C.~M., {Coppi}, P., {et~al.} 2013, \apj, 779, 100,
  \dodoi{10.1088/0004-637X/779/2/100}

\bibitem[{{Jorstad} {et~al.}(2005){Jorstad}, {Marscher}, {Lister}, {Stirling},
  {Cawthorne}, {Gear}, {G{\'o}mez}, {Stevens}, {Smith}, {Forster}, \&
  {Robson}}]{Jorstad2005}
{Jorstad}, S.~G., {Marscher}, A.~P., {Lister}, M.~L., {et~al.} 2005, \aj, 130,
  1418, \dodoi{10.1086/444593}

\bibitem[{Kataoka {et~al.}(2008)Kataoka, Madejski, Sikora, Roming, Chester,
  Grupe, Tsubuku, Sato, Kawai, Tosti, Impiombato, Kovalev, Kovalev, Edwards,
  Wagner, Moderski, Stawarz, Takahashi, \& Watanabe}]{Kataoka2008}
Kataoka, J., Madejski, G., Sikora, M., {et~al.} 2008, The Astrophysical
  Journal, 672, 787, \dodoi{10.1086/523093}

\bibitem[{{Kova{\v{c}}evi{\'c}} {et~al.}(2010){Kova{\v{c}}evi{\'c}},
  {Popovi{\'c}}, \& {Dimitrijevi{\'c}}}]{Kovacevic2010}
{Kova{\v{c}}evi{\'c}}, J., {Popovi{\'c}}, L.~{\v{C}}., \& {Dimitrijevi{\'c}},
  M.~S. 2010, \apjs, 189, 15, \dodoi{10.1088/0067-0049/189/1/15}

\bibitem[{{Le{\'o}n-Tavares} {et~al.}(2011){Le{\'o}n-Tavares}, {Valtaoja},
  {Tornikoski}, {L{\"a}hteenm{\"a}ki}, \& {Nieppola}}]{LeonTavares2011}
{Le{\'o}n-Tavares}, J., {Valtaoja}, E., {Tornikoski}, M.,
  {L{\"a}hteenm{\"a}ki}, A., \& {Nieppola}, E. 2011, \aap, 532, A146,
  \dodoi{10.1051/0004-6361/201116664}

\bibitem[{Le\'on-Tavares {et~al.}(2013)Le\'on-Tavares, Chavushyan,
  {Pati{\~n}o-{\'A}lvarez}, Valtaoja, Arshakian, Popović, Tornikoski, Lobanov,
  nana, Carrasco, \& Lähteenmäki}]{LeonTavares2013}
Le\'on-Tavares, J., Chavushyan, V., {Pati{\~n}o-{\'A}lvarez}, V.~M., {et~al.}
  2013, The Astrophysical Journal Letters, 763, L36,
  \dodoi{10.1088/2041-8205/763/2/L36}

\bibitem[{{MAGIC Collaboration} {et~al.}(2018){MAGIC Collaboration}, {Acciari},
  {Ansoldi}, {Antonelli}, {Arbet Engels}, {Arcaro}, {Baack}, {Babi{\'c}},
  {Banerjee}, {Bangale}, {Barres de Almeida}, {Barrio}, {Bednarek},
  {Bernardini}, {Berti}, {Besenrieder}, {Bhattacharyya}, {Bigongiari},
  {Biland}, {Blanch}, {Bonnoli}, {Carosi}, {Ceribella}, {Cikota}, {Colak},
  {Colin}, {Colombo}, {Contreras}, {Cortina}, {Covino}, {D'Elia}, {da Vela},
  {Dazzi}, {de Angelis}, {de Lotto}, {Delfino}, {Delgado}, {di Pierro}, {Do
  Souto Espi{\~n}era}, {Dom{\'\i}nguez}, {Dominis Prester}, {Dorner}, {Doro},
  {Einecke}, {Elsaesser}, {Fallah Ramazani}, {Fattorini},
  {Fern{\'a}ndez-Barral}, {Ferrara}, {Fidalgo}, {Foffano}, {Fonseca}, {Font},
  {Fruck}, {Galindo}, {Gallozzi}, {Garc{\'\i}a L{\'o}pez}, {Garczarczyk},
  {Gaug}, {Giammaria}, {Godinovi{\'c}}, {Guberman}, {Hadasch}, {Hahn},
  {Hassan}, {Herrera}, {Hoang}, {Hrupec}, {Inoue}, {Ishio}, {Iwamura}, {Kubo},
  {Kushida}, {Kuve{\v{z}}di{\'c}}, {Lamastra}, {Lelas}, {Leone}, {Lindfors},
  {Lombardi}, {Longo}, {L{\'o}pez}, {L{\'o}pez-Oramas}, {Maggio}, {Majumdar},
  {Makariev}, {Maneva}, {Manganaro}, {Mannheim}, {Maraschi}, {Mariotti},
  {Mart{\'\i}nez}, {Masuda}, {Mazin}, {Minev}, {Miranda}, {Mirzoyan}, {Molina},
  {Moralejo}, {Moreno}, {Moretti}, {Munar-Adrover}, {Neustroev}, {Niedzwiecki},
  {Nievas Rosillo}, {Nigro}, {Nilsson}, {Ninci}, {Nishijima}, {Noda},
  {Nogu{\'e}s}, {Paiano}, {Palacio}, {Paneque}, {Paoletti}, {Paredes},
  {Pedaletti}, {Pe{\~n}il}, {Peresano}, {Persic}, {Prada Moroni}, {Prandini},
  {Puljak}, {Garcia}, {Rhode}, {Rib{\'o}}, {Rico}, {Righi}, {Rugliancich},
  {Saha}, {Saito}, {Satalecka}, {Schweizer}, {Sitarek}, {{\v{S}}nidari{\'c}},
  {Sobczynska}, {Somero}, {Stamerra}, {Strzys}, {Suri{\'c}}, {Tavecchio},
  {Temnikov}, {Terzi{\'c}}, {Teshima}, {Torres-Alb{\`a}}, {Tsujimoto}, {van
  Scherpenberg}, {Vanzo}, {Vazquez Acosta}, {Vovk}, {Ward}, {Will},
  {Zari{\'c}}, {Fermi-Lat Collaboration}, {Becerra Gonz{\'a}lez}, {Raiteri},
  {Sandrinelli}, {Hovatta}, {Kiehlmann}, {Max-Moerbeck}, {Tornikoski},
  {L{\"a}hteenm{\"a}ki}, {Tammi}, {Ramakrishnan}, {Thum}, {Agudo}, {Molina},
  {G{\'o}mez}, {Fuentes}, {Casadio}, {Traianou}, {Myserlis}, \&
  {Kim}}]{MAGIC2018}
{MAGIC Collaboration}, {Acciari}, V.~A., {Ansoldi}, S., {et~al.} 2018, \aap,
  619, A159, \dodoi{10.1051/0004-6361/201833618}

\bibitem[{{Malmrose} {et~al.}(2011){Malmrose}, {Marscher}, {Jorstad},
  {Nikutta}, \& {Elitzur}}]{Malmrose2011}
{Malmrose}, M.~P., {Marscher}, A.~P., {Jorstad}, S.~G., {Nikutta}, R., \&
  {Elitzur}, M. 2011, \apj, 732, 116, \dodoi{10.1088/0004-637X/732/2/116}

\bibitem[{{Maraschi} {et~al.}(1992){Maraschi}, {Ghisellini}, {Celotti}, \&
  {Maraschi}}]{Maraschi1992}
{Maraschi}, L., {Ghisellini}, G., {Celotti}, A., \& {Maraschi}, L. 1992, The
  Astrophysical Journal Letters, 397, L5, \dodoi{10.1086/186531}

\bibitem[{Marscher {et~al.}(2010)Marscher, Jorstad, Larionov, Aller, Aller,
  Lähteenmäki, Agudo, Smith, Gurwell, Hagen-Thorn, Konstantinova, Larionova,
  Larionova, Melnichuk, Blinov, Kopatskaya, Troitsky, Tornikoski, Hovatta,
  Schmidt, D’Arcangelo, Bhattarai, Taylor, Olmstead, Manne-Nicholas,
  Roca-Sogorb, Gómez, McHardy, Kurtanidze, Nikolashvili, Kimeridze, \&
  Sigua}]{Marscher2010}
Marscher, A.~P., Jorstad, S.~G., Larionov, V.~M., {et~al.} 2010, The
  Astrophysical Journal Letters, 710, L126,
  \dodoi{10.1088/2041-8205/710/2/L126}

\bibitem[{{Nalewajko} {et~al.}(2012){Nalewajko}, {Sikora}, {Madejski}, {Exter},
  {Szostek}, {Szczerba}, {Kidger}, \& {Lorente}}]{Nalewajko2012}
{Nalewajko}, K., {Sikora}, M., {Madejski}, G.~M., {et~al.} 2012, \apj, 760, 69,
  \dodoi{10.1088/0004-637X/760/1/69}

\bibitem[{{Orienti} {et~al.}(2011){Orienti}, {Venturi}, {Dallacasa},
  {D'Ammando}, {Giroletti}, {Giovannini}, {Vercellone}, \&
  {Tavani}}]{Orienti2011}
{Orienti}, M., {Venturi}, T., {Dallacasa}, D., {et~al.} 2011, \mnras, 417, 359,
  \dodoi{10.1111/j.1365-2966.2011.19272.x}

\bibitem[{{Orienti} {et~al.}(2013){Orienti}, {Koyama}, {D'Ammando},
  {Giroletti}, {Kino}, {Nagai}, {Venturi}, {Dallacasa}, {Giovannini},
  {Angelakis}, {Fuhrmann}, {Hovatta}, {Max-Moerbeck}, {Schinzel}, {Akiyama},
  {Hada}, {Honma}, {Niinuma}, {Gasparrini}, {Krichbaum}, {Nestoras},
  {Readhead}, {Richards}, {Riquelme}, {Sievers}, {Ungerechts}, \&
  {Zensus}}]{Orienti2013}
{Orienti}, M., {Koyama}, S., {D'Ammando}, F., {et~al.} 2013, \mnras, 428, 2418,
  \dodoi{10.1093/mnras/sts201}

\bibitem[{Oshlack {et~al.}(2002)Oshlack, Webster, Whiting, \&
  Oshlack}]{Oshlack2002}
Oshlack, A. Y. K.~N., Webster, R.~L., Whiting, M.~T., \& Oshlack, A. Y. K.~N.
  2002, The Astrophysical Journal, 576, 81, \dodoi{10.1086/341729}

\bibitem[{{Paliya} {et~al.}(2018){Paliya}, {Zhang}, {B{\"o}ttcher}, {Ajello},
  {Dom{\'\i}nguez}, {Joshi}, {Hartmann}, \& {Stalin}}]{Paliya2018}
{Paliya}, V.~S., {Zhang}, H., {B{\"o}ttcher}, M., {et~al.} 2018, \apj, 863, 98,
  \dodoi{10.3847/1538-4357/aad1f0}

\bibitem[{Park {et~al.}(2019)Park, Lee, Kim, Hodgson, Trippe, Kim, Algaba,
  Kino, Zhao, Lee, \& Gurwell}]{Park2019}
Park, J., Lee, S.-S., Kim, J.-Y., {et~al.} 2019, The Astrophysical Journal,
  877, 106, \dodoi{10.3847/1538-4357/ab1b27}

\bibitem[{{Park} \& {Trippe}(2014)}]{ParkAndTrippe2014}
{Park}, J.-H., \& {Trippe}, S. 2014, \apj, 785, 76,
  \dodoi{10.1088/0004-637X/785/1/76}

\bibitem[{{Pati{\~n}o-{\'A}lvarez}
  {et~al.}(2013{\natexlab{a}}){Pati{\~n}o-{\'A}lvarez}, {Carrami{\~n}ana},
  {Carrasco}, \& {Chavushyan}}]{PatinoAlvarez2013Variability}
{Pati{\~n}o-{\'A}lvarez}, V.~M., {Carrami{\~n}ana}, A., {Carrasco}, L., \&
  {Chavushyan}, V. 2013{\natexlab{a}}, in 4th Fermi Symposium, eConf
  Proceedings C121028, Monterey, CA, 0--6, \dodoi{10.48550/arXiv.1303.1898}

\bibitem[{{Pati{\~n}o-{\'A}lvarez}
  {et~al.}(2013{\natexlab{b}}){Pati{\~n}o-{\'A}lvarez}, {Chavushyan},
  {Le{\'o}n-Tavares}, {Vald{\'e}s}, {Carrami{\~n}ana}, {Carrasco}, \&
  {Torrealba}}]{PatinoAlvarez2013Monitoring}
{Pati{\~n}o-{\'A}lvarez}, V.~M., {Chavushyan}, V., {Le{\'o}n-Tavares}, J.,
  {et~al.} 2013{\natexlab{b}}, in 4th Fermi Symposium, eConf Proceedings
  C121028, Monterey, CA, arXiv:1303.1893, \dodoi{10.48550/arXiv.1303.1893}

\bibitem[{{Pati{\~n}o-{\'A}lvarez} {et~al.}(2016){Pati{\~n}o-{\'A}lvarez},
  {Torrealba}, {Chavushyan}, {Cruz Gonz{\'a}lez}, {Arshakian}, {Le{\'o}n
  Tavares}, \& {Popovic}}]{PatinoAlvarez2016}
{Pati{\~n}o-{\'A}lvarez}, V.~M., {Torrealba}, J., {Chavushyan}, V., {et~al.}
  2016, Frontiers in Astronomy and Space Sciences, 3, 19,
  \dodoi{10.3389/fspas.2016.00019}

\bibitem[{{Peterson}(1997)}]{Peterson1997}
{Peterson}, B.~M. 1997, {An Introduction to Active Galactic Nuclei}

\bibitem[{{Poutanen} \& {Stern}(2010)}]{Poutanen2010}
{Poutanen}, J., \& {Stern}, B. 2010, \apjl, 717, L118,
  \dodoi{10.1088/2041-8205/717/2/L118}

\bibitem[{Prince {et~al.}(2019)Prince, Gupta, Nalewajko, \&
  Prince}]{Prince2019}
Prince, R., Gupta, N., Nalewajko, K., \& Prince, R. 2019, The Astrophysical
  Journal, 883, 137, \dodoi{10.3847/1538-4357/ab3afa}

\bibitem[{{Rakshit}(2020)}]{Rakshit2020}
{Rakshit}, S. 2020, A\&A, 642, A59, \dodoi{10.1051/0004-6361/202038324}

\bibitem[{{Rani} {et~al.}(2010){Rani}, {Gupta}, {Strigachev}, {Bachev},
  {Wiita}, {Semkov}, {Ovcharov}, {Mihov}, {Boeva}, {Peneva}, {Spassov},
  {Tsvetkova}, {Stoyanov}, \& {Valcheva}}]{Rani2010}
{Rani}, B., {Gupta}, A.~C., {Strigachev}, A., {et~al.} 2010, \mnras, 404, 1992,
  \dodoi{10.1111/j.1365-2966.2010.16419.x}

\bibitem[{{Rani} {et~al.}(2011){Rani}, {Gupta}, {Bachev}, {Strigachev},
  {Semkov}, {D'Ammando}, {Wiita}, {Gurwell}, {Ovcharov}, {Mihov}, {Boeva}, \&
  {Peneva}}]{Rani2011}
{Rani}, B., {Gupta}, A.~C., {Bachev}, R., {et~al.} 2011, \mnras, 417, 1881,
  \dodoi{10.1111/j.1365-2966.2011.19373.x}

\bibitem[{Richards {et~al.}(2011)Richards, Max-Moerbeck, Pavlidou, Readhead,
  Pearson, King, Reeves, Stevenson, \& Shepherd}]{Richards2011}
Richards, J.~L., Max-Moerbeck, W., Pavlidou, V., {et~al.} 2011, Radio
  Variability Studies of Gamma-Ray Blazars with the OVRO 40 m Telescope.
\newblock \doarXiv{1111.0318}

\bibitem[{{Romero} {et~al.}(2017){Romero}, {Boettcher}, {Markoff}, \&
  {Tavecchio}}]{Romero2017}
{Romero}, G.~E., {Boettcher}, M., {Markoff}, S., \& {Tavecchio}, F. 2017, \ssr,
  207, 5, \dodoi{10.1007/s11214-016-0328-2}

\bibitem[{{Schlafly} \& {Finkbeiner}(2011)}]{SchlaflyAndFinkbeiner2011}
{Schlafly}, E.~F., \& {Finkbeiner}, D.~P. 2011, \apj, 737, 103,
  \dodoi{10.1088/0004-637X/737/2/103}

\bibitem[{Shaw {et~al.}(2012)Shaw, Romani, Cotter, Healey, Michelson, Readhead,
  Richards, Max-Moerbeck, King, \& Potter}]{Shaw2012}
Shaw, M.~S., Romani, R.~W., Cotter, G., {et~al.} 2012, The Astrophysical
  Journal, 748, 49, \dodoi{10.1088/0004-637X/748/1/49}

\bibitem[{Sher(1968)}]{Sher1968}
Sher, D. 1968, Journal of the Royal Astronomical Society of Canada, Vol. 62, p.
  105, 62, 105

\bibitem[{{Sikora} {et~al.}(1994){Sikora}, {Begelman}, \& {Rees}}]{Sikora1994}
{Sikora}, M., {Begelman}, M.~C., \& {Rees}, M.~J. 1994, \apj, 421, 153,
  \dodoi{10.1086/173633}

\bibitem[{{Smith} {et~al.}(2009){Smith}, {Montiel}, {Rightley}, {Turner},
  {Schmidt}, \& {Jannuzi}}]{Smith2009C}
{Smith}, P.~S., {Montiel}, E., {Rightley}, S., {et~al.} 2009, in Fermi
  Symposium, eConf Proceedings C091122, Washington, D.C, 0--6,
  \dodoi{10.48550/arXiv.0912.3621}

\bibitem[{Stroh \& Falcone(2013)}]{StrohAndFalcone2013}
Stroh, M.~C., \& Falcone, A.~D. 2013, The Astrophysical Journal Supplement
  Series, 207, 28, \dodoi{10.1088/0067-0049/207/2/28}

\bibitem[{{Tadhunter} {et~al.}(1993){Tadhunter}, {Morganti}, {di Serego
  Alighieri}, {Fosbury}, \& {Danziger}}]{Tadhunter1993}
{Tadhunter}, C.~N., {Morganti}, R., {di Serego Alighieri}, S., {Fosbury},
  R.~A.~E., \& {Danziger}, I.~J. 1993, \mnras, 263, 999,
  \dodoi{10.1093/mnras/263.4.999}

\bibitem[{{Tavecchio} {et~al.}(2010){Tavecchio}, {Ghisellini}, {Bonnoli}, \&
  {Ghirlanda}}]{Tavecchio2010}
{Tavecchio}, F., {Ghisellini}, G., {Bonnoli}, G., \& {Ghirlanda}, G. 2010,
  \mnras, 405, L94, \dodoi{10.1111/j.1745-3933.2010.00867.x}

\bibitem[{{Tody}(1986)}]{Tody1986}
{Tody}, D. 1986, in Society of Photo-Optical Instrumentation Engineers (SPIE)
  Conference Series, Vol. 627, Instrumentation in astronomy VI, ed. D.~L.
  {Crawford}, 733, \dodoi{10.1117/12.968154}

\bibitem[{{Tody}(1993)}]{Tody1993}
{Tody}, D. 1993, in Astronomical Society of the Pacific Conference Series,
  Vol.~52, Astronomical Data Analysis Software and Systems II, ed. R.~J.
  {Hanisch}, R.~J.~V. {Brissenden}, \& J.~{Barnes}, 173

\bibitem[{Tresse {et~al.}(1999)Tresse, Maddox, Loveday, \&
  Singleton}]{Tresse1999}
Tresse, L., Maddox, S., Loveday, J., \& Singleton, C. 1999, Monthly Notices of
  the Royal Astronomical Society, 310, 262,
  \dodoi{10.1046/j.1365-8711.1999.02977.x}

\bibitem[{{Trippe} {et~al.}(2011){Trippe}, {Krips}, {Pi{\'e}tu}, {Neri},
  {Winters}, {Gueth}, {Bremer}, {Salome}, {Moreno}, {Boissier}, \&
  {Fontani}}]{Trippe2011}
{Trippe}, S., {Krips}, M., {Pi{\'e}tu}, V., {et~al.} 2011, \aap, 533, A97,
  \dodoi{10.1051/0004-6361/201015558}

\bibitem[{Urry \& Padovani(1995)}]{UrryAndPadovani1995}
Urry, C.~M., \& Padovani, P. 1995, Publications of the Astronomical Society of
  the Pacific, 107, 803, \dodoi{10.1086/133630}

\bibitem[{{Urry} {et~al.}(1997){Urry}, {Treves}, {Maraschi}, {Marshall}, {Kii},
  {Madejski}, {Penton}, {Pesce}, {Pian}, {Celotti}, {Fujimoto}, {Makino},
  {Otani}, {Sambruna}, {Sasaki}, {Shull}, {Smith}, {Takahashi}, \&
  {Tashiro}}]{Urry1997}
{Urry}, C.~M., {Treves}, A., {Maraschi}, L., {et~al.} 1997, \apj, 486, 799,
  \dodoi{10.1086/304536}

\bibitem[{{Urry}(1999)}]{Urry1999}
{Urry}, M. 1999, in Astronomical Society of the Pacific Conference Series, Vol.
  159, BL Lac Phenomenon, ed. L.~O. {Takalo} \& A.~{Sillanp{\"a}{\"a}}, 3,
  \dodoi{10.48550/arXiv.astro-ph/9812420}

\bibitem[{{Virtanen} {et~al.}(2020){Virtanen}, {Gommers}, {Oliphant},
  {Haberland}, {Reddy}, {Cournapeau}, {Burovski}, {Peterson}, {Weckesser},
  {Bright}, {van der Walt}, {Brett}, {Wilson}, {Millman}, {Mayorov}, {Nelson},
  {Jones}, {Kern}, {Larson}, {Carey}, {Polat}, {Feng}, {Moore}, {VanderPlas},
  {Laxalde}, {Perktold}, {Cimrman}, {Henriksen}, {Quintero}, {Harris},
  {Archibald}, {Ribeiro}, {Pedregosa}, {van Mulbregt}, \& {SciPy 1. 0
  Contributors}}]{Virtanen2020}
{Virtanen}, P., {Gommers}, R., {Oliphant}, T.~E., {et~al.} 2020, Nature
  Methods, 17, 261, \dodoi{10.1038/s41592-019-0686-2}

\bibitem[{{Xie} {et~al.}(2005){Xie}, {Liu}, {Cha}, {Zhou}, {Ma}, {Xie}, \&
  {Chen}}]{Xie2005}
{Xie}, G.~Z., {Liu}, H.~T., {Cha}, G.~W., {et~al.} 2005, \aj, 130, 2506,
  \dodoi{10.1086/497163}

\bibitem[{{Yuan} {et~al.}(2023){Yuan}, {Kushwaha}, {Gupta}, {Tripathi},
  {Wiita}, {Zhang}, {Liu}, {L{\"a}hteenm{\"a}ki}, {Tornikoski}, {Tammi},
  {Ramakrishnan}, {Cui}, {Wang}, {Gu}, {Bambi}, \& {Volvach}}]{Yuan2023}
{Yuan}, Q., {Kushwaha}, P., {Gupta}, A.~C., {et~al.} 2023, \apj, 953, 47,
  \dodoi{10.3847/1538-4357/acdd74}

\end{thebibliography}
\bibliographystyle{aasjournal}

\clearpage
\appendix
\section*{Online Material:\\ Cross-Correlation Figures}\label{sec:apexcc}
\begin{figure}[h!]
\begin{tabularx}{1in}{c} 
{\includegraphics[width=\columnwidth]{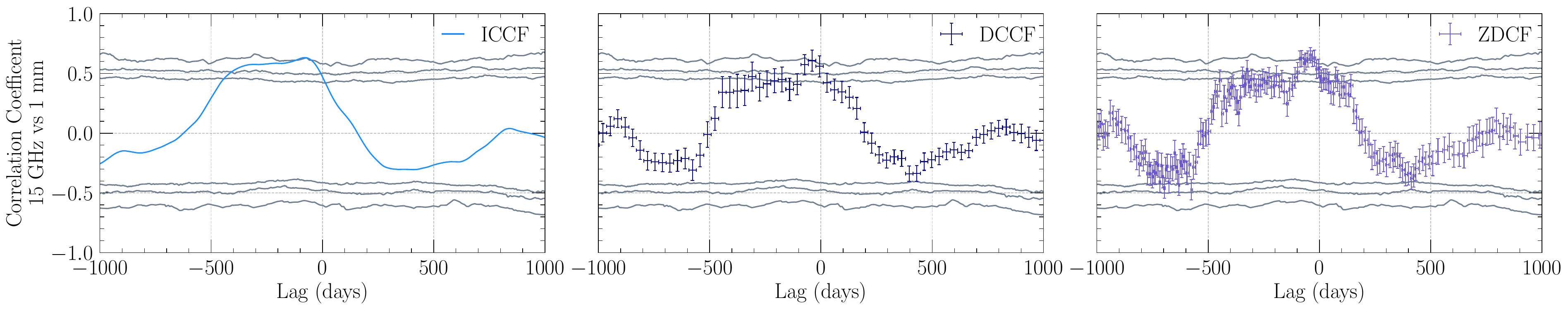}} \\
{\includegraphics[width=\columnwidth]{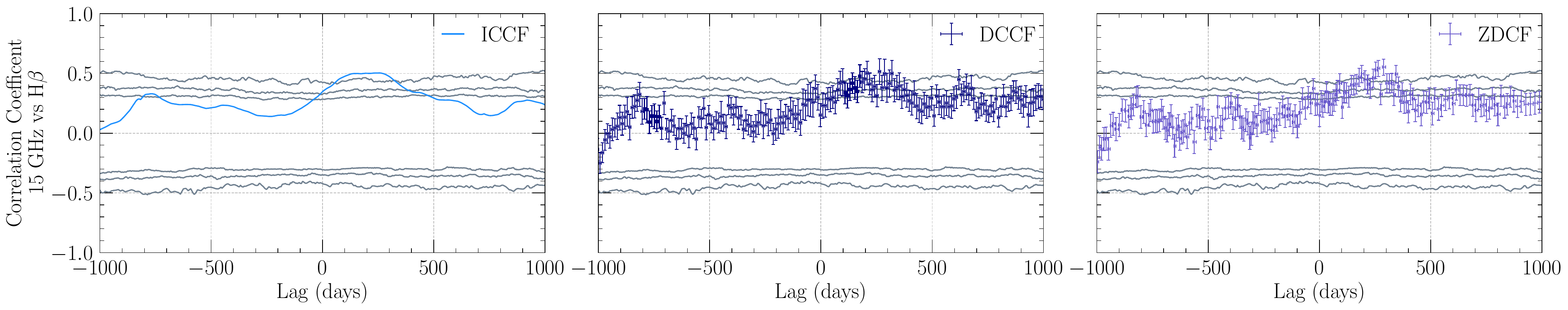}} \\
{\includegraphics[width=\columnwidth]{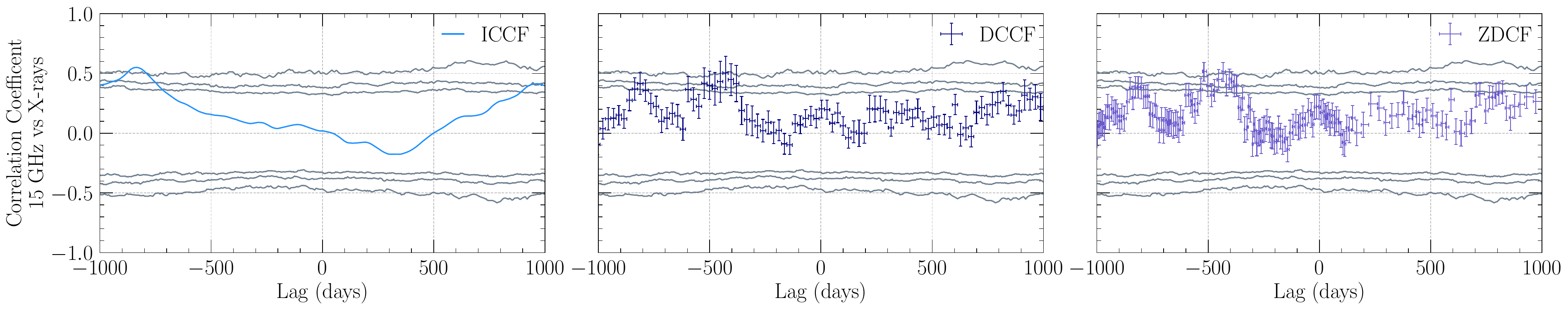}} \\
{\includegraphics[width=\columnwidth]{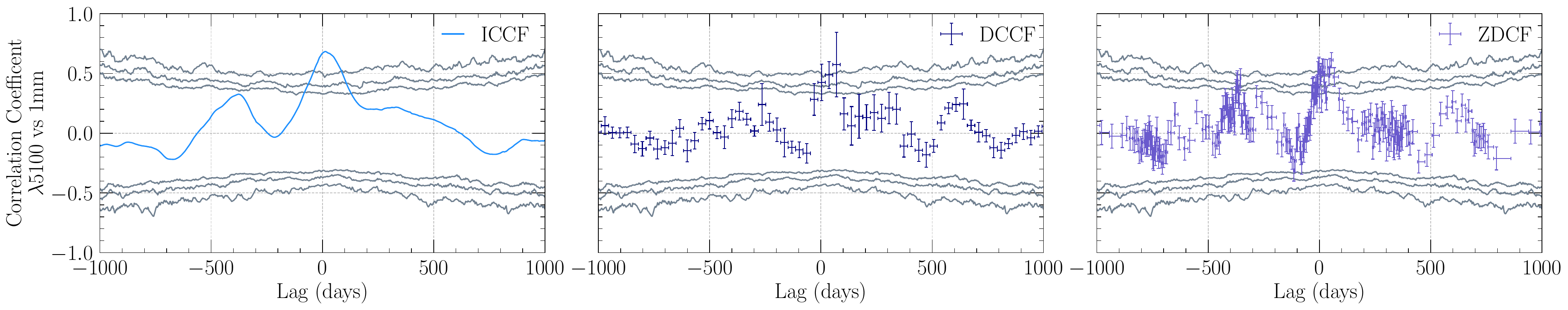}} \\
\end{tabularx}
\caption{Cross-correlation functions detailed in Section 3 are presented in rows, with each row depicting an analysis between two respective features using three methods: the cross-correlated interpolation function (ICCF), the discrete cross-correlation function (DCCF), and the Z-transformed cross-correlation function (ZDCF). Gray lines indicate significance levels at $90\%$, $95\%$, and $99\%$. Time delays, identified as non-spurious, are listed in \autoref{tab:delays}.}
\label{fig:cc1}
\end{figure}

\begin{figure}[h!]
\begin{tabularx}{1in}{c} 
{\includegraphics[width=\columnwidth]{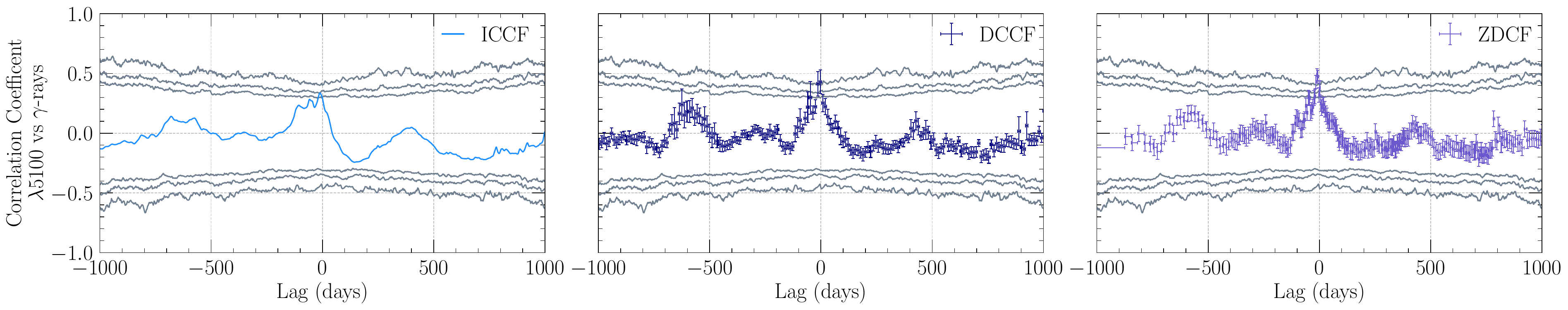}} \\
{\includegraphics[width=\columnwidth]{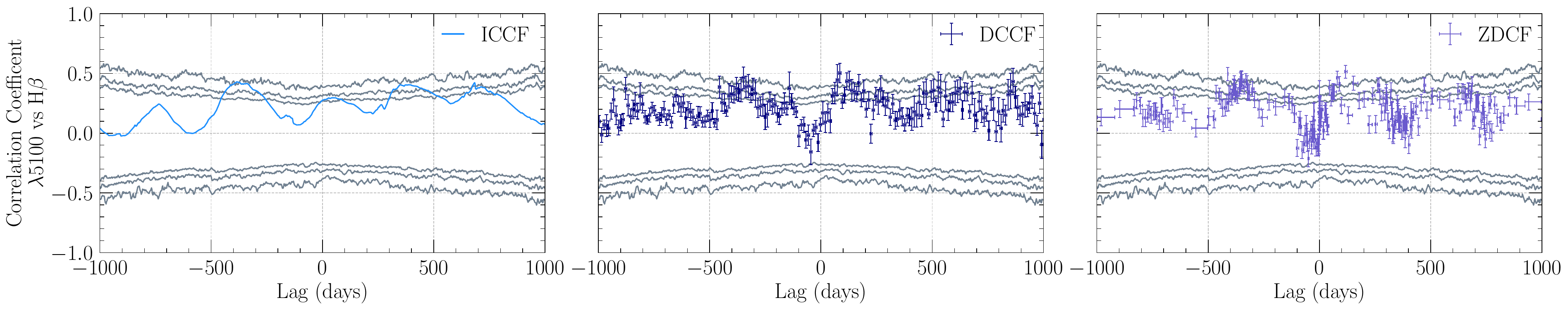}} \\
{\includegraphics[width=\columnwidth]{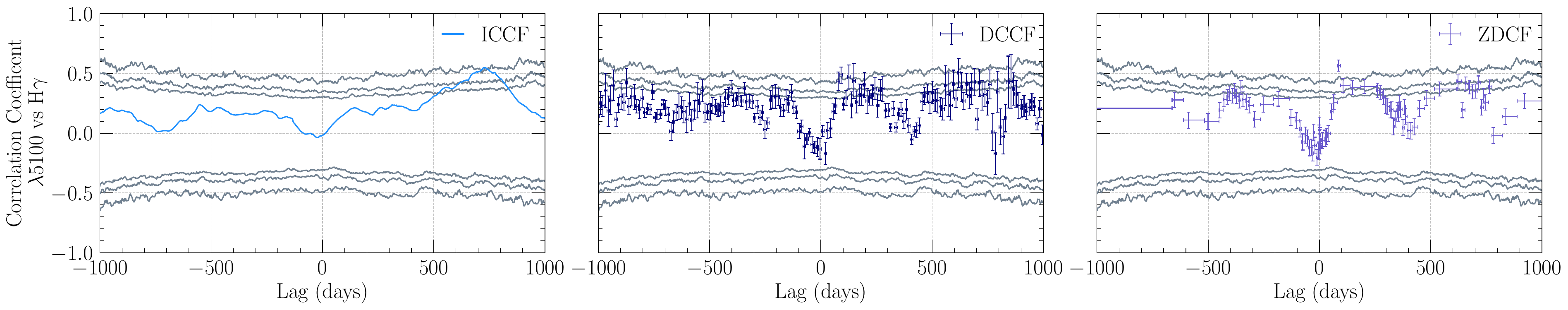}} \\
{\includegraphics[width=\columnwidth]{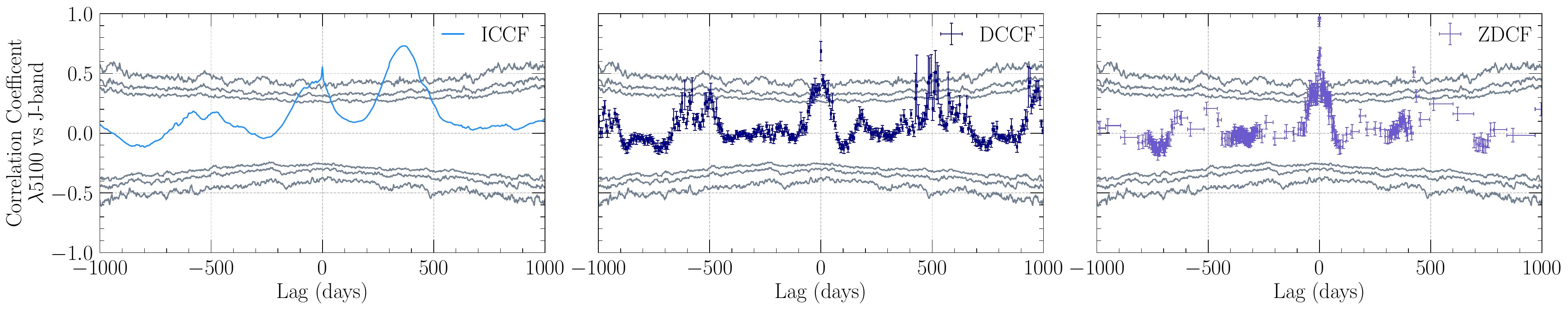}} \\
{\includegraphics[width=\columnwidth]{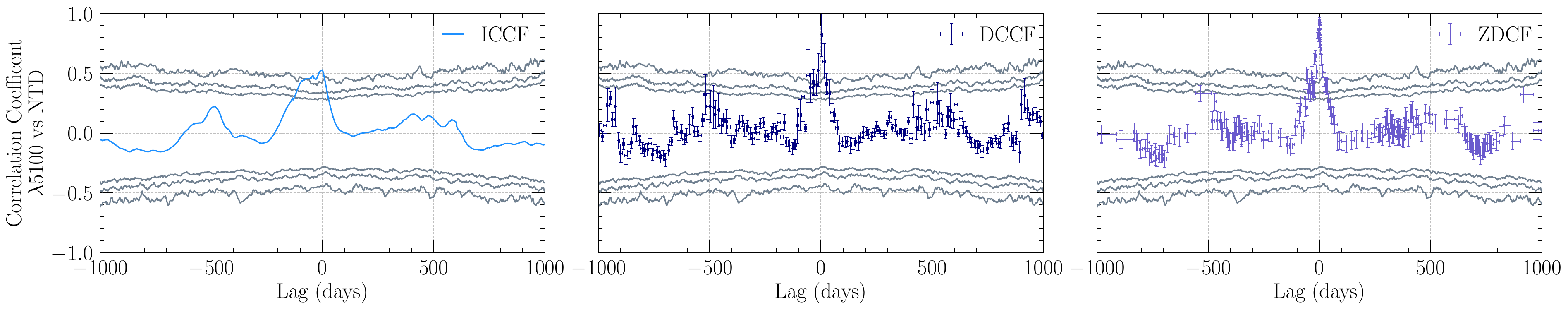}} \\
{\includegraphics[width=\columnwidth]{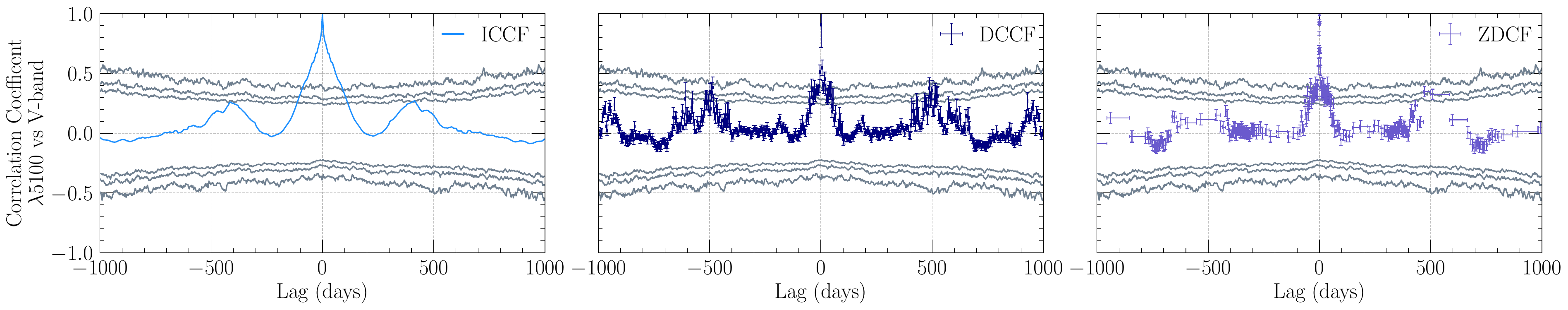}} \\
\end{tabularx}
\caption{Cross-correlation Functions. The lines represent the same as in \autoref{fig:cc1}.}
\label{fig:cc2}
\end{figure}

\begin{figure}[h!]
\begin{tabularx}{1in}{c} 
{\includegraphics[width=\columnwidth]{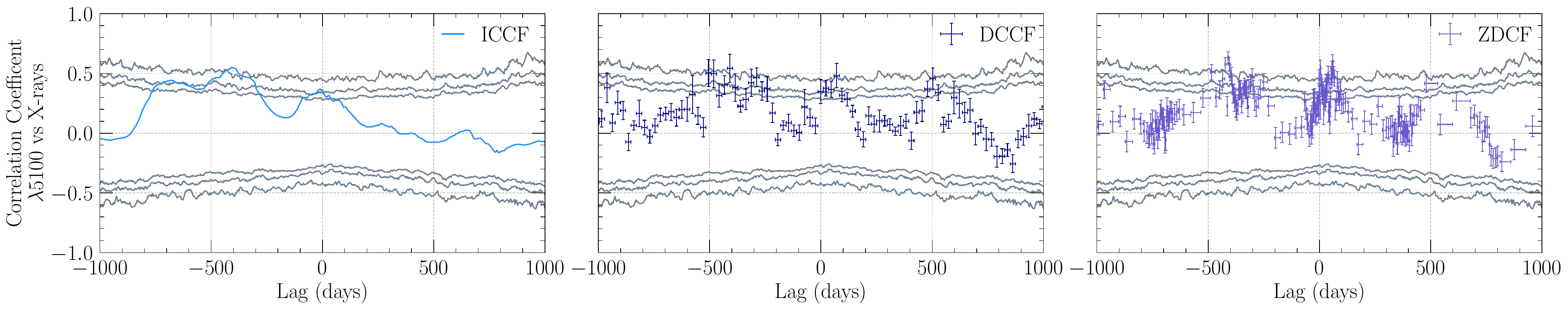}} \\
{\includegraphics[width=\columnwidth]{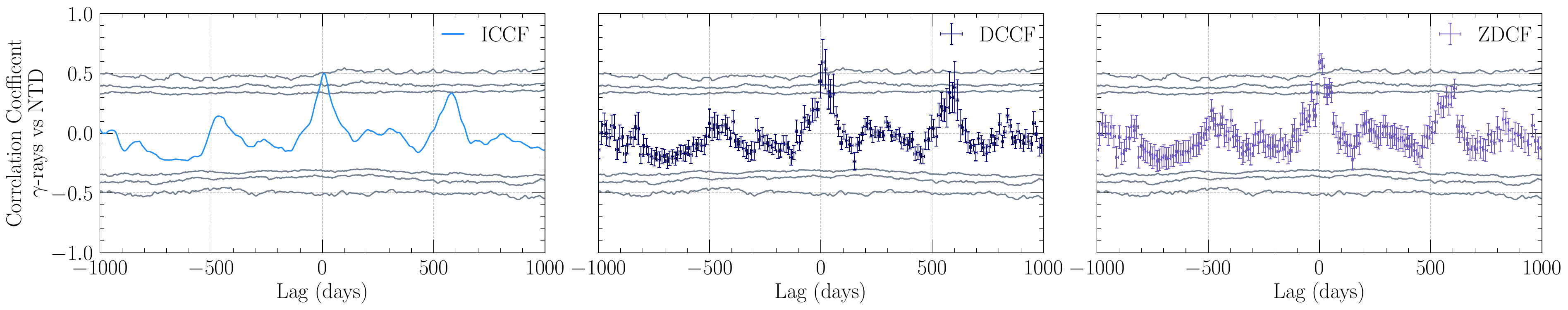}} \\
{\includegraphics[width=\columnwidth]{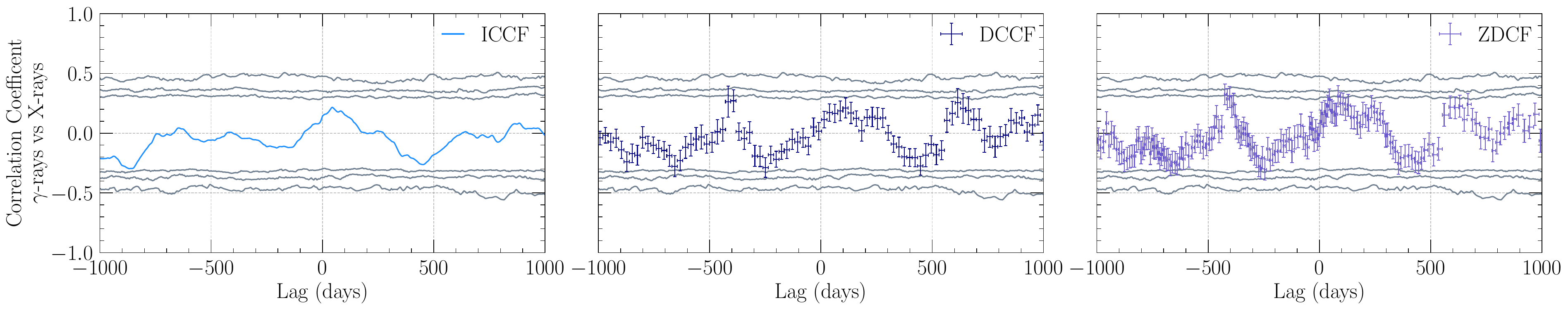}} \\
{\includegraphics[width=\columnwidth]{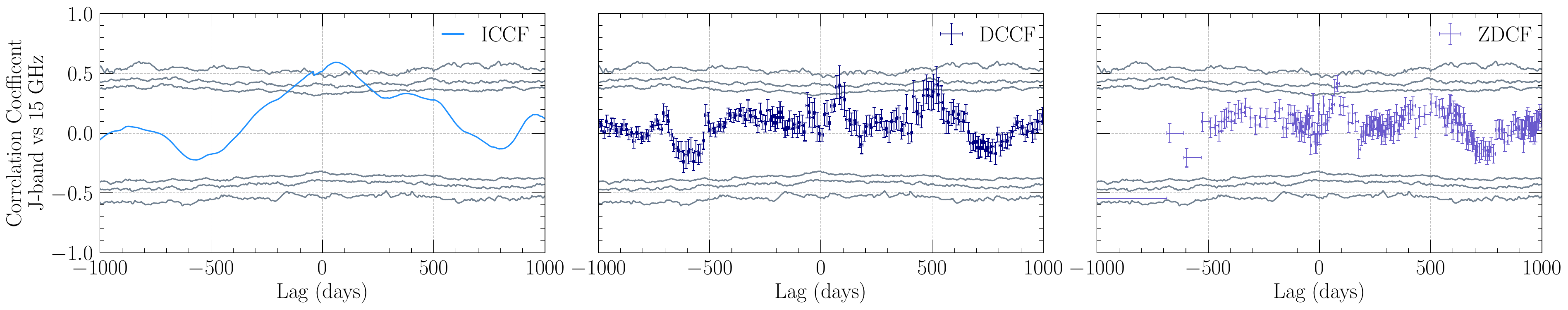}} \\
{\includegraphics[width=\columnwidth]{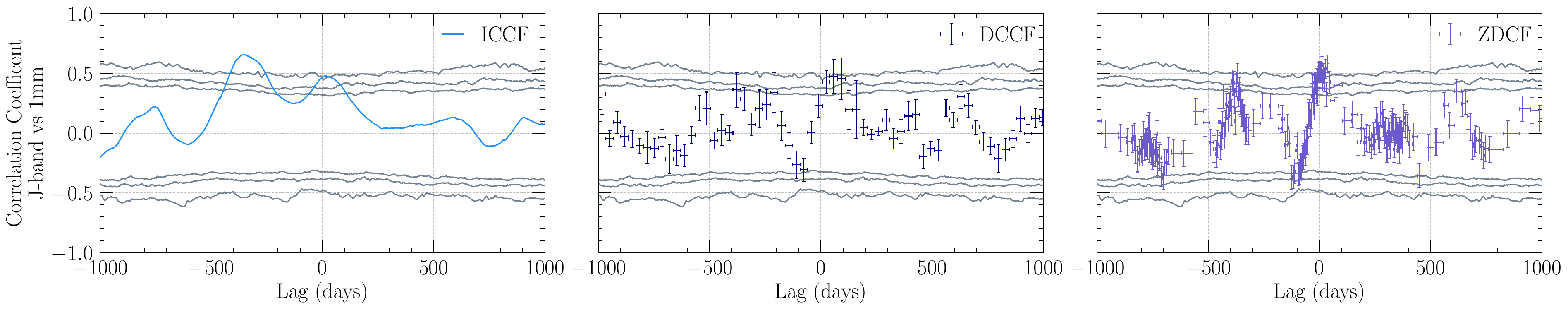}} \\
{\includegraphics[width=\columnwidth]{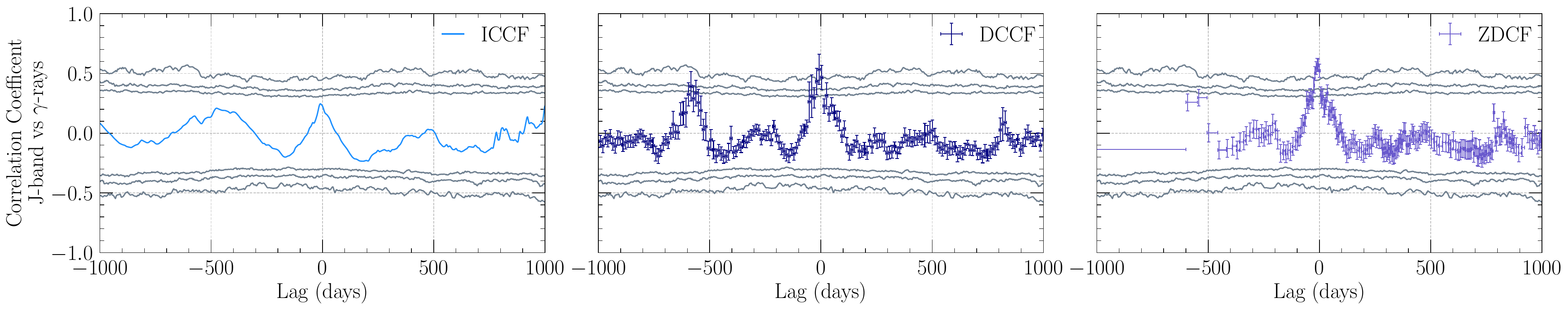}} \\
\end{tabularx}
\caption{Cross-correlation Functions. The lines represent the same as in \autoref{fig:cc1}.}
\label{fig:cc3}
\end{figure}

\begin{figure}[h!]
\begin{tabularx}{1in}{c} 
{\includegraphics[width=\columnwidth]{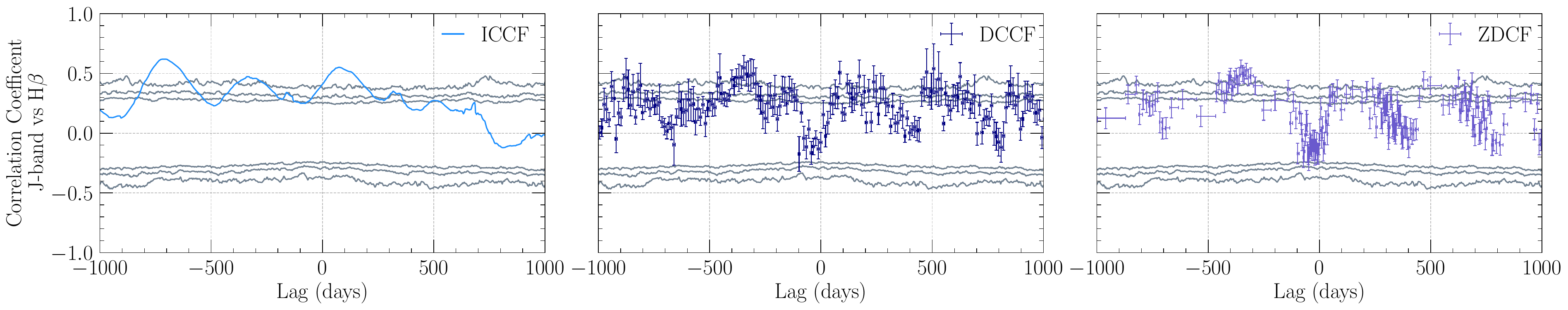}} \\
{\includegraphics[width=\columnwidth]{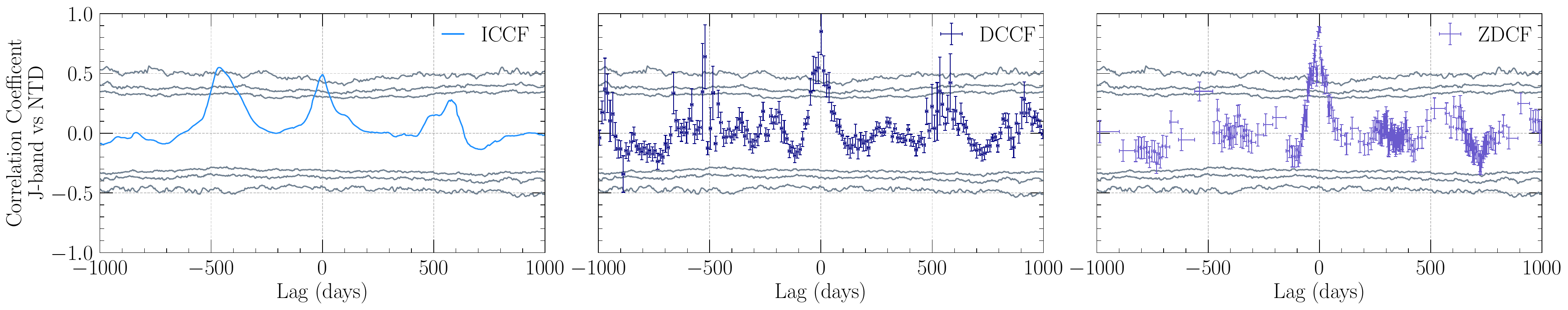}} \\
{\includegraphics[width=\columnwidth]{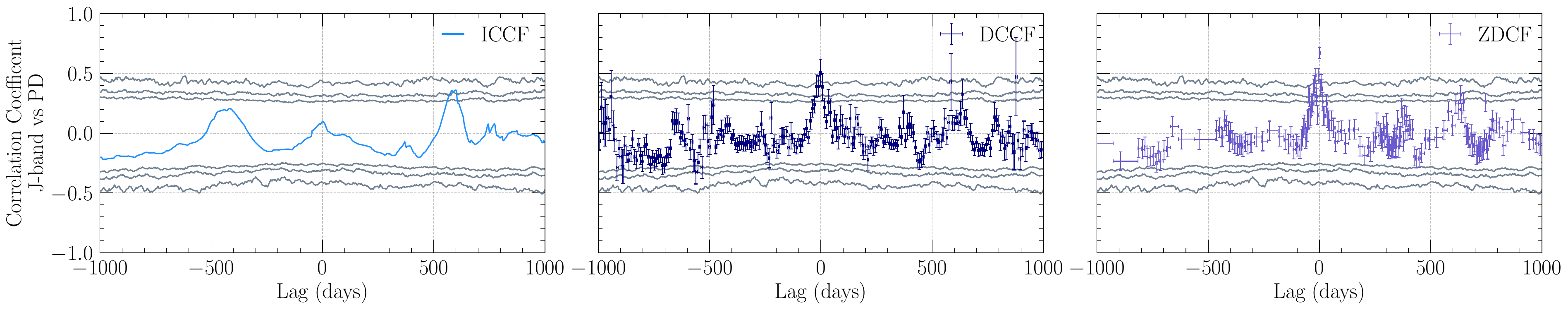}} \\
{\includegraphics[width=\columnwidth]{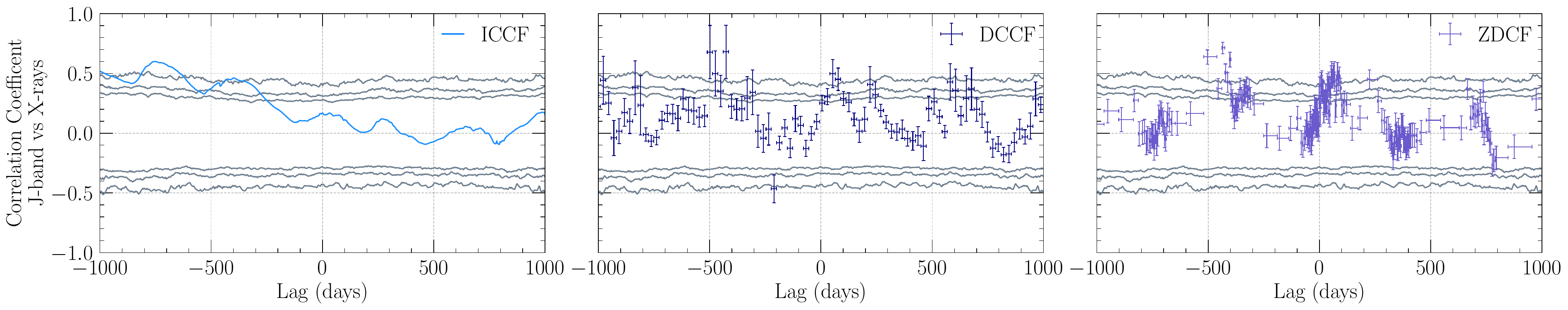}} \\
{\includegraphics[width=\columnwidth]{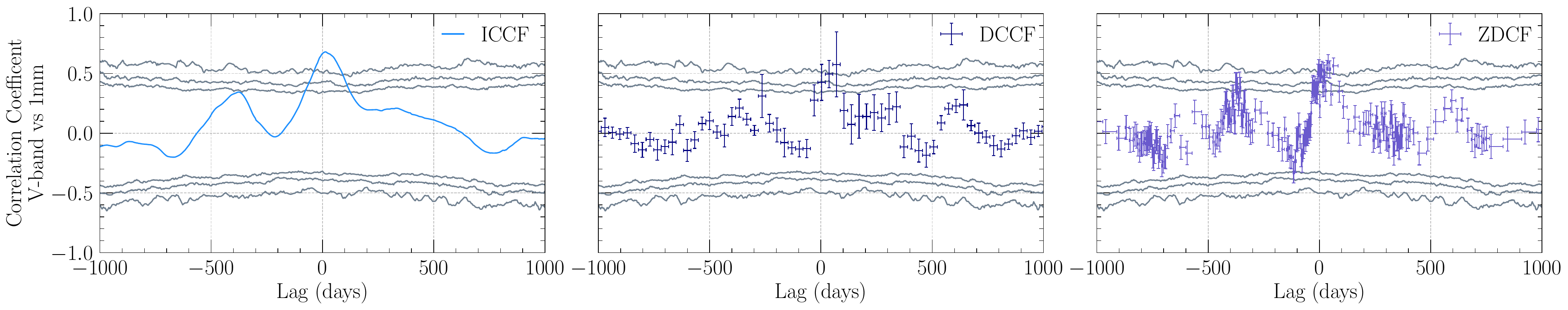}} \\
{\includegraphics[width=\columnwidth]{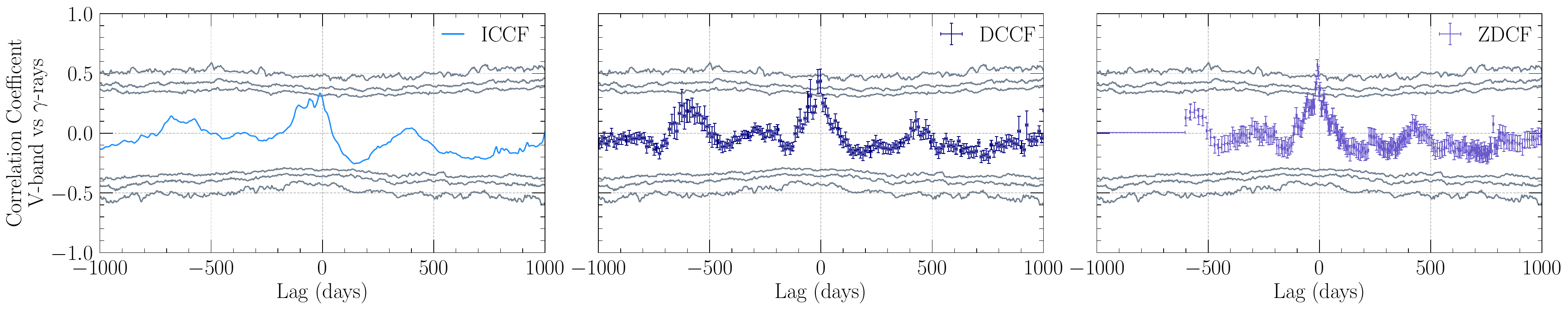}} \\
\end{tabularx}
\caption{Cross-correlation Functions. The lines represent the same as in \autoref{fig:cc1}.}
\label{fig:cc4}
\end{figure}

\begin{figure}[h!]
\begin{tabularx}{1in}{c} 
{\includegraphics[width=\columnwidth]{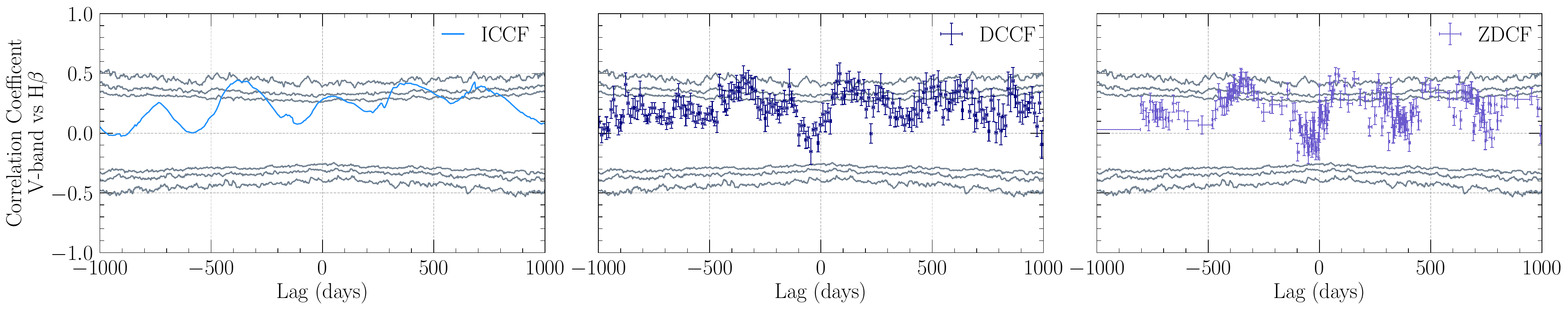}} \\
{\includegraphics[width=\columnwidth]{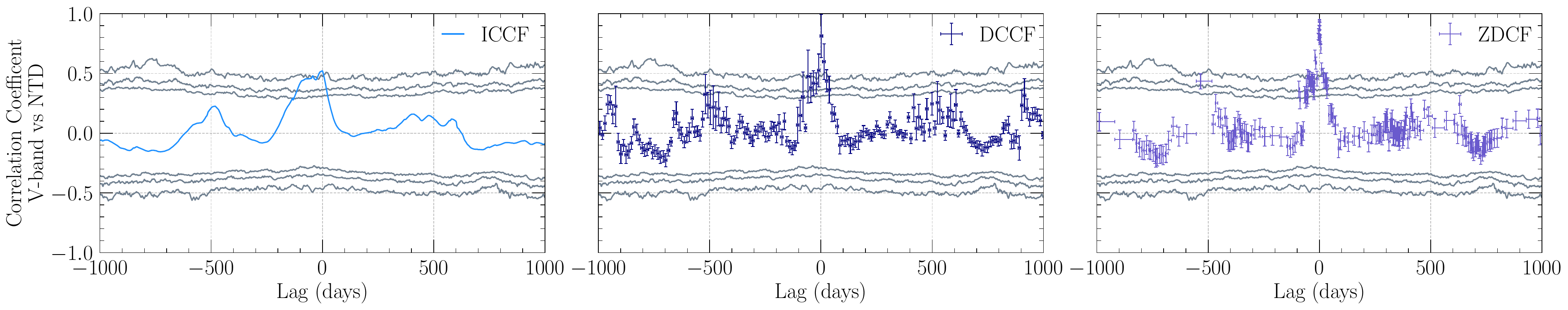}} \\
{\includegraphics[width=\columnwidth]{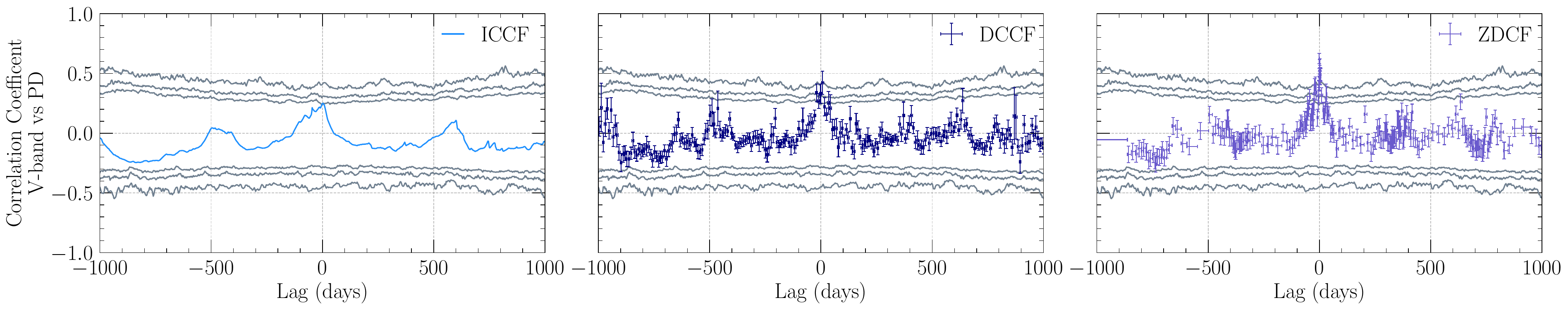}} \\
{\includegraphics[width=\columnwidth]{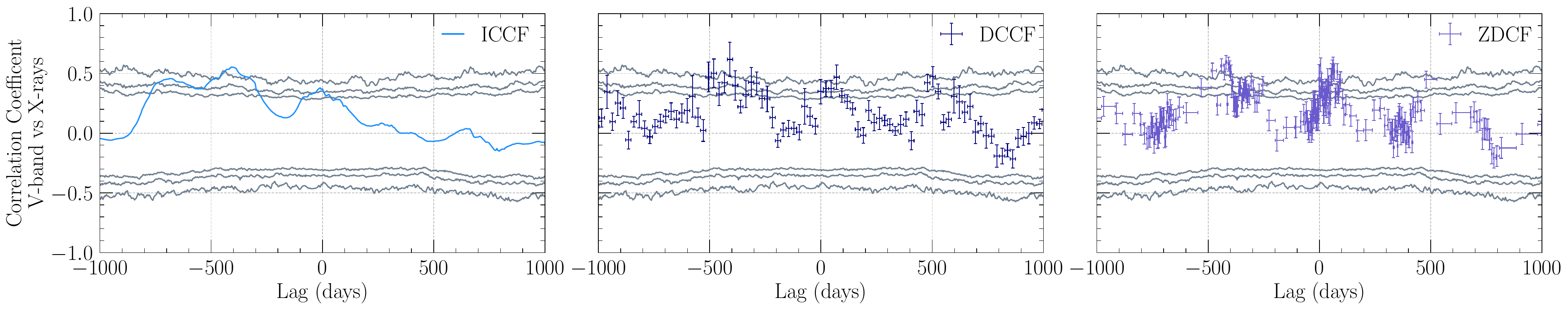}} \\
\end{tabularx}
\caption{Cross-correlation Functions. The lines represent the same as in \autoref{fig:cc1}.}
\label{fig:cc5}
\end{figure}

The series of figures corresponding to the cross-correlation analysis outlined in \autoref{sec:variability} are presented in \autoref{fig:cc1} through \autoref{fig:cc5}. Demonstrating significant correlations that are not spurious are shown.

\end{document}